\documentclass[aps,prx,twocolumn,amsfonts,amsmath,amssymb,footinbib,showpacs,superscriptaddress,nobalancelastpage,floatfix,10pt]{revtex4-2}

\usepackage[english]{babel}
\usepackage[utf8]{inputenc}
\usepackage[T1]{fontenc}
\usepackage{amsmath}
\usepackage{amssymb}
\usepackage{nccmath}
\usepackage{bm}
\usepackage{physics}
\usepackage{braket}
\usepackage{graphicx}
\usepackage[dvipsnames]{xcolor}
\usepackage{hyperref}
\usepackage[all]{hypcap}
\usepackage{cleveref}

\hypersetup{
    colorlinks=true,
    breaklinks=true,
    urlcolor=black,
    linkcolor=black,
    citecolor=NavyBlue,
    bookmarksopen=true,
    pdftoolbar=false,
    pdfmenubar=false,
}

\crefname{equation}{Eq.}{Eqs.}
\Crefname{equation}{Equation}{Equations}
\crefname{figure}{Fig.}{Figs.}
\Crefname{figure}{Figure}{Figures}
\crefname{section}{Sec.}{Secs.}
\Crefname{section}{Section}{Sections}
\crefname{subsection}{Sec.}{Secs.}
\Crefname{subsection}{Section}{Sections}
\crefname{subsubsection}{Sec.}{Secs.}
\Crefname{subsubsection}{Section}{Sections}
\crefname{appendix}{Appendix}{Appendices}
\Crefname{appendix}{Appendix}{Appendices}
\crefname{subappendix}{Appendix}{Appendices}
\crefname{subsubappendix}{Appendix}{Appendices}

\newcommand{\hneg}{\mspace{-1.5mu}}
\newcommand{\Ha}{\bm{H}}
\newcommand{\R}{\bm{R}}
\newcommand{\D}{\bm{D}}
\newcommand{\Id}{\bm{I}}
\newcommand{\hE}{\bm{E}}
\newcommand{\hPi}{\bm{\Pi}}
\newcommand{\ha}{\bm{a}}
\newcommand{\hf}{\bm{f}}
\newcommand{\hx}{\bm{x}}
\newcommand{\bL}{\bm{L}}
\newcommand{\K}{\bm{K}}
\newcommand{\C}{\bm{C}}
\newcommand{\J}{\bm{J}}
\newcommand{\Z}{\bm{Z}}
\newcommand{\U}{\bm{U}}
\newcommand{\hsigma}{\bm{\sigma}}
\newcommand{\hrho}{\bm{\rho}}
\newcommand{\Rm}{\R_m}
\newcommand{\Df}{\D_f}
\newcommand{\Rf}{\R_f}
\newcommand{\CU}{\bm{C\hneg\U}}
\newcommand{\CR}{\bm{C\!R}}
\newcommand{\CD}{\bm{C\!D}}
\newcommand{\ECD}{\bm{EC\!D}}
\newcommand{\FCU}{\bm{F\hneg C U}}
\newcommand{\FCR}{\bm{F\hneg C\!R}}
\newcommand{\FCD}{\bm{F\hneg C\!D}}
\newcommand{\CRR}{\bm{C \! R}_{mf}}
\newcommand{\ftheta}{\bar{\theta}}
\newcommand{\flag}{\zeta}
\newcommand{\ie}{i.e. }

\begin{document}

\title{Protecting bosonic codes from ancilla-induced errors with continuous-variable flags}

\author{Thomas Décultot}
\affiliation{Alice \& Bob, Paris, France}
\affiliation{Laboratoire de Physique de l'\'Ecole Normale Sup\'erieure, Inria, ENS, Mines ParisTech, Universit\'e PSL, Sorbonne Universit\'e, Paris, France}

\author{Ronan Gautier}
\affiliation{Alice \& Bob, Paris, France}

\author{Mazyar Mirrahimi}
\affiliation{Laboratoire de Physique de l'\'Ecole Normale Sup\'erieure, Inria, ENS, Mines ParisTech, Universit\'e PSL, Sorbonne Universit\'e, Paris, France}
\affiliation{RIKEN Center for Quantum Computing (RQC), Wako, Saitama 351-0198, Japan}

\begin{abstract}
Bosonic quantum error correction represents a route towards significant hardware overhead reduction for fault-tolerant quantum computation. However, controlling quantum harmonic oscillators that encode the quantum information comes with its own challenges. In many platforms such as superconducting circuits and trapped ions, the bosonic degree of freedom is controlled by an ancillary qubit whose decay during a qubit-oscillator gate propagates to the oscillator as a continuous error. Such errors generally lie beyond the scope of errors that the bosonic code is capable of correcting.
We introduce the \textit{continuous-variable flag}: an auxiliary oscillator coupled to the ancilla whose phase-space position encodes the continuous error corrupting the storage mode. After the gate, a heterodyne measurement of the flag estimates this continuous-variable error and a subsequent feedback control undoes it up to a bosonic code stabilizer. The scheme requires no new ingredients, as the flag is driven by the same dispersive coupling, displacements and rotations that implement the gates it is supposed to protect. We construct sequences protecting conditional rotations to all orders in ancilla decay with one flag, and conditional displacements to first order with one flag and to all orders with two. Simulations with realistic circuit-QED parameters predict more than three orders of magnitude fewer logical errors for the parity measurement of a four-component cat code, and a finite-energy GKP logical lifetime near its noiseless-ancilla bound. By discretizing continuous-variable errors, the flags introduced in this work are a useful tool for any hybrid oscillator-qubit processor, and open the route towards fault-tolerant simulation and computation with bosonic modes.
\end{abstract}

\date{\today}
\maketitle


\section{Introduction}\label{sec:intro}

Bosonic codes offer a hardware-efficient route to quantum error correction. By encoding a logical qubit directly in the infinite dimensional Hilbert space of a single harmonic oscillator, they achieve error correction with a dramatically reduced hardware footprint~\cite{Terhal_2020,Joshi_2021,Cai_2021,Ma_2021}. Several encodings have been developed along these lines, the most prominent being the Gottesman-Kitaev-Preskill (GKP) grid states~\cite{Gottesman_2001}, the Schr\"odinger cat codes~\cite{Cochrane_1999,Leghtas_2013,Mirrahimi_2014,Leghtas_2015, Lescanne_2020, Berdou_2023,Reglade_2024}, the binomial codes~\cite{Michael_2016}, and the dual-rail encoding~\cite{Teoh_2023,Kubica_2023,Levine_2024,Chou_2024}, each tailored to a different dominant noise process~\cite{Albert_2018}. Over the past decade these ideas have matured into experiments across superconducting circuits, trapped ions~\cite{Fluhmann_2019,deNeeve_2022} and propagating light~\cite{Konno_2024}, with several platforms now demonstrating error corrected logical qubits~\cite{Hu_2019,Campagne_Ibarcq_2020} that reach or exceed the break-even point of quantum error correction~\cite{Ofek_2016,Sivak_2023,Ni_2023,Putterman_2025}. Bosonic quantum error correction has thus become a mature and rapidly advancing pillar of the effort toward fault-tolerant quantum computing.

Manipulating information encoded in a linear oscillator nevertheless requires a source of nonlinearity, which these architectures often supply through an ancillary qubit, typically a superconducting transmon artificial atom under dispersive interaction~\cite{Blais_2004,Schuster_2007,Blais_2021}. This interaction provides the standard toolbox for oscillator control, generating the conditional rotation and conditional displacement gates that lie at the heart of syndrome extraction~\cite{Sun_2014,Ofek_2016}, entangling operations, and universal control of the mode~\cite{Leghtas-PRA-2013,Krastanov_2015,Heeres_2015,Heeres_2017,Eickbusch_2022,Gautier_2023,Diringer_2024}. During such a gate, however, a spontaneous decay of the ancilla propagates to the oscillator as an error whose magnitude depends on when the decay occurred, as detailed in~\cref{subsec:intro_CR,subsec:intro CD}. This makes ancilla decay one of the leading error mechanisms limiting present-day experiments~\cite{Rosenblum_2018,Reinhold_2020,Sivak_2023,Putterman_2025}. Existing mitigation strategies operate largely at the encoding or at the hardware level, either distributing the logical information over several bosonic modes so that multimode encodings are intrinsically more robust to the propagation of ancilla relaxation errors~\cite{Royer_2022}, engineering longer-lived ancillas~\cite{Place_2021,Somoroff_2023}, or exploiting higher qubit levels to render the operation insensitive to relaxation at first order~\cite{Rosenblum_2018,Reinhold_2020,Ma_2020,Elder_2020,Tsunoda_2023}. The latter approach engineers a particular type of coupling between the ancilla and the bosonic mode, such that the mode undergoes the same path despite the decay of the ancilla. It relies on delicate matching conditions, whose complexity limits its scalability to the large-scale systems required for fault-tolerant quantum computation. It also remains fundamentally limited to first-order errors, since higher-order decay cascades restore a continuous, uncorrected error channel.

In this work, we adopt a conceptually different approach: rather than preventing the error imposed on the memory, we make it detectable and correctable. We introduce an auxiliary oscillator mode, the \emph{continuous-variable flag}, coupled to the same ancilla via a dispersive interaction, such that its state becomes correlated with the perturbation induced on the memory by the decay of the ancilla. This information is encoded in the flag’s phase-space position. Measuring the flag then provides an estimate of the perturbation, which can be compensated on the memory through feedback. When the perturbation is sufficiently well resolved, the residual operation approaches either the identity or a stabilizer of the bosonic code, and is therefore harmless: the encoded information is preserved. We apply this framework to the two elementary conditional rotation and conditional displacement gates, constructing explicit flagging sequences, and benchmarking them on the parity measurement of a four-component cat code~\cite{Leghtas_2013, Ofek_2016} and the small-Big-small (sBs) stabilization of a finite energy GKP code~\cite{Royer_2020, Sivak_2023, Lachance-Quirion_2024}.

The paper is organized as follows. We first review, in~\cref{sec:context}, how ancilla decay generates continuous error channels on the memory oscillator. \Cref{sec:cv_flags} introduces the general continuous-variable flag framework, and~\cref{sec:CV_flag_for_CR} and~\cref{sec:CV_flag_for_CD} develop explicit protocols for the conditional rotation and conditional displacement gates respectively, and benchmark them on the two representative error correction tasks. \Cref{sec:limitations} examines the robustness of these protocols against the main experimental imperfections.

\section{Ancilla-induced errors during conditional gates}\label{sec:context}

In this work, we focus on the two most common ancilla-assisted operations in bosonic quantum error-correction architectures, namely the \textit{conditional rotation} and \textit{conditional displacement}. They are defined as
\begin{subequations}
\begin{align}
\CR(\theta)&=\R(\theta) \otimes \ketbra{e} + \R(-\theta) \otimes \ketbra{g}, \label{eq:conditional_rotation}\\
\CD(\beta)&=\D(\beta) \otimes \ketbra{e} + \D(-\beta) \otimes \ketbra{g}, \label{eq:conditional-displacement}
\end{align}
\end{subequations}
where $\ket{g}$ and $\ket{e}$ denote the ground and excited states of a two-level ancilla, and where $\R(\pm\theta)=e^{\pm i\theta \ha^\dagger \ha}$ and $\D(\pm \beta)=e^{\pm(\beta \ha^\dagger - \beta^* \ha)}$ are, respectively, the rotation and displacement operators of an oscillator with annihilation operator $\ha$. These two operations are typically engineered by combining precisely timed drives and rotations with a dispersive coupling between the oscillator and the ancilla, $\Ha_{\text{disp}}/\hbar = -\frac{\chi}{2} \ha^\dagger \ha \hsigma_z$ where $\hsigma_z = \ketbra{e} - \ketbra{g}$. Owing to their simple implementations, these gates constitute fundamental building blocks for entangling operations and qubit-controlled manipulations of bosonic modes. 

\subsection{Ancilla errors during conditional rotations}\label{subsec:intro_CR}

\begin{figure}[t]
\centering
\includegraphics[width=\columnwidth]{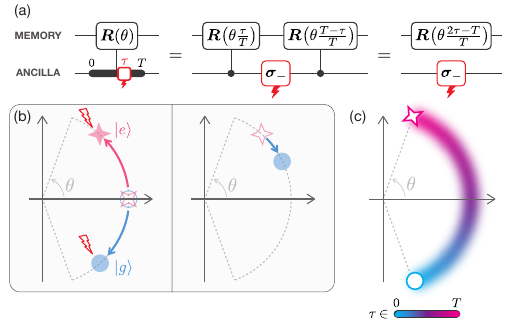}
\caption{
    \textbf{Ancilla-induced continuous error channel for the conditional rotation.}
    (a) Representation of a conditional gate interrupted by an ancilla decay $\hsigma_-$ (red symbol) at time $\tau \in [0,T]$, with $T$ the gate duration. The decay leaves the memory rotated by $\theta \tfrac{2\tau-T}{T}$.
    (b) Phase-space trajectories of the memory components correlated with $\ket{g}$ (blue circle) and $\ket{e}$ (red star), for an initial coherent state. At $\tau$, the $\ket{g}$ component vanishes and the $\ket{e}$ component relaxes to $\ket{g}$, reversing its rotation.
    (c) Final memory position after a jump, with the colorscale indicating $\tau$. The circle and the star mark the no-jump endpoints.
}
\label{fig:CR_cont_err_channel}
\end{figure}
The conditional rotation gate of~\cref{eq:conditional_rotation} is typically implemented with dispersive interactions, which generate the unitary evolution $e^{-i \Ha_{\text{disp}} t / \hbar} = \CR(\chi t / 2)$. During this evolution, ancilla errors, most notably spontaneous decay, can propagate to the oscillator. 

To illustrate this effect, consider a $\CR(\theta)$ gate applied to an initial state $\ket{\psi_0} \ket{+}$, where $\ket{\psi_0}$ is an arbitrary oscillator state, $\ket{+}=(\ket{g}+\ket{e})/\sqrt{2}$, and suppose that the ancilla decays at time $\tau \in [0,T]$, with $T=2\theta/\chi$ the total gate duration. The state resulting from this process is
\begin{equation}\label{eq:traj_decay_during_CR}
\begin{aligned}
    \ket{\Psi_\tau} &\propto \CR\!\left(\theta \tfrac{T-\tau}{T}\right) \hsigma_- \CR\!\left(\theta \tfrac{\tau}{T}\right) \ket{\psi_0}\ket{+} \\
    &\propto \R\!\left(\theta \tfrac{2\tau-T}{T}\right) \ket{\psi_0} \ket{g}
\end{aligned}
\end{equation}
where $\hsigma_- = \ketbra{g}{e}$ is the ancilla decay jump operator. This splitting of the interrupted gate at the jump time, together with the notation used throughout the paper for a conditional gate interrupted by an ancilla decay, is shown in~\cref{fig:CR_cont_err_channel}(a).

\Cref{fig:CR_cont_err_channel}(b) schematically illustrates the corresponding oscillator evolution. For $t<\tau$, the conditional rotation proceeds as intended. At $t=\tau$, the ancilla decay makes the $\ket{g}$ component of the oscillator vanish while the $\ket{e}$ component relaxes to $\ket{g}$. This effectively reverses the direction of rotation of this second component midway through the gate. As a result, the oscillator undergoes an \textit{unconditional} rotation by an angle that depends on the jump time $\tau$. The resulting final rotations of the memory state, for all jump times, are collected in~\cref{fig:CR_cont_err_channel}(c). If, for simplicity, one assumes that the decay time is uniformly distributed over the gate duration, then a random ancilla decay induces a random oscillator rotation with angle in the interval $[-\theta,\theta]$.

More generally, the noisy $\CR(\theta)$ gate in the presence of spontaneous ancilla decay can be described using its Kraus-map representation. This representation is obtained by evolving an arbitrary density matrix under the dispersive Hamiltonian $\Ha_\text{disp}$ together with the loss operator $L = \sqrt{\kappa} \hsigma_-$. It reads $\mathcal{N}_R \circ \mathcal{CR}_\theta$ where $\mathcal{CR}_\theta(\cdot) = \CR(\theta) \cdot \CR^\dagger(\theta)$ denotes the ideal conditional rotation channel, and $\mathcal{N}_R$ is an effective noise channel, derived in~\cref{subsubsec-apdx:CR_error_channel}, that takes the form
\begin{equation} \label{eq:CR_cont_err_channel}
\displaystyle
\mathcal{N}_{R}(\hrho) = \R^\mathrm{nj}_0 \hrho (\R^\mathrm{nj}_0)^\dagger + \int_{-2\theta}^{0}  d\phi \; F_{ge}(\phi) \, \R_\phi^{ge} \hrho (\R_\phi^{ge})^\dagger
\end{equation}
with Kraus operators
\begin{equation} \label{eq:CR_cont_err_channel_Kraus_Op}
\begin{aligned}
\R^\mathrm{nj}_\phi &= \R(\phi)\otimes\left(\ketbra{g}+e^{-\frac{\kappa T}{2}}\ketbra{e}\right) \\
\R^{ge}_\phi &=  \R(\phi) \otimes \ketbra{g}{e}, \\
\end{aligned}
\end{equation}
and with the weight function 
\begin{equation}
\label{eq:weight_function_uncorr_rot}
F_{ge}(\phi)=\frac{\kappa T}{2\theta}e^{-\kappa T\left(1+\frac{\phi}{2\theta}\right)}
\end{equation}
almost uniform over $[-2\theta,0]$. The first term of the noise channel~\cref{eq:CR_cont_err_channel} describes the no-jump trajectory, while the second one accounts for trajectories with an ancilla decay. Crucially, different jump times $\tau$ correspond to different final oscillator rotations. We therefore refer to $\mathcal{N}_R$ as a \textit{continuous error channel}: the errors induced on the oscillator are continuous rotations of the oscillator's phase space.

In bosonic quantum error correction with rotation-symmetric codes~\cite{Grimsmo_2020}, including cat codes~\cite{Leghtas_2013} and binomial codes~\cite{Michael_2016}, error syndromes are commonly extracted through repeated parity measurements implemented with $\CR(\pi/2)$ gates~\cite{Sun_2014,Ofek_2016}. In this process, the above continuous and random oscillator rotations can take values beyond what is correctable by such codes. Indeed, these undesired ancilla-induced errors represent the main limitation in recent experimental performances~\cite{Sun_2014,Ofek_2016,Hu_2019,Gertler-Nature-2021,Shirol-PRX-2026}. 

The purpose of the continuous-variable flag introduced in this work is to transform the noise channel $\mathcal{N}_R$ of~\cref{eq:CR_cont_err_channel} into a new noise channel of the form
\begin{multline} \label{eq:CR_cont_err_channel_fl}
\tilde{\mathcal{N}}_R(\hrho) = \int d\phi \;  \tilde F_{\mathrm{nj}}(\phi) \, \R_\phi^{\mathrm{nj}} \hrho (\R_\phi^{\mathrm{nj}})^\dagger \\
+ \int d\phi \;  \tilde F_{ge}(\phi) \, \R_\phi^{ge} \hrho (\R_\phi^{ge})^\dagger
\end{multline}
where the non-negative weight functions $\tilde F_{\mathrm{nj}}$ and $\tilde F_{ge}$ replace, respectively, the Dirac peak at $\phi=0$ and the almost uniform density $F_{ge}$ of the bare gate. The no-jump weight $\tilde F_{\mathrm{nj}}$, which carries almost all of the probability, remains concentrated around $\phi=0$. The jump weight $\tilde F_{ge}$ is no longer spread over the whole interval $[-2\theta,0]$, but instead splits into two peaks of nearly equal weight located at its endpoints, $\phi=0$ and $\phi=-2\theta$: decays occurring in the second half of the gate are undone, while those occurring in the first half are completed into a full rotation by $-2\theta$. The width of these peaks shrinks as the flag amplitude increases. We refer to this protocol as a \textit{flag-mediated discretization} of the continuous error channel.

Two remarks are in order. First, this is almost as good as one can do to remove the ancilla-induced errors. Indeed, one can note that, even in the case of the application of a perfect conditional rotation gate $\CR(\theta)$, a decay of the qubit just before the operation would propagate to the bosonic mode as an additional rotation of angle $-2\theta$. More precisely, a perfect $\CR(\theta)$ gate preceded by a qubit decay can be seen as a perfect $\CR(\theta)$ followed by a qubit decay and a bosonic mode rotation of angle $-2\theta$. Second, for the purpose of QEC with rotation-symmetric codes this modification of the ancilla induced noise is enough to ensure the fault-tolerance. Indeed, in this picture the operator $\R(-2\theta)$ represents a stabilizer of the bosonic code and therefore acts as identity on the code subspace. Therefore, the mere effect of the noise channel are the rotations at a small random angle which remain correctable by the rotation-symmetric code.

\subsection{Ancilla errors during conditional displacements}\label{subsec:intro CD}

\begin{figure}[t]
\centering
\includegraphics[width=\columnwidth]{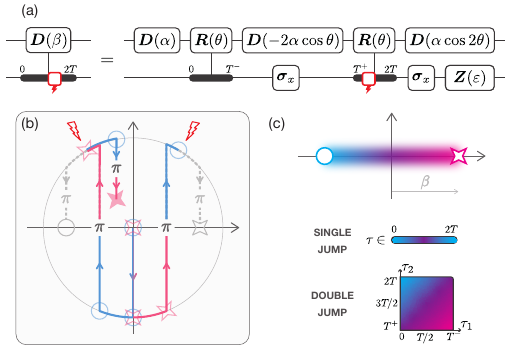}
\caption{
    \textbf{Ancilla-induced continuous error channel for the conditional displacement.}
    (a) ECD implementation of $\CD(\beta)$, with $\CR(\theta)$ gates of duration $T$ separated by a qubit echo $\hsigma_x$ and interleaved with instantaneous memory displacements. For readability, the $\lambda$ subscripts are dropped in this and the following circuit diagrams, with $\alpha,\theta,\varepsilon$ standing for $\alpha_\lambda,\theta_\lambda,\varepsilon_\lambda$ of~\cref{eq:ECD_def,eq:link_CD_and_ECD}. In this example, the qubit decays at $\tau \in (T,2T]$.
    (b) Memory trajectories for an initial vacuum state under the ECD sequence of panel~(a), with the qubit decay at $\tau$ (solid lines) and without error (grey dashed lines). Here, $\pi$ labels the echo that interchanges the two components.
    (c) Final memory positions in the large intermediate displacement  limit, $\lambda \gg 1$, for which the error is a pure displacement along $\beta$. Top, single jump at $\tau \in [0,2T]$. Bottom, double jump at $(\tau_1,\tau_2) \in [0,T^-]\times[T^+,2T]$.
}
\label{fig:CD_cont_err_channel}
\end{figure}

A similar continuous error channel arises for the conditional displacement gate of~\cref{eq:conditional-displacement}. This gate is typically realized via the Echoed Conditional Displacement (ECD) protocol~\cite{Campagne_Ibarcq_2020,Eickbusch_2022}, which combines conditional rotations with intermediate oscillator displacements and an ancilla $\pi$-pulse $\hsigma_x$ acting as an echo, as illustrated in~\cref{fig:CD_cont_err_channel}(a). Under the standard approximation that displacements and ancilla rotations are instantaneous compared to the slower conditional rotations, the ECD sequence reads
\begin{equation}\label{eq:ECD_def}
\begin{aligned}
\ECD(\alpha,\theta) &= \D(\alpha\cos 2\theta)\,\CR(\theta)\,\D(-\alpha\cos\theta) \\
&\quad\; \times \hsigma_x\,\D(-\alpha\cos\theta)\,\CR(\theta)\,\D(\alpha).
\end{aligned}
\end{equation}
By commuting the operators, one verifies that this sequence realizes the $\CD(\beta)$ gate up to ancilla rotations, provided $\beta = i\alpha \sin 2\theta$. For a given target displacement $\beta$, this relation is satisfied by a family of ECD parameters $(\alpha_\lambda, \theta_\lambda)$, indexed by $\lambda \in \mathbb{R}$ with $|\lambda| \ge 1$, and defined by $\alpha_\lambda = -i\lambda\beta$ and $\theta_\lambda = \frac{1}{2}\arcsin(1/\lambda)$. The resulting identity
\begin{equation}\label{eq:link_CD_and_ECD}
    \CD(\beta)=\Z_q(\varepsilon_\lambda) \hsigma_x\ECD(\alpha_\lambda,\theta_\lambda),
\end{equation}
where $\Z_q(\varepsilon) = e^{-i\varepsilon\hsigma_z/2}$ denotes an ancilla rotation about its $Z$ axis by the residual angle $\varepsilon_\lambda=2(\lambda - \sqrt{\lambda^2-1})|\beta|^2$, reveals that $\lambda$ controls the trade-off between the intermediate displacement amplitude $|\alpha_\lambda| = |\lambda\beta|$ and the conditional rotation angle $\theta_\lambda$. In particular, increasing $\lambda$ reduces the conditional rotation angle and thus shortens the total gate duration to below the $2\pi/\chi$ timescale defined by the dispersive coupling, enabling rapid universal control of the oscillator~\cite{Eickbusch_2022}.

As with the conditional rotation, ancilla decay during the conditional rotations of the ECD propagates errors to the oscillator. Unlike the $\CR(\theta)$ case, the presence of the $\pi$-pulse, which flips the ancilla state midway through the sequence, gives rise to several distinct decay configurations. The ancilla may decay during the first conditional rotation at time $\tau \in [0,T]$ (before the echo), during the second at time $\tau \in [T,2T]$ (after the echo), or undergo two successive decay events at times $(\tau_1, \tau_2)$, one before and one after the echo. \Cref{fig:CD_cont_err_channel}(b) illustrates the phase-space trajectory of the memory components for a decay after the echo. In each case, the resulting oscillator state acquires a displacement whose amplitude depends continuously on the jump time(s). \Cref{fig:CD_cont_err_channel}(c) collects the final displacements of the memory state in the limit of large intermediate displacements ($\lambda \gg 1$), for single-jump trajectories as a function of $\tau$ (top) and for double-jump trajectories as a function of $(\tau_1,\tau_2)$ (bottom).

Substituting the Kraus map of the noisy $\CR(\theta)$ into~\cref{eq:ECD_def}, one finds that the noisy $\CD(\beta)$ gate takes the form $\mathcal{N}_D \circ \mathcal{CD}_\beta$, where $\mathcal{CD}_\beta(\cdot) = \CD(\beta) \cdot \CD^\dagger(\beta)$ denotes the ideal conditional displacement channel. The continuous error channel $\mathcal{N}_D$, derived in~\cref{subsec-apdx:error_channels}, induces oscillator displacements of order $|\beta|$ in phase space, accompanied by residual rotations that depend on $\lambda$. In the limit of large intermediate displacements ($\lambda \gg 1$), these residual rotations become negligible and the channel reduces to 

\begin{equation} \label{eq:CD_cont_err_channel}
\begin{aligned}
&\mathcal{N}_{D}(\hrho) = F_{\mathrm{nj}}~\D^\mathrm{nj}_0 \hrho (\D^\mathrm{nj}_0)^\dagger \\&
~~+ \int_{-2}^{-1} ds \; F_{ge}(s) \, \D^{ge}_{s\beta} \hrho (\D^{ge}_{s\beta})^\dagger\\
&~~
+ \int_{0}^{1} ds \;  F_{eg}(s) \, \D^{eg}_{s\beta} \hrho (\D^{eg}_{s\beta})^\dagger\\
&+ \int_{0}^{1} ds_1\int_{1}^2 ds_2 \; F_{ee}(s_1+s_2) \, \D^{ee}_{(s_1-s_2)\beta} \hrho (\D^{ee}_{(s_1-s_2)\beta})^\dagger
\end{aligned}
\end{equation}

with Kraus operators and associate weight functions

\begin{equation} \label{eq:CD_cont_err_channel_Kraus_Ops}
\begin{array}{rll}
\D^\mathrm{nj}_{\alpha}&=\D(\alpha), & F_\mathrm{nj} = e^{-\kappa T}\\
\D^{ge}_{\alpha}&=\D(\alpha)\otimes\ketbra{g}{e},& F_{ge}(s)=\kappa T e^{-\kappa T(s+3)}\\
\D^{eg}_{\alpha}&=\D(\alpha)\otimes\ketbra{e}{g},& F_{eg}(s)=\kappa T e^{-\kappa T(1-s)}\\
\D^{ee}_{\alpha}&=\D(\alpha)\otimes\ketbra{e}{e},& F_{ee}(s)=(\kappa T)^2 e^{-\kappa T(s-1)}.
\end{array}
\end{equation}
Here, $T=2\theta / \chi$ denotes the duration of a single $\CR(\theta)$ gate, so that the full ECD sequence lasts $2T$. The four terms of~\cref{eq:CD_cont_err_channel} correspond to the possible decay configurations. The first term describes the dominant no-jump trajectory, which leaves the oscillator undisplaced. The next two terms, of order $\kappa T$, describe a single decay occurring either before or after the echo pulse. A decay before the echo selects the $\ket{e}$ branch and displaces the oscillator by $s\beta$ with $s$ almost uniformly distributed over $[-2,-1]$ (weight $F_{ge}$), whereas a decay after the echo gives $s$ almost uniformly distributed over $[0,1]$ (weight $F_{eg}$) on the $\ket{g}$ branch. The last term, of order $(\kappa T)^2$, describes the rarer trajectories with one decay on each side of the echo, whose displacements are spread over the whole interval $s\in[-2,0]$ on the $\ket{e}$ branch.

These random displacements are the translational counterpart of the rotation errors of~\cref{subsec:intro_CR}, and limit the bosonic error correction performance for the class of translation-symmetric GKP codes in much the same way. Indeed, in GKP codes, the $\CD(\beta)$ gate is the key building block for syndrome extraction and entangling operations~\cite{Terhal_2016, Royer_2020, Campagne_Ibarcq_2020, Sivak_2023}. An ancilla decay during such a gate can displace the encoded state beyond the correctable region of the GKP lattice, resulting in a logical error.

Similarly to the case of rotation-symmetric codes, the performance of recent experimental implementations of QEC with GKP codes in superconducting circuits has been limited by such ancilla-induced random and continuous displacements~\cite{Campagne_Ibarcq_2020,Sivak_2023,Lachance-Quirion_2024}. We show in this paper that the continuous-variable flag approach allows one to discretize the continuous-displacement error channel $\mathcal{N}_D$ into a new noise channel of the form
\begin{equation} \label{eq:CD_cont_err_channel_fl}
\begin{aligned}
\tilde{\mathcal{N}}_{D}(\hrho) &= \int ds \, \tilde{F}_{\mathrm{nj}}(s) \; \D_{s\beta}^{\mathrm{nj}} \, \hrho \, (\D_{s\beta}^{\mathrm{nj}})^\dagger \\
&+ \int ds \, \tilde{F}_{ge}(s) \; \D_{s\beta}^{ge} \hrho \, (\D_{s\beta}^{ge})^\dagger \\
&+ \int ds \, \tilde{F}_{eg}(s) \; \D_{s\beta}^{eg} \hrho \, (\D_{s\beta}^{eg})^\dagger \\
&+
\int ds \, \tilde{F}_{ee}(s) \;
\D_{s\beta}^{ee} \hrho \, (\D_{s\beta}^{ee})^\dagger
\end{aligned}
\end{equation}
where the weight functions $\tilde F_{\mathrm{nj}}(s)$, $\tilde F_{ge}(s)$, $\tilde F_{eg}(s)$, $\tilde F_{ee}(s)$ concentrate the displacement errors around the multiple $s=0$ and $s=-2$ of the targeted displacement $\beta$. Noting, once again that in the context of the translation-symmetric codes, $\D(-2\beta)$ represents a stabilizer, such displacements are not harmful and remain correctable by the code. 

More specifically, we show first that using a single flag mode, one is able to concentrate the weights at first order in ancilla decay, with  $\tilde F_{\mathrm{nj}}(s)$ and $\tilde F_{eg}(s)$ concentrated around $s=0$, $\tilde F_{ge}(s)$ concentrated around $s=-2$, while the weight related to second order trajectories $\tilde F_{ee}(s)$ remains spread over $s\in[-2,0]$. Then, we show that introducing a second flag mode, one is able to detect errors at all order, and further concentrate $\tilde F_{ee}(s)$ around $s=-2$ and $s=0$.

\section{Continuous-variable flags: general perspective}\label{sec:cv_flags}

\begin{figure*}[t]
\centering
\includegraphics[width=\textwidth]{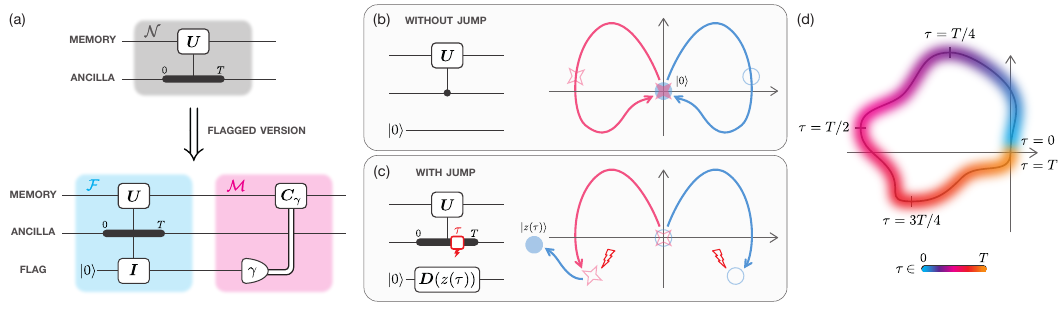}
\caption{
    \textbf{Continuous-variable flag.}
    (a) General protocol, replacing the noise channel $\mathcal{N}$ by the effective channel $\tilde{\mathcal{N}} \equiv\mathcal{M} \circ \mathcal{F}$. The flagging protocol $\mathcal{F}$ implements $\FCU(\flag)$, in which the flag, initialized in vacuum, undergoes the identity $\Id$ in the absence of ancilla decay. The measure-and-feedback protocol $\mathcal{M}$ reads out the flag by heterodyne detection and applies the correction $\C_\gamma$ to the memory, with $\gamma$ the measurement outcome.
    (b) Flag trajectories without a jump: the $\ket{g}$ and $\ket{e}$ components recombine in vacuum, as required by~\cref{eq:flag_disentangling_condition}.
    (c) With a decay at $\tau$, the flag ends in the coherent state $\ket{z(\tau)}$.
    (d) Position distribution of $z(\tau)$ for single jumps, with the colorscale indicating $\tau \in [0,T]$. Under the correlation condition of~\cref{eq:correlation_condition}, the flag position faithfully records the memory error.
}
\label{fig:protocol}
\end{figure*}

In a continuous error channel, knowing that a quantum jump occurred is not enough to correct the resulting error: one must also know when it occurred. Consequently, flag-qubit schemes used in discrete-variable codes~\cite{Yoder_2017, Chao_2018, Chamberland_2018, Chao_2020, Chamberland_2020}, which are designed to detect the occurrence of a jump, can at best provide erasure information~\cite{Grassl_1997, Bennett_1997, Kubica_2023, Teoh_2023, Levine_2024, Chou_2024}.

In this work, we introduce the idea of the \textit{continuous-variable flag}: an auxiliary oscillator mode whose phase-space position, after the gate, encodes not only whether a jump occurred, but also which error it induced on the memory. To implement this, we extend the setup of~\cref{sec:context} by adding a second oscillator, the flag mode, dispersively coupled to the ancilla. While we focus here on a single memory mode and a single flag mode for clarity, the construction readily generalizes to multiple modes. The dispersive Hamiltonian reads
\begin{equation} \label{eq:H_disp_with_flag}
    \Ha_{\text{disp}}/\hbar = - \frac{\chi_m}{2} \ha^\dagger \ha \hsigma_z - \frac{\chi_f}{2} \hf^\dagger \hf \hsigma_z,
\end{equation}
where $\ha$ and $\hf$ denote the annihilation operators of the memory and flag modes respectively, and generates over a duration $T$ the joint conditional rotation
\begin{equation} \label{eq:CRR_def}
    \CRR(\theta,\ftheta) \equiv e^{-i \Ha_{\text{disp}} T / \hbar} = \CR_m(\theta)\,\CR_f(\ftheta),
\end{equation}
where $\theta = \chi_m T / 2$ and $\ftheta = \chi_f T / 2$, where $\CR_m$ and $\CR_f$ denote conditional rotations of the memory and flag alone, so that the memory is rotated by $\pm\theta$ and the flag by $\pm\ftheta$ according to the ancilla state. The full system Hamiltonian $\Ha$ also contains drives on the two oscillators and on the ancilla, generating displacements and qubit rotations. We treat them as instantaneous compared to the dispersive dynamics so that ancilla errors only occur during conditional rotations. This approximation is well justified in practice since single-mode and qubit operations last 5--50\,ns against 200--1000\,ns for conditional rotations~\cite{Eickbusch_2022}. Nevertheless, as discussed in~\cref{subsec:finite_duration}, the results of this paper hold without this approximation.

The protection scheme targets an arbitrary ancilla-controlled operation $\CU$, such as $\CR(\theta)$ or $\CD(\beta)$, with ideal channel $\mathcal{CU}(\cdot) = \CU \cdot \CU^\dagger$ and noisy channel $\mathcal{N} \circ\,\mathcal{CU}$, where $\mathcal{N}$ is the continuous error channel of~\cref{sec:context}. As sketched in~\cref{fig:protocol}(a), it proceeds in two stages. The \textit{flagging protocol} $\mathcal{F}$ couples the flag to the system during the implementation of $\CU$, replacing $\mathcal{N} \circ\,\mathcal{CU}$ with a flagged evolution $\mathcal{F} \circ\,\mathcal{CU}$ in which the error is recorded in the phase-space position of the flag. The \textit{measure-and-feedback protocol} $\mathcal{M}$ then reads out the flag by heterodyne detection and applies a conditional correction to the memory. The total channel $\mathcal{M} \circ \mathcal{F} \circ \mathcal{CU}$ replaces $\mathcal{N}$ with an effective channel $\tilde{\mathcal{N}}$ close to a discrete one, in the sense of~\cref{sec:context}.

\textit{Flagging protocol.}---The flag is prepared in a coherent state, coupled to the ancilla throughout the sequence implementing $\CU$, and operated so as to disentangle from the memory and the ancilla at the end of the sequence. Up to initial and final displacements, we may assume that the flag starts and ends in the vacuum state $\ket{0}$, so that the flagged operation $\FCU$ acts on it as an identity and satisfies
\begin{equation} \label{eq:flag_disentangling_condition}
    \FCU \left(\bullet \otimes \ket{0}\right) = \CU\!\left(\bullet \right) \otimes \ket{0}.
\end{equation}

\Cref{fig:protocol}(b) illustrates the phase-space mechanism behind this condition: the flag components correlated with $\ket{g}$ and $\ket{e}$ follow distinct trajectories and recombine in vacuum, so that the flag carries no information in the absence of decay. An ancilla decay at time $\tau \in [0,T]$ instead corrupts both the operation $\CU$ and the identity on the flag, which ends in a coherent state $\ket{z(\tau)}$ away from vacuum, as illustrated in~\cref{fig:protocol}(c).

More generally, the dynamics under $\FCU$ in the presence of ancilla decay at rate $\kappa$ follows a Lindblad master equation whose solution admits an expansion at increasing orders of $\kappa$. The flagged channel $\mathcal{F} = \sum_{p} \mathcal{F}^{(p)}$ is a statistical mixture over trajectories, organized by their number $p$ of decay events. In the regime of rare errors, $\kappa T \ll 1$, the contribution of $p$ jumps is of order $(\kappa T)^p$, and to first order
\begin{equation} \label{eq:flagged_channel}
\begin{aligned}
    \mathcal{F}(\hrho) &= \K_{\mathrm{nj}} \hrho \K_{\mathrm{nj}}^\dagger \otimes \ketbra{0} \\
    &\quad + \int_{0}^{T} \! \kappa d\tau \; \K_\tau \hrho \K_\tau^\dagger \otimes \ketbra{z(\tau)},
\end{aligned}
\end{equation}
where $\K_{\mathrm{nj}}$ and $\K_\tau$ are the memory-ancilla Kraus operators of the no-jump trajectory and of a single decay at time $\tau$. In the sequences considered in this paper, they coincide, up to ancilla rotations, with those of the bare channel $\mathcal{N}$ of~\cref{sec:context}: the flag records the error without altering it. Trajectories with several decays at times $\vec{\tau} = (\tau_1, \dots, \tau_p)$ contribute similarly, with Kraus operators $\K_{\vec{\tau}}$ and flag positions $z(\vec{\tau})$.

The flag is useful to the extent that its final position tells which error the memory has suffered. We therefore require the memory error to be \textit{perfectly correlated} with the flag position: any two trajectories, whatever their numbers of jumps, that leave the flag at the same position must induce the same memory error,
\begin{equation} \label{eq:correlation_condition}
    z(\vec{\tau}) = z(\vec{\tau}\,') \;\Longrightarrow\; \bm{E}_{\vec{\tau}} \simeq \bm{E}_{\vec{\tau}\,'},
\end{equation}
where the no-jump trajectory is included with $z = 0$ and $\bm{E} = \Id$, and where $\simeq$ denotes equality up to an operation that is harmless for the task at hand, such as a stabilizer of the bosonic code. The memory error is then a function $\bm{E}(z)$ of the flag position alone, and knowing $z$ suffices to undo it. \Cref{fig:protocol}(d) shows an example illustration of the flag positions $z(\tau)$ for single-jump trajectories. This condition must hold jointly across trajectories with different numbers of jumps. We say that the flag provides a protection up to order $n$, if the above perfect correlation condition is satisfied for trajectories up to $n$ jumps. Several flag modes may be used to lift  degeneracies, the flag position becoming a vector $(z_1(\vec{\tau}), \dots, z_d(\vec{\tau})) \in \mathbb{C}^d$.

\textit{Measure-and-feedback protocol.}---If the flag is ignored, tracing it out returns the bare channel, $\mathcal{N}(\hrho) = \Tr_{\hf}[\mathcal{F}(\hrho)]$. Measuring it instead conditions the noise on the outcome. Since the error is recorded in the position of a coherent state, we use heterodyne detection, described at unit efficiency by the POVM $\{\hPi(\gamma) = \frac{1}{\pi} \ketbra{\gamma}\}_{\gamma \in \mathbb{C}}$; for a flag in the coherent state $\ket{z}$, the outcome follows the Gaussian distribution $p(\gamma|z) = \frac{1}{\pi} e^{-|\gamma-z|^2}$. Given an outcome $\gamma$, obtained with probability density $p(\gamma) = \Tr[\hPi(\gamma)\mathcal{F}(\hrho)]$, the noise channel is updated to $\mathcal{N}_{|\gamma} \circ \mathcal{CU}$, with
\begin{equation} \label{eq:conditional-noise-channel}
    \mathcal{N}_{|\gamma} (\hrho)= \frac{1}{p(\gamma)}\Tr_{\hf}[ \hPi(\gamma)\mathcal{F}(\hrho)],
\end{equation}
in which each trajectory of~\cref{eq:flagged_channel} is reweighted by the likelihood $p(\gamma|z)/p(\gamma)$ of its flag position given the outcome. Under the correlation condition, and for flag positions well separated compared to the width of $p(\gamma|z)$, this conditional channel reduces to a single memory error $\bm{E}(\gamma)$, which the correction $\C_\gamma = \bm{E}(\gamma)^\dagger$ undoes. At finite flag amplitude, neighboring positions overlap, in particular near the vacuum where the dominant no-jump trajectory sits, and the outcome only partially discriminates the trajectories.

The correction $\C_\gamma$ then depends on the operation being protected, on the relevant input states and on the figure of merit. For bosonic codes, the objective is to minimize the probability of a logical error, and we choose
\begin{equation} \label{eq:optimal_correction}
    \C_\gamma = \underset{\C \text{ unitary}}{\operatorname{argmin}}~~\max_{\hrho_0\in A} J[\C\mathcal{N}_{|\gamma}(\hrho_0) \C^\dagger],
\end{equation}
where $J$ is a cost functional, typically the probability that the corrected state lies outside the code space, $A$ is the set of relevant input states, typically the code space with the ancilla in its initial state, and the $\operatorname{argmin}$ runs over the physically realizable corrections, memory rotations or displacements in what follows. Averaging over outcomes, the full protocol reads
\begin{equation} \label{eq:full_protocol}
\tilde{\mathcal{N}}(\hrho) \equiv\mathcal{M} \circ \mathcal{F}(\hrho) = \int d^2\gamma \; p(\gamma) \, \C_\gamma \mathcal{N}_{|\gamma}(\hrho) \C_\gamma^\dagger .
\end{equation}
As shown in~\cref{sec:CV_flag_for_CR,sec:CV_flag_for_CD} for the conditional rotation and the conditional displacement, this yields effective noise channels $\tilde{\mathcal{N}}_R$ and $\tilde{\mathcal{N}}_D$ concentrated on discrete rotation and displacement errors, as announced in~\cref{sec:context}.

\section{Conditional rotation gate} \label{sec:CV_flag_for_CR}

\begin{figure*}[t]
\centering
\includegraphics[width=\textwidth]{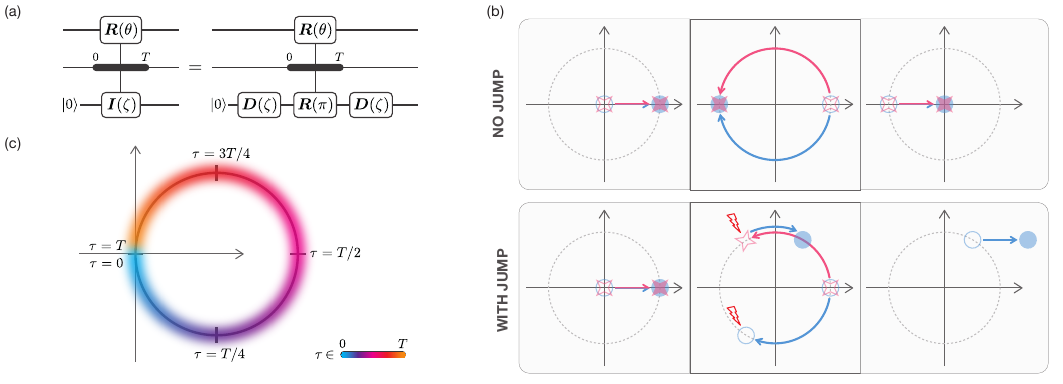}
\caption{
    \textbf{Flagging sequence for the $\CR(\theta)$ gate under $\chi$-matching.}
    (a) Flagging sequence of~\cref{eq:protocol_CR_matching}, with $\chi_f=\pi\chi_m/\theta$. The joint conditional rotation $\CRR(\theta,\pi)$ rotates the memory by $\pm\theta$ and the flag by $\pm\pi$, and is framed by two flag displacements $\Df(\flag)$.
    (b) Flag trajectories, with the symbols of~\cref{fig:CR_cont_err_channel}. Top, without a jump, the two components recombine at $\ket{-\flag}$ and the flag returns to vacuum. Bottom, with a decay at $\tau$, the $\ket{g}$ component vanishes and the $\ket{e}$ component relaxes to $\ket{g}$, reversing its rotation and leaving the flag away from vacuum.
    (c) Final flag positions $z(\tau)$ of~\cref{eq:CR_flag_position} for single jumps, with the colorscale indicating $\tau \in [0,T]$. They lie on the circle of radius $|\flag|$ centered at $\flag$, whose polar angle encodes the memory rotation.
}
\label{fig:protocol_CR_matching}
\end{figure*}

We now apply the framework of~\cref{sec:cv_flags} to the case of conditional rotation operation $\CR(\theta)$, for which a single continuous-variable flag provides protection against ancilla relaxation to all orders in $\kappa T$. Following~\cref{sec:cv_flags}, the construction requires a flagging sequence that implements $\CR(\theta)$ while returning the flag to vacuum, a flag position $z(\tau)$ perfectly correlated with the memory error, and a correction unitary that optimizes a given functional. In this section, we construct these three ingredients in turn to optimize the performance of a rotation-symmetric code.

\subsection{Flagging of ancilla decay}\label{subsec:flag_CR}

\subsubsection{Flagging with $\chi$-matching}\label{subsubsec:chi-match}

We first assume the $\chi$-matching condition given by $\chi_f = \pi \chi_m / \theta$, which will be relaxed in the following subsection. Under this condition, the flagging sequence illustrated in~\cref{fig:protocol_CR_matching}(a) reads
\begin{equation} \label{eq:protocol_CR_matching}
    \FCR(\theta,\flag) = \Df(\flag) \, \CRR(\theta,\pi) \, \Df(\flag),
\end{equation}
where $\Df(\flag)$ corresponds to a displacement of the flag mode by $\flag\in\mathbb{C}$. Over the interaction time $T = 2\theta / \chi_m$, $\CRR(\theta,\pi)$ rotates the memory by $\pm\theta$ and the flag by $\pm\pi$, conditionally on the ancilla state.

In the absence of decay, the flag is first displaced to a coherent state $\ket{\flag}$; its $\ket{e}$ and $\ket{g}$ components then rotate by $\pm\pi$ and recombine at $\ket{-\flag}$, before the final displacement returns the flag to vacuum. The corresponding trajectory is shown in the top row of~\cref{fig:protocol_CR_matching}(b). The flag thus ends in vacuum, disentangled from the rest of the system, while the memory and ancilla undergo the ideal $\CR(\theta)$ gate, so that the sequence satisfies the disentangling condition of~\cref{eq:flag_disentangling_condition}.

An ancilla decay at time $\tau\in[0,T]$ instead interrupts this sequence with a jump $\hsigma_-$. The memory-ancilla system is then left exactly as in the unflagged case of~\cref{eq:traj_decay_during_CR}, but the flag no longer returns to vacuum: before the jump, its two components rotate by $\pm\tfrac{\tau}{T}\pi$; the decay then projects $\ket{e}\to\ket{g}$, and the rotation is reversed for the remaining time $T-\tau$, as illustrated in the bottom row of~\cref{fig:protocol_CR_matching}(b). The flag therefore ends in the coherent state $\ket{z(\tau)}$, with
\begin{equation} \label{eq:CR_flag_position}
    z(\tau) = \flag \left(1 - e^{i2\pi\tau/T}\right).
\end{equation}
As $\tau$ spans $[0,T]$, these states trace a circle of radius $|\flag|$ centered at $\flag$, see~\cref{fig:protocol_CR_matching}(c), whose polar angle uniquely identifies the jump time. The only exceptions occur at the endpoints $\tau=0$ and $\tau=T$, for which the flag returns to vacuum, rendering these events indistinguishable from the no-jump case. These events leave the memory rotated by $-2\theta$ or $0$, respectively, which is harmless for a rotation-symmetric code whose stabilizer is $\R(-2\theta)$, see~\cref{subsec:CR_cat_parity}. The flag's coherent state therefore uniquely determines the error channel acting on the memory mode.

Rigorously, the flagging sequence of~\cref{eq:protocol_CR_matching} subject to ancilla decay yields the flagged channel (see~\cref{subsubsec-apdx:CR_flagging} for the derivation)
\begin{equation}\label{eq: Kraus map protocol CR}
\begin{split}
    \mathcal{F}(\hrho) &= \R^{\mathrm{nj}}_0 \hrho (\R^{\mathrm{nj}}_0)^\dagger \otimes \ketbra{0} \\
    & +\int_{-2\theta}^{0} d\phi F_{ge}(\phi)~\R_\phi^{ge} \hrho (\R_\phi^{ge})^\dagger \otimes \ketbra{\tilde z(\phi)},
\end{split}
\end{equation}
where 
\begin{equation}\label{eq:map_chi_match}
\tilde z(\phi)=z\left(T\left(1+\frac{\phi}{2\theta}\right)\right)=\flag\left(1-e^{i\pi\phi/\theta}\right).
\end{equation}
The map $\tilde z$ is injective on $(-2\theta,0)$ and vanishes at both endpoints, $\tilde z(-2\theta)=\tilde z(0)=0$, as in the no-jump case. A heterodyne detection of the flag can therefore reveal the random rotation angle of the memory. In practice, the precision of this estimate is limited by the quantum fluctuations of the coherent states, as well as by the classical noise of the heterodyne detection. In the ideal case, it can be improved by increasing the flag amplitude $|\flag|$. 

Interestingly, for the $\chi$-matched flag, a similar calculation of the flagged channel in the presence of thermal excitations of the ancilla qubit shows that the memory rotation remains encoded in the flag, up to a stabilizer, at all orders, see~\cref{sec-apdx:decay_excitation}. More precisely, the memory rotation induced by any sequence of decay and excitation events of the ancilla is encoded in a unique coherent state of the flag. This rotation can thus be corrected by the same heterodyne detection of the flag, followed by the appropriate feedback.

\subsubsection{Flagging without $\chi$-matching}\label{subsubsec:CR-no-chi-matching}

\begin{figure}[t]
\centering
\includegraphics[width=\columnwidth]{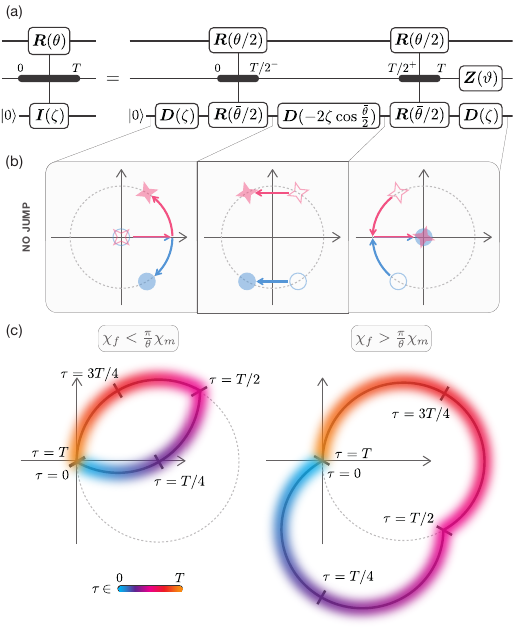}
\caption{
    \textbf{Flagging sequence for the $\CR(\theta)$ gate without $\chi$-matching.}
    (a) Flagging sequence of~\cref{eq:CR_unmatched}, for an arbitrary flag rotation $\ftheta=\theta\chi_f/\chi_m$. The two halves $\CRR(\theta/2,\ftheta/2)$ are separated by the flag displacement $\Df(-2\flag\cos\tfrac{\ftheta}{2})$, and the final ancilla rotation compensates the residual phase $\vartheta=2|\flag|^2 \sin\bar\theta$.
    (b) Flag trajectories without a jump, for $\ftheta<\pi$. The two components recombine after the second half, and the final displacement returns the flag to vacuum.
    (c) Final flag positions $z(\tau)$ of~\cref{eq:CR_unmatched_flags} for single jumps, with the colorscale indicating $\tau\in[0,T]$. Left, $\chi_f<\pi\chi_m/\theta$. Right, $\chi_f>\pi\chi_m/\theta$. The circle of the matched case becomes a closed curve of two circular arcs, which still encodes the memory rotation injectively as long as $\ftheta<2\pi$.}
\label{fig:CR_unmatched}
\end{figure}

The protocol introduced so far relies on the $\chi$-matching condition $\chi_f=\frac{\pi}{\theta} \chi_m$, for which the flag rotation angle is exactly $\pi$ during the memory conditional rotation. This condition poses experimental challenges: dispersive couplings are determined by the physical design and cannot be easily adjusted in situ, and the condition depends on the gate angle $\theta$, so that a device matched for one conditional rotation is mismatched for any other. Away from this condition, a memory rotation of $\pm\theta$ is accompanied by a flag rotation of $\pm\ftheta$ with $\ftheta=\frac{\chi_f}{\chi_m}\theta$, and the two flag components of the sequence~\cref{eq:protocol_CR_matching} no longer recombine at the end of the gate. The flag then remains entangled with the memory-ancilla system even in the absence of decay, so that $\CR_m(\theta)$ is no longer faithfully implemented: tracing out the flag dephases the ancilla, and the flag itself becomes a source of errors.

Disentanglement can nevertheless be restored by inserting an unconditional displacement of the flag midway through the gate, leading to the sequence \begin{equation} \label{eq:CR_unmatched} \begin{aligned} \FCR(\theta,\flag) &= \Z_q(2|\flag|^2\sin{\ftheta}) \Df(\flag)\CRR(\tfrac{\theta}{2},\tfrac{\ftheta}{2}) \\ &\quad\times \Df(-2\flag\cos\tfrac{\ftheta}{2}) \CRR(\tfrac{\theta}{2},\tfrac{\ftheta}{2}) \Df(\flag), \end{aligned} \end{equation} illustrated in~\cref{fig:CR_unmatched}(a). Geometrically, this intermediate displacement reshapes the two flag trajectories in phase space, effectively shortening them for $\ftheta<\pi$ and extending them for $\ftheta>\pi$, so that both components recombine at the end of the protocol, as illustrated for $\ftheta<\pi$ in~\cref{fig:CR_unmatched}(b). The additional qubit rotation $\Z_q$ ensures a perfect implementation of the flagged operation.

The flagged channel then keeps the form of~\cref{eq: Kraus map protocol CR}, but the jump states no longer lie on a circle. Instead, they trace a closed curve composed of two circular arcs, 
\begin{equation} \label{eq:CR_unmatched_flags}
    \left\{
    \begin{aligned}
        z(\tau)&=\flag \left(e^{-i\ftheta\frac{T-2\tau}{T}}- e^{-i\ftheta}\right), && \tau\in[0,T/2],\\
        z(\tau)&=\flag\left(1 - e^{-2i\ftheta\frac{T-\tau}{T}}\right), && \tau\in[T/2,T],
    \end{aligned}
    \right.
\end{equation}
shown in~\cref{fig:CR_unmatched}(c), left for $\chi_f<\pi\chi_m/\theta$ and right for $\chi_f>\pi\chi_m/\theta$. This corresponds to a mapping of the memory's random rotation angle as follows
\begin{equation} \label{eq:map_unmatched_flags}
\tilde z(\phi)=
\begin{cases}
\flag\left(e^{i(\ftheta+\bar\phi)}-e^{-i\ftheta}\right), & \phi\in[-2\theta,-\theta],\\
\flag\left(1-e^{i\bar\phi}\right), & \phi\in[-\theta,0],
\end{cases}
\end{equation}
where $\bar\phi=\phi\,\chi_f/\chi_m$. As long as $\ftheta<2\pi$, the map $\phi\mapsto \tilde z(\phi)$ remains injective, so the ancilla-induced random rotation angles of the memory mode remain unambiguously encoded in the complex amplitude of the flag's coherent state.

In practice, however, for small values of $\ftheta$, the curve becomes elongated and locally compressed, as can already be seen in the left panel of~\cref{fig:CR_unmatched}(c), which reduces the signal-to-noise ratio of the jump time estimate in heterodyne detection. This phenomenon can be mitigated by shaping the trajectory in phase space using additional intermediate displacements, or more generally by optimizing the displacement envelopes beyond the instantaneous displacement approximation.

\begin{figure*}[t]
\centering
\includegraphics[width=\textwidth]{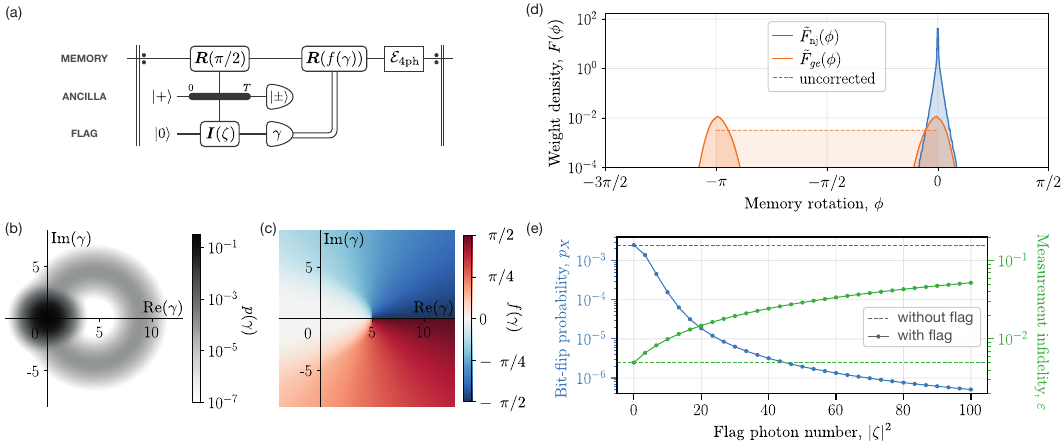}
\caption{
    \textbf{Flagged parity measurement for the four-component cat code.}
    Parameters: $\chi_f=2\chi_m$, $\chi_m/2\pi = 500~\mathrm{kHz}$, $T_1^{\text{qubit}}=T_2^{\text{qubit}}=100~\mu\mathrm{s}$, $|\alpha|^2 = 9$, $T_1^{\text{flag}} = 1~\mathrm{ms}$, $\eta = 0.5$ and $|\flag|^2=20$.
    (a) One cycle of the flagged parity measurement of $\ket{0_L}$, with the flagging sequence of~\cref{eq:protocol_CR_matching}, the rotation feedback $\Rm(f(\gamma))$ of~\cref{eq:cat_optimal_feedback} and the four-photon dissipation $\mathcal{E}_{4\mathrm{ph}}$.
    (b) Probability density $p(\gamma)$ of the heterodyne outcomes, with the no-jump peak at the origin and the ring of jump outcomes along $z(\tau)$.
    (c) Optimal feedback angle $f(\gamma)$.
    (d) Weight functions of the memory rotation $\phi$, before (dashed, $F_{ge}$) and after (solid,~\cref{eq:CR_corrected_weights}) correction. The corrected weights concentrate on $\phi=0$ and $-\pi$.
    (e) Bit-flip probability per cycle $p_X$ (blue, left axis) and parity-measurement infidelity $\varepsilon$ (green, right axis) versus $|\flag|^2$, with the unflagged values dashed.
}
\label{fig:cat_parity}
\end{figure*}

\subsection{Application to error correction with a rotation-symmetric code}\label{subsec:CR_cat_parity}

We now apply the flagged conditional rotation to a concrete error-correction task, the parity measurement of a four-component cat code~\cite{Leghtas_2013,Ofek_2016}. We first show how a qubit decay during the parity gate turns into an uncorrectable (logical) error and how the flag identifies it. We then derive the optimal rotation feedback and the error channel it leaves behind and finally evaluate the bit-flip probability it achieves in simulation.
 
\subsubsection{Parity measurement and four-component cat code}

The four-component cat code~\cite{Leghtas_2013,Ofek_2016} is a member of the family of rotation-symmetric bosonic codes~\cite{Grimsmo_2020}. Its stabilizers are given by the photon-number parity operator, $\Pi=e^{i\pi\ha^\dag\ha}$, and the four-photon dissipation operator, $\ha^4-\alpha^4\Id$. Its logical states are the even-parity cat states,
\begin{align}\label{eq:4legcat_logical}
\ket{+_L}&=\frac{1}{\mathcal{N}_+}\sum_{k=0}^{3}e^{i\frac{k\pi}{2}\ha^\dag\ha}\ket{\alpha},\\
\ket{-_L}&=\frac{1}{\mathcal{N}_-}\sum_{k=0}^{3}(-1)^ke^{i\frac{k\pi}{2}\ha^\dag\ha}\ket{\alpha},\notag
\end{align}
with $\mathcal{N}_{\pm}$ normalization constants, and $\ket{0_L/1_L}=(\ket{+_L}\pm\ket{-_L})/\sqrt{2}$.
While small phase shifts can be corrected by a four-photon driven-dissipative mechanism as in~\cite{Mirrahimi_2014}, or through a teleportation protocol~\cite{Grimsmo_2020}, photon jumps can be tracked and corrected through the readout of the photon-number parity stabilizer~\cite{Sun_2014}. Although, for simplicity, we specialize the discussion to the four-component cat code, the analysis carries over to any rotation-symmetric bosonic code, as photon jumps are corrected through generalized photon-number parity measurements. This also includes concatenated constructions such as repetition cat-codes, where we require the joint-parity readout of two-component cat codes~\cite{Cohen_2017,Guillaud_2019,Chamberland_2022,Putterman_2025}.

In practice, the parity is mapped onto an ancillary qubit (e.g. a superconducting transmon qubit)~\cite{Davidovich-97,Bertet-2002}: the qubit is prepared in $\ket{+}$, a conditional rotation $\CR(\pi/2)$ correlates the memory parity with the qubit state, and the qubit is finally read out in the $\{\ket{+},\ket{-}\}$ basis~\cite{Sun_2014, Ofek_2016}. A spontaneous decay of the qubit at time $\tau$ propagates to the memory as a rotation error $\R(\varphi)$ by an angle $\varphi \in [-2\theta,0]=[-\pi,0]$, with the almost uniform density $F_{ge}$ of~\cref{eq:weight_function_uncorr_rot}. The two endpoints of this interval are harmless, $\R(0)$ being the identity and $\R(-2\theta)=\R(-\pi)$ the parity stabilizer. Intermediate angles can cause logical bit flips: after four-photon driven dissipation, any angle in $[-3\pi/4,-\pi/4]$ leads to one with a probability close to 1, and other angles with a lower probability.
 
Here, we benchmark the flagging protocol on a dissipatively stabilized four-component cat code~\cite{Mirrahimi_2014}, in which four-photon driven dissipation (denoted $\mathcal{E}_{4\text{ph}}$ in~\cref{fig:cat_parity}(a)) steers the stabilizer $\ha^4-\alpha^4$ to 0. More precisely, in this error-correction protocol, we alternate the photon-number parity stabilizer measurements with four-photon driven dissipative mechanism. During this four-photon stabilization, the effective logical bit-flip probability can be directly read off from the conserved quantities~\cite{Albert_2014}, without resolving the full evolution. Indeed, the four-photon driven dissipation admits a set of invariants $\J_{ij}$, $i,j \in \{+,-\}$ (observables conserved by the dissipative evolution) whose expectation values in the initial state determine the asymptotic state reached within the cat code manifold~\cite{Mirrahimi_2014}. In particular, the combination $\J_1=(\J_{++}+\J_{--}-\J_{+-}-\J_{-+})/2$ is the invariant associated with the logical population $\ketbra{1_L}$, so that for any initial memory state $\hrho$ the population of $\ket{1_L}$ after stabilization equals $\Tr[\J_1\hrho]$. The per-round logical bit-flip probability therefore reads
\begin{equation} \label{eq:cat_bitflip_probability}
    p_{X} = \Tr[\J_1\, \mathcal{C}(\ketbra{0_L} \otimes \ketbra{+})],
\end{equation}
with $\mathcal{C}=\mathcal{N}_R$ for the bare gate and $\mathcal{C}=\mathcal{M}\circ\mathcal{F}$ for the flagged protocol, the ideal $\CR(\pi/2)$ being factored out in both cases. This $p_X$ is the natural figure of merit to be minimized in a parity measurement round.
 
\subsubsection{Flagged parity measurement and optimal feedback}
 
Because the syndrome measurement reduces to a single $\CR(\pi/2)$ gate, it can be protected using the continuous-variable flag solution presented in this paper. To simplify the discussion, we assume in this subsection a flagging protocol under the $\chi$-matching condition of~\cref{subsubsec:chi-match}. As illustrated in~\cref{fig:cat_parity}(a), we thus modify each round of parity measurement to replace the bare $\CR(\pi/2)$ by the flagging sequence $\FCR(\pi/2,\flag)$ of~\cref{eq:protocol_CR_matching}, operated under the condition $\chi_f=2\chi_m$. The qubit readout is left unaltered, and the flag is read out by heterodyne detection at the end of the measurement round. Following~\cref{sec:cv_flags}, this measurement is converted into a corrective memory rotation $\C_\gamma=\Rm(f(\gamma))$ applied before the stabilization by four-photon driven dissipation.
 
The rotation angle $f(\gamma)$ follows from the general correction rule of~\cref{eq:optimal_correction}, with the bit-flip probability of~\cref{eq:cat_bitflip_probability} as cost functional $J$, evaluated on the code space with the ancilla prepared in $\ket{+}$. Restricting the feedback to memory rotations, the rule gives
\begin{equation} \label{eq:cat_optimal_feedback}
    f(\gamma) = \underset{\mu}{\operatorname{argmin}} \Tr[\J_1 \Rm(\mu)\, \hrho_{|\gamma} \,\Rm(\mu)^\dagger],
\end{equation}
where $\hrho_{|\gamma} = \mathcal{N}_{|\gamma}(\ketbra{0_L} \otimes \ketbra{+})$ is the conditional state of~\cref{eq:conditional-noise-channel}. Note that in~\cref{eq:optimal_correction}, before taking the minimum on the angle of the rotation, we maximize the cost functional over all states in the code space. Here however, by symmetry, the bit-flip probability is the same when initializing the code in $\ket{1_L}$. We thus simplify the optimization problem of~\cref{eq:optimal_correction} to that of~\cref{eq:cat_optimal_feedback}. \Cref{fig:cat_parity}(b)-(d) display the outcome of this optimization for a flag activated at mean photon number $|\flag|^2=20$.
 
\Cref{fig:cat_parity}(b) shows the probability density $p(\gamma)$ of the heterodyne outcomes, which exhibits the two structures expected from the flagging channel of~\cref{eq: Kraus map protocol CR}. The dominant Gaussian peak at the origin is the no-jump contribution, and the faint ring carrying the remaining weight collects the jump trajectories, spread along the circle $z(\tau)$ of~\cref{eq:CR_flag_position}, whose polar angle encodes the jump time. The two structures merge near the origin, where the endpoint degeneracy $z(0)=z(T)=0$ precludes any discrimination, but these are precisely the trajectories that need no correction, as their residual rotation is already $0$ or $-\pi$, under which the code space is invariant.
 
\Cref{fig:cat_parity}(c) shows the corresponding optimal feedback $f(\gamma)$ of~\cref{eq:cat_optimal_feedback}. Near the origin, where the no-jump hypothesis dominates, little to no correction is applied. Away from it, the correction varies with the polar angle of the outcome, following the jump time it points to, and interpolates smoothly in between, reflecting how a finite flag amplitude trades off the competing no-jump and jump hypotheses.

\Cref{fig:cat_parity}(d) shows the effect of this feedback on the error channel, in the form of~\cref{eq:CR_cont_err_channel_fl}. Composing the conditional channel of~\cref{eq: Kraus map protocol CR} with the flag measurement and the feedback $\C_\gamma=\Rm(f(\gamma))$, and averaging over outcomes, a no-jump trajectory is left rotated by $\phi=f(\gamma)$, with $\gamma$ drawn from $p(\gamma|0)$. A jump trajectory at time $\tau$, initially rotated by $\varphi$, is left rotated by $\phi=\varphi+f(\gamma)$, with $\gamma$ drawn from $p(\gamma|\tilde z(\varphi) )$. The weight functions of~\cref{eq:CR_cont_err_channel_fl} therefore read
\begin{equation} \label{eq:CR_corrected_weights}
\begin{aligned}
    \tilde F_{\mathrm{nj}}(\phi) &= \int d^2\gamma \; p(\gamma|0)\; \delta_{\phi-f(\gamma)}, \\
    \tilde F_{ge}(\phi) &= \int\! d^2\gamma \int_{-2\theta}^{0} \!\! d\varphi \, F_{ge}(\varphi)\, p(\gamma|\tilde{z}(\varphi))\, \delta_{\phi-\varphi-f(\gamma)}.
\end{aligned}
\end{equation}
Before correction, the no-jump weight is a delta at $\phi=0$ and the jump weight $F_{ge}$ is almost uniform over $[-\pi,0]$ (orange dashed line). After correction, an outcome on the ring of panel (b) identifies the jump time through its polar angle, and panel (c) shows that the feedback rotates $\varphi$ to the nearest angle leaving the code invariant. Jumps in the first half of the gate are thus completed to $\phi=-\pi$ and those in the second half undone to $\phi=0$. The switch happens at the mid-gate jump, $\varphi=-\pi/2$, visible as the discontinuity of $f$ in panel (c). $\tilde F_{ge}$ therefore splits into two lobes of nearly equal weight at $-\pi$ and $0$, broadened by the heterodyne noise that blurs the ring. Increasing the flag amplitude improves the precision with which a trajectory is identified, and therefore narrows these lobes. The small feedback applied near the origin also gives $\tilde F_{\mathrm{nj}}$ a finite width around $0$, as the optimization slightly rotates the dominant trajectory to better recenter the jump contributions it cannot separate from it. The remaining logical error comes from the tails of the three peaks.
 
\subsubsection{Numerical results}
 
\emph{Numerical method.} Since the flagged round is described analytically by the flagging channel of~\cref{eq: Kraus map protocol CR} and is followed by a code restabilization, the bit-flip probability of~\cref{eq:cat_bitflip_probability} can be evaluated semi-analytically, without Monte-Carlo sampling. We discretize the heterodyne outcomes on a grid of the complex plane and, for each $\gamma$, build the conditional memory state $\hrho_{|\gamma}$ from~\cref{eq:conditional-noise-channel} together with the outcome probability density $p(\gamma)$. The feedback angle $f(\gamma)$ of~\cref{eq:cat_optimal_feedback} is then obtained by sweeping over possible rotation angles $\mu$, where the minimum yields the conditional bit-flip probability $p_{X|\gamma}$. Averaging over outcomes finally gives $p_X = \int d^2\gamma \; p(\gamma)\, p_{X|\gamma}$. The same computation produces~\cref{fig:cat_parity}(b)-(d).
 
\emph{Parameters.} We consider a qubit of lifetime and coherence time $T_1^{\text{qubit}} = T_2^{\text{qubit}}=100~\mu\mathrm{s}$, dispersively coupled at $\chi_m/2\pi = 500~\mathrm{kHz}$ to a memory holding a four-component cat code of amplitude $|\alpha|^2 = 9$. Note also that with the notations of the present paper, we call $\kappa=1/T_1^{\text{qubit}}$. The $\CR(\pi/2)$ gate, realizing the parity measurement, then lasts $T = \pi/\chi_m = 1~\mu\mathrm{s}$, corresponding to a decay probability per gate $(1 - e^{-\kappa T})/2 \approx 5 \times 10^{-3}$. The flag is operated under the $\chi$-matching condition, $\chi_f = 2\chi_m$, \ie $\chi_f/2\pi = 1~\mathrm{MHz}$, and its mean photon number is swept over $|\flag|^2 \in [0,100]$ in~\cref{fig:cat_parity}(e), the other panels using $|\flag|^2 = 20$. To account for the limitations discussed in~\cref{sec:limitations}, the flag mode is taken with lifetime $T_1^{\text{flag}} = 1~\mathrm{ms}$ and the heterodyne measurement is performed with efficiency $\eta = 0.5$.
 
\emph{Results.} Consider first the unflagged parity measurement. A qubit decay, occurring with probability $(1-e^{-\kappa T})/2$ for an ancilla prepared in $\ket{+}$, both corrupts the parity outcome and rotates the memory by a random angle $\varphi$, distributed essentially uniformly over $[-\pi,0]$. About half of these rotations fall in $[-3\pi/4,-\pi/4]$ and end as a logical bit flip, which gives the per-round bit-flip probability $p_X \approx (1 - e^{-\kappa T})/4 \approx 2.5\times 10^{-3}$, the dashed blue line of~\cref{fig:cat_parity}(e).
 
Flagging changes this picture qualitatively: the jump time is now recorded in the flag, and the feedback rotation brings the memory back to $0$ or $-\pi$. The residual bit-flip probability (solid blue) therefore decreases monotonically with the flag mean photon number, which sets the precision of the jump-time inference, reaching $\approx 5\times 10^{-7}$ at $|\flag|^2 = 100$, more than three orders of magnitude below the unprotected value. The residual bit-flip probability is essentially the weight of the three peaks of~\cref{eq:CR_corrected_weights}, illustrated in~\cref{fig:cat_parity}(d), lying inside the undesired interval $[-3\pi/4,-\pi/4]$. The width of these peaks is set by $\sqrt{\eta}|\flag|$. A flag outcome away from the origin moreover heralds a qubit decay, so the syndrome bit of that round can be marked as unreliable and discarded at the decoding stage.
 
This gain is not for free, and the green curve of~\cref{fig:cat_parity}(e) shows its counterpart. Indeed, because of the flag mode's decay, the infidelity of the parity readout increases with the flag size $|\flag|^2$. This is further analyzed in~\cref{sec:limitations}. 
 
The two curves thus quantify a trade-off between errors of very different severity. A degraded parity contrast is a syndrome-measurement error. It does not propagate to the memory and is naturally handled by the decoder, at the cost of a few extra measurement rounds to resolve a syndrome. On the other hand, the suppressed bit-flip errors are uncorrectable logical errors. Trading three orders of magnitude of logical bit-flip probability against the increase of syndrome infidelity shown in~\cref{fig:cat_parity}(e) is therefore strongly favorable.

\section{Conditional displacement gate} \label{sec:CV_flag_for_CD}

We now turn to the protection of the conditional displacement $\CD(\beta)$, realized by the ECD sequence of~\cref{eq:ECD_def,eq:link_CD_and_ECD}. The construction follows the same lines as for the conditional rotation of~\cref{sec:CV_flag_for_CR}, with one distinctive feature: by repopulating the excited ancilla state midway through the gate, the echo pulse gives rise to double-decay trajectories in the continuous error channel of~\cref{eq:CD_cont_err_channel}.

Since the ECD sequence contains two conditional-rotation gates separated by the echo pulse, a straightforward strategy would be to protect each conditional rotation independently with the protocol of~\cref{sec:CV_flag_for_CR}. Such an approach, however, fails precisely where protection is most needed: a decay occurring close to the echo sits at the endpoint of both conditional rotations, where each individual flag returns to vacuum by the endpoint degeneracy discussed in~\cref{subsec:flag_CR}, yet the memory suffers a displacement by $\pm\beta$ (see \cref{fig:CD_cont_err_channel}(c)) that typically results in a logical error for a translation-symmetric code.

In this section, we instead introduce continuous-variable flagging protocols tailored to the full ECD sequence. We first construct a one-flag protocol protecting the gate at first order in ancilla decay (\cref{subsec:CD-one-flag}), and benchmark it on the sBs stabilization of a GKP code (\cref{subsec:CD-gkp-sbs}). We then add a second flag to reach all-order protection (\cref{subsec:CD-two-flags}).

\subsection{First-order flagging protocol}\label{subsec:CD-one-flag}

\begin{figure*}[t]
\centering
\includegraphics[width=\textwidth]{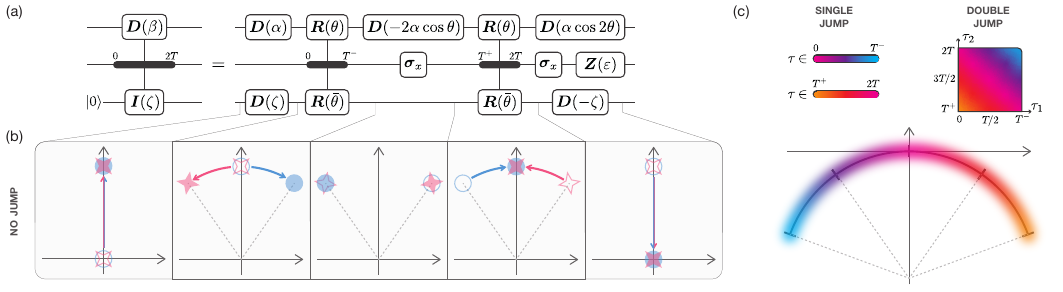}
\caption{
    \textbf{Flagging sequence for the $\CD(\beta)$ gate with one flag.}
    (a) Flagging sequence of~\cref{eq:protocol_ECD_one_flag}, with the notation of~\cref{fig:CD_cont_err_channel}(a). The conditional rotations of the ECD sequence become joint rotations $\CRR(\theta,\bar\theta)$, with arbitrary $\bar\theta=\theta\chi_f/\chi_m$, framed by the flag displacements $\Df(\pm\flag)$.
    (b) Flag trajectories without a jump. The echo $\hsigma_x$ interchanges the two components, which retrace their paths and recombine at $\ket{\flag}$.
    (c) Final flag positions, on an arc of curvature radius $|\flag|$ and angular width $4\bar\theta$. Single jumps $z(\tau)$ of~\cref{eq:CD_flag_position}, before ($\tau\in[0,T^-]$) or after ($\tau\in[T^+,2T]$) the echo, and double jumps $z(\vec{\tau})$ of~\cref{eq:CD_twojump_flag_position} lie on the same arc, which limits the one-flag protection to first order.
}
\label{fig:CD_oneflag}
\end{figure*}

To protect the ECD implementation of $\CD(\beta)$ against ancilla relaxation, we make two minimal modifications: each conditional rotation $\CR(\theta)$ is promoted to the joint memory-flag rotation $\CRR(\theta,\ftheta)$ of~\cref{eq:CRR_def}, with $\ftheta=\theta \chi_f / \chi_m$ set by the ratio of the dispersive couplings, and the whole sequence is sandwiched between two opposite displacements of the flag. Denoting by $\ECD_{mf}(\alpha,\theta)$ the echoed sequence of~\cref{eq:ECD_def} with its conditional rotations upgraded in this way, the flagging sequence reads
\begin{equation} \label{eq:protocol_ECD_one_flag}
\begin{split}
    \FCD(\beta,\flag) &= \Z_q(\varepsilon_\lambda)\hsigma_x \\
    &\quad \times \Df(-\flag)\ECD_{mf}(\alpha_\lambda,\theta_\lambda)\,\Df(\flag).
\end{split}
\end{equation}
Here, the ECD parameters $(\alpha_\lambda,\theta_\lambda)$, and the residual qubit rotation $\Z_q(\varepsilon_\lambda)$, are those of~\cref{eq:link_CD_and_ECD}, such that the sequence reduces to $\CD(\beta)$ when the flag is ignored. This protocol is illustrated in~\cref{fig:CD_oneflag}(a). 

The decay-free dynamics of the flag, depicted in~\cref{fig:CD_oneflag}(b), proceed as follows. The flag is first displaced to the coherent state $\ket{\flag}$. Its $\ket{e}$ and $\ket{g}$ components then rotate by $\pm\ftheta$ during the first conditional rotation. They are next swapped by the echo, and retrace their paths during the second conditional rotation, so that both components recombine at $\ket{\flag}$. The final displacement brings the flag back to vacuum. The disentangling condition of~\cref{eq:flag_disentangling_condition} is therefore fulfilled, with the memory-ancilla system undergoing the ideal $\CD(\beta)$ gate whenever no decay occurs. Remarkably, no $\chi$-matching condition is required here: since, in the absence of intermediate displacements of the flag, $\CR_f(\ftheta)\,\hsigma_x\,\CR_f(\ftheta)=\hsigma_x$ for any rotation angle $\ftheta$, the echo itself supplies the recombination mechanism.

Suppose now that the ancilla relaxes at time $\tau\in[0,2T]$, with $T$ the duration of a single conditional rotation. While the memory-ancilla system is affected exactly as in the unflagged gate of~\cref{subsec:intro CD}, the jump $\hsigma_-$ also prevents the flag trajectory from closing. If the decay precedes the echo, the flag component entangled with $\ket{e}$ has rotated by $+\ftheta\times\tau/T$ at the jump, then counter-rotates until the echo, accumulating a net angle in $[-\ftheta,\ftheta]$. The echo then returns it to $\ket{e}$, and it rotates by a further $+\ftheta$, so the flag ends at a polar angle in $[0,2\ftheta]$. A decay after the echo similarly lands at a polar angle in $[-2\ftheta,0]$. Altogether, after such first-order decay events, the flag is left in the coherent state $\ket{z(\tau)}$, with
\begin{equation} \label{eq:CD_flag_position}
\left\{
\begin{aligned}
z(\tau) &= \flag\left(e^{2i\ftheta\tau/T}-1\right), && \tau\in[0,T),\\
z(\tau) &= \flag\left(e^{2i\ftheta(\tau-2T)/T}-1\right), && \tau\in(T,2T].
\end{aligned}
\right.
\end{equation}
When $\tau$ runs over the gate duration, the jump states sweep an arc of angular width $4\ftheta$ on the circle of radius $|\flag|$ centered at $-\flag$, as illustrated in~\cref{fig:CD_oneflag}(c). As long as $\ftheta<\pi/2$, the pre-echo branch, with polar angles in $(0,2\ftheta)$, and the post-echo branch, with polar angles in $(-2\ftheta,0)$, do not overlap, and the map $\tau\mapsto z(\tau)$ is injective except at the endpoints $\tau=0$ and $\tau=2T$, for which the flag returns to vacuum. As in~\cref{subsubsec:chi-match}, these events are harmless, since they leave the memory displaced by $-2\beta$, a stabilizer of translation-symmetric codes, or not displaced at all, see~\cref{subsec:intro CD}.

Trajectories with two decays, one before the echo pulse and one after, with jump times $\vec{\tau}=(\tau_1, \tau_2) \in [0,T) \times (T,2T]$, leave the flag in the coherent state $\ket{z(\vec{\tau})}$, with
\begin{equation} \label{eq:CD_twojump_flag_position}
    z(\vec{\tau}) = \flag \left(e^{2i\bar\theta \frac{\tau_1+\tau_2-2T}{T}} - 1\right),
\end{equation}
which lies on the same arc.

More precisely, as discussed in~\cref{subsec-apdx:flagging_channels},
the flagging sequence of~\cref{eq:protocol_ECD_one_flag} leads, in the limit of large intermediate displacements ($\lambda \gg 1$), to the channel
\begin{equation} \label{eq:CD_oneflag_flagging_channel}
\begin{aligned}
\mathcal{F}_{D}(\hrho) &= F_{\mathrm{nj}}\,\mathcal{D}^{\mathrm{nj}}_{0}(\hrho)\otimes\ketbra{0} \\
&+ \int_{-2}^{-1}\! ds\, F_{ge}(s)\,\mathcal{D}^{ge}_{s\beta}(\hrho)\otimes\ketbra{\tilde z_{ge}(s)} \\
&+ \int_{0}^{1}\! ds\, F_{eg}(s)\,\mathcal{D}^{eg}_{s\beta}(\hrho)\otimes\ketbra{\tilde z_{eg}(s)} \\
&+ \int_{0}^{1}\!\! ds_1\!\int_{1}^{2}\!\! ds_2\, F_{ee}(s_1+s_2)\,\mathcal{D}^{ee}_{(s_1-s_2)\beta}(\hrho) \\
&\qquad\qquad\otimes\ketbra{\tilde z_{ee}(s_1+s_2)},
\end{aligned}
\end{equation}
where the weight functions are those of~\cref{eq:CD_cont_err_channel_Kraus_Ops}, $\mathcal{D}^{k}_{s\beta}(\hrho)=\D^{k}_{s\beta}\,\hrho\,(\D^{k}_{s\beta})^\dagger$ denotes the displacement channel associated with the Kraus operator $\D^{k}_{s\beta}$ of the same equation, and the flag's associated coherent state amplitudes are given by
\begin{equation} \label{eq:CD_oneflag_values}
\begin{aligned}
\tilde z_{ge}(s)&=\flag(e^{2i\bar\theta(s+2)}-1),\\
\tilde z_{eg}(s)&=\flag(e^{-2i\bar\theta s}-1),\\
\tilde z_{ee}(s)&=\flag(e^{2i\bar\theta(s-2)}-1).
\end{aligned}
\end{equation}
For single-decay trajectories, $\tilde z_{ge}$ and $\tilde z_{eg}$ map the intervals $(-2,-1)$ and $(0,1)$ injectively onto the two branches of the arc, so that the correlation condition of~\cref{eq:correlation_condition} holds at first order. The situation is different for trajectories with one decay on each side of the echo. The flag then ends in the coherent state $\ket{\tilde z_{ee}(s_1+s_2)}$, which only depends on the sum $s_1+s_2$, whereas the memory is displaced by $(s_1-s_2)\beta$, which depends on their difference. Two such trajectories can therefore leave the flag at the same position while displacing the memory by different amounts, and they moreover share the arc of the single-decay trajectories, so that the correlation condition of~\cref{eq:correlation_condition} is no longer satisfied at second order. The one-flag protocol thus protects the conditional displacement against first-order decay events only, a limitation that we lift in~\cref{subsec:CD-two-flags} by adding a second flag.

\subsection{Application to error correction with a translation-symmetric code}\label{subsec:CD-gkp-sbs}
 
We now apply the flagged conditional displacement to a concrete error-correction protocol with the square-lattice GKP code, as an example of a translation-symmetric code. We consider the small-Big-small (sBs) protocol proposed in~\cite{Royer_2020} and experimentally demonstrated in~\cite{Sivak_2023,Lachance-Quirion_2024}. We first recall the protocol and how a qubit decay propagates to the bosonic mode as a logical error. We then show how the flagging protocol captures this error to first order and derive an optimal feedback strategy to undo it. Finally, we numerically evaluate the performance of the protocol.

\begin{figure*}[t]
\centering
\includegraphics[width=\textwidth]{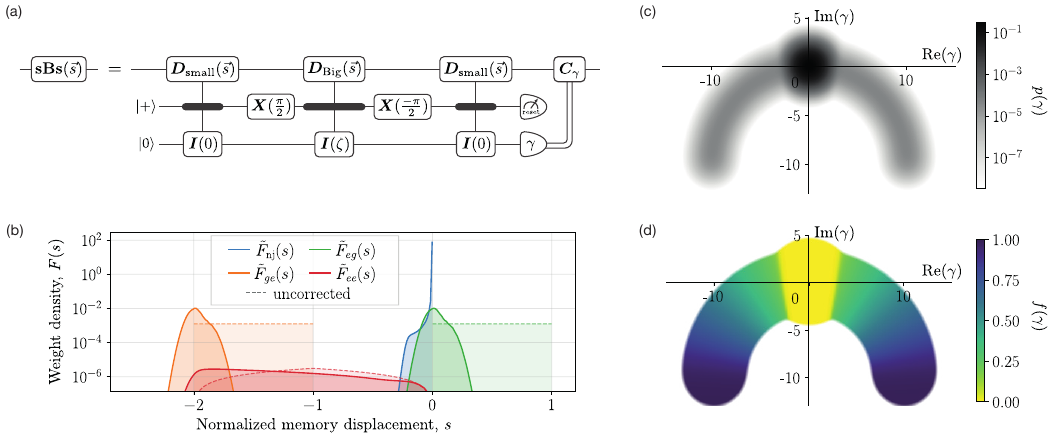}
\caption{
    \textbf{Flag protection of the sBs-stabilization protocol of a finite-energy GKP code.}
    Parameters: $1/\kappa = 100~\mu\mathrm{s}$, $\chi_m/2\pi = 50~\mathrm{kHz}$, $\chi_f/2\pi = 1~\mathrm{MHz}$, $\Delta = 0.25$, Big and small conditional displacements of $500$ and $250~\mathrm{ns}$ respectively ($\bar\theta = \pi/4$) and $|\flag|^2 = 100$. In (b)-(d), the memory errors are taken in the pure-displacement limit $\lambda\to\infty$ of~\cref{eq:CD_oneflag_flagging_channel}, while the jump statistics and flag positions are those of the $500~\mathrm{ns}$ Big conditional displacement.
    (a) One flagged sBs round. Only the Big conditional displacement is flagged, with the sequence of~\cref{eq:protocol_ECD_one_flag}, and the outcome $\gamma$ sets the feedback $\D(-f(\gamma)\beta)$.
    (b) Weight functions of the memory displacement $s\beta$, before (dashed) and after (solid,~\cref{eq:gkp_corrected_weights}) correction.
    (c) Probability density $p(\gamma)$ of the outcomes, with the no-jump peak at the origin and the arc of pre-echo (left) and post-echo (right) decays along $z(\tau)$.
    (d) Optimal correction amplitude $f(\gamma)$ of~\cref{eq:gkp_optimal_feedback}, in units of $-\beta$.
}\label{fig:gkp_sbs_protocol}
\end{figure*}
 
\subsubsection{GKP code and the sBs protocol}
 
The GKP code~\cite{Gottesman_2001} encodes a qubit in periodic superpositions of position eigenstates, protecting the logical information against small displacements of the oscillator. Its stabilizers are the displacements by one period of the square lattice, its logical Paulis the displacements by half a period, and the correctable region around each code word accordingly extends over a quarter of a period. Its ideal code words are infinitely squeezed grid states of infinite energy, hence unphysical, and any realization relies on the finite-energy approximation obtained by applying a Gaussian envelope $\hE_\Delta = e^{-\Delta^2 \ha^\dagger \ha}$ that regularizes the grid over an extent set by $\Delta$~\cite{Matsuura_2020, Royer_2020}. These finite-energy code words are not preserved by the free evolution of a noisy oscillator and must therefore be actively stabilized, both to return the memory back to the lattice and to restore the envelope, which noise would otherwise diffuse. In current experiments this stabilization is realized by an engineered dissipative process, implemented through an ancillary qubit dispersively coupled to the memory. Several such protocols have been proposed, most notably the small-Big-small (sBs), Sharpen-Trim and Big-small-Big schemes~\cite{Campagne_Ibarcq_2020,Royer_2020, deNeeve_2022, Sivak_2023}, and we focus here on sBs, the discussion below extending straightforwardly to the others.
 
Each sBs round stabilizes one quadrature by interleaving qubit-controlled conditional displacements of the memory with qubit rotations~\cite{Royer_2020, Eickbusch_2022, Sivak_2023}, as recalled in~\cref{fig:gkp_sbs_protocol}(a). The sequence comprises two \textit{small} conditional displacements $\CD_{\mathrm{small}}$, of amplitude $\propto \Delta^2$, which enforce the finite-energy envelope, framing one \textit{Big} conditional displacement $\CD_{\mathrm{Big}}$, which pumps the state back onto the grid. The qubit is prepared in $\ket{+}$ and reset at the end of the round. The amplitude of the Big conditional displacement is set by the lattice: its two conditional branches are separated by $2\beta$, one lattice period, so that $\beta$ is half a period. A by-product of the sequence, rather than an intended feature of it, is that each round leaves the memory with an extra logical Pauli. Being deterministic, it is simply tracked in the frame of the encoded qubit.
 
The two sizes of the two conditional displacements propagate qubit decay very differently. As shown in~\cref{subsec:intro CD}, a decay during $\CD(\beta)$ displaces the memory by $s\beta$, with $s\in[-2,-1]$ before the echo and $s\in[0,1]$ after it. During the small conditional displacements the induced shift stays of order $\mathcal{O}(\Delta^2)$ and remains well within the correctable region. But the one during the Big displacement sweeps a full lattice period. Since $s$ is almost uniformly distributed, about half of the single-decay trajectories of the Big conditional displacement therefore end as a logical error. It is thus the decay during the \textit{Big} conditional displacement that limits the logical lifetime, and this is the operation the continuous-variable flag must primarily protect.
 
\subsubsection{Flagged displacement and optimal feedback}
 
One round of the flagged protocol is shown in~\cref{fig:gkp_sbs_protocol}(a). The bare Big conditional displacement is replaced by the flagging sequence $\FCD(\beta,\flag)$ of~\cref{eq:protocol_ECD_one_flag}, while the two small displacements are left untouched and ideally act as the identity on the flag. The ancilla qubit reset closing the bare round is complemented by a heterodyne readout of the flag. Since the disentangling condition of~\cref{eq:flag_disentangling_condition} ensures that the flag records nothing other than the decay, the round continues to apply the same stabilization map on the memory, while its outcome $\gamma$ now labels the error. Following~\cref{sec:cv_flags}, this outcome is converted into a corrective memory displacement $\C_\gamma$ applied before the next round.
 
The correction to be applied for a given outcome follows from the general prescription of~\cref{eq:optimal_correction}, whose cost functional $J$ is here the probability of a logical error at the end of the round. We evaluate it from the logical-$Z$ observable~\cite{Shaw-PRXQ-24,Sellem-PRX-25}
\begin{equation} \label{eq:gkp_logicalZ}
\bar{Z} = \operatorname{sign}\!\left(\cos(\sqrt{\pi} \hx)\right),
\end{equation}
where $\hx$ denotes the memory position quadrature $(\ha+\ha^\dagger)/\sqrt{2}$, so that the ideal code words of the square GKP lattice are supported on even multiples of $\sqrt{\pi}$ for $\ket{+\bar{Z}}$ and on odd multiples for $\ket{-\bar{Z}}$. Rather than measuring the exact physical position, this observable identifies the logical cell of the lattice in which the memory is located. It is thus insensitive to the finite energy envelope as well as to displacements caused by a stabilizer, while changing sign under the action of a logical Pauli operator. This quantity therefore evaluates the residual displacement exactly as the code itself does: by assigning no error for a shift $s$ close to an even integer, and a full logical error for an $s$ close to an odd integer. The round considered here ideally maps $\ket{+\bar{Z}}$ onto $\ket{-\bar{Z}}$, owing to the logical Pauli applied by its Big displacement, so that the cost functional reads $J[\hrho] = \tfrac{1}{2}(1+\Tr[\bar{Z}\hrho])$.
 
Optimizing this cost on $\ket{+\bar{Z}}$ alone already solves the maximization over the code space in \mbox{\cref{eq:optimal_correction}}. In the limit $\lambda \gg 1$ the conditional channel $\mathcal{N}_{|\gamma}$ of \mbox{\cref{eq:conditional-noise-channel}} is a mixture of displacements along the axis of $\beta$, and the logical Pauli of that axis is itself such a displacement. It therefore commutes with every error of the mixture up to a phase, and it anticommutes with $\bar{Z}$, so that the probability of ending in the wrong cell starting from $\ket{-\bar{Z}}$ equals the one starting from $\ket{+\bar{Z}}$. Any other code state being a superposition of the two, and the corrected observable remaining diagonal in that basis, the cost takes the same value on the whole code space, exactly as in \mbox{\cref{subsec:CR_cat_parity}}.
 
Restricting the feedback to displacements along the axis of $\beta$, which is the class that is optimal in the semiclassical limit, the correction rule of~\cref{eq:optimal_correction} gives $\C_\gamma = \D(-f(\gamma)\beta)$ with
\begin{equation} \label{eq:gkp_optimal_feedback}
    f(\gamma) = \underset{\mu}{\operatorname{argmin}} \Tr\!\left[\bar{Z}\, \D_m(-\mu\beta)\, \hrho_{|\gamma}\, \D_m(-\mu\beta)^\dagger\right],
\end{equation}
where $\hrho_{|\gamma} = \mathcal{N}_{|\gamma}\circ\,\mathcal{CD}_\beta(\ketbra{+\bar{Z}} \otimes \ketbra{+})$ is the conditional state of~\cref{eq:conditional-noise-channel}, in which the ideal Big displacement $\mathcal{CD}_\beta$ is kept, since it sets the logical Pauli entering $J$. \Cref{fig:gkp_sbs_protocol}(b)-(d) display the outcome of this optimization for a flag activated at mean photon number $|\flag|^2 = 100$.
 
\Cref{fig:gkp_sbs_protocol}(c) shows the probability density $p(\gamma)$ of the heterodyne outcomes, which exhibits the two structures expected from the flagged channel of~\cref{eq:CD_oneflag_flagging_channel}. The dominant Gaussian peak at the origin is the no-jump contribution. The faint arc carrying the remaining weight collects the decay trajectories, spread along the locus $z(\tau)$ of~\cref{eq:CD_flag_position}. At dominant order, its left branch is the pre-echo family $\D^{ge}_{s\beta}$ and its right branch the post-echo family $\D^{eg}_{s\beta}$, the polar angle along each branch encoding $s$. The two structures merge near the origin, where the endpoint degeneracy $z(0)=z(2T)=0$ precludes any discrimination. These are, however, precisely the trajectories that need no correction, as their residual displacement is already $s=-2$ or $s=0$. Conversely, the ends of the arc, farthest from the no-jump peak and thus best resolved, carry the decays adjacent to the echo, that is the most harmful errors $s=\mp 1$.
 
\Cref{fig:gkp_sbs_protocol}(d) shows the corresponding correction amplitude $f(\gamma)$ of~\cref{eq:gkp_optimal_feedback}, which follows this structure. It vanishes near the origin, where the no-jump hypothesis dominates, and grows monotonically along either branch away from the origin, \ie as the inferred jump time approaches the echo, up to $f=1$ at the ends of the arc. The smooth interpolation in between reflects how a finite flag amplitude trades off the competing no-jump and decay hypotheses. A single nonnegative amplitude covers both branches because the feedback only has to return the state to the code space up to a stabilizer, as anticipated in~\cref{sec:cv_flags}. A post-echo decay, with $s \in [0,1]$, is undone by $\D(-s\beta)$, whereas a pre-echo decay, with $s \in [-2,-1]$, is completed into the stabilizer $\D(-2\beta)$ by $\D(-(2+s)\beta)$. Both corrections thus have an amplitude in $[0,1]$.

\Cref{fig:gkp_sbs_protocol}(b) shows the effect of this feedback on the error channel, in the form of~\cref{eq:CD_cont_err_channel_fl}. Composing the flagging channel of~\cref{eq:CD_oneflag_flagging_channel} with the flag measurement and feedback $\C_\gamma = \D(-f(\gamma)\beta)$ and averaging over outcomes, a trajectory that left the memory displaced by $s'\beta$ and the flag in $\ket{\tilde z_{k}(s')}$ ends displaced by $s = s' - f(\gamma)$, with $\gamma$ drawn from $p(\gamma|\tilde z_{k}(s'))$. The weight functions therefore read
\begin{equation} \label{eq:gkp_corrected_weights}
\begin{aligned}
\tilde F_{\mathrm{nj}}(s) &= F_{\mathrm{nj}} \int d^2\gamma\, p(\gamma|0) \delta_{s+f(\gamma)}, \\
\tilde F_{k}(s) &= \int d^2\gamma \int ds' F_{k}(s') p\big(\gamma|\tilde z_k(s')\big) \delta_{s-s'+f(\gamma)},
\end{aligned}
\end{equation}
for $k \in \{ge,eg,ee\}$, where $p(\gamma|\cdot)$ is the heterodyne likelihood and the flag positions are given by~\cref{eq:CD_oneflag_values}. Before correction, the no-jump weight is a delta at $s=0$, and the single-decay weights $F_{ge}$ and $F_{eg}$ are almost uniform over $[-2,-1]$ and $[0,1]$ (dashed lines). After correction, $\tilde F_{ge}$ and $\tilde F_{eg}$ are concentrated around $s=-2$ and $s=0$. These peaks are broadened by the heterodyne noise that blurs the arc of panel (c). Increasing the flag amplitude improves the precision with which a trajectory is identified, and therefore narrows them. The no-jump weight $\tilde F_{\mathrm{nj}}$ remains sharply peaked at $s=0$, but the tails of the no-jump Gaussian peak of the flag in~\cref{fig:gkp_sbs_protocol}(c) extend beyond the region where $f\simeq 0$. Since the feedback is nonnegative, they acquire a small spread toward negative $s$ only. This is the price the optimization pays to better correct the decays it cannot separate from the no-jump peak.

The double-decay weight $\tilde F_{ee}$ shows the limit of the one-flag protocol. Since these trajectories share the arc of the single-decay ones, the outcome cannot tell whether one or two decays occurred, and the feedback applies the correction inferred from the single-decay hypothesis. The triangular density $F_{ee}$, peaked at $s=-1$, is thus flattened rather than concentrated. $\tilde F_{ee}$ remains spread over $[-2,0]$ and keeps a finite density around the logical error $s=-1$. The single flag therefore resolves the first-order decays but leaves the $\mathcal{O}((\kappa T)^2)$ component unresolved. The remaining logical error is the weight that the four components leave closer to an odd than to an even value of $s$, and it is dominated, as the flag amplitude increases, by this second-order contribution.

\begin{figure*}[t]
\centering
\includegraphics[width=\textwidth]{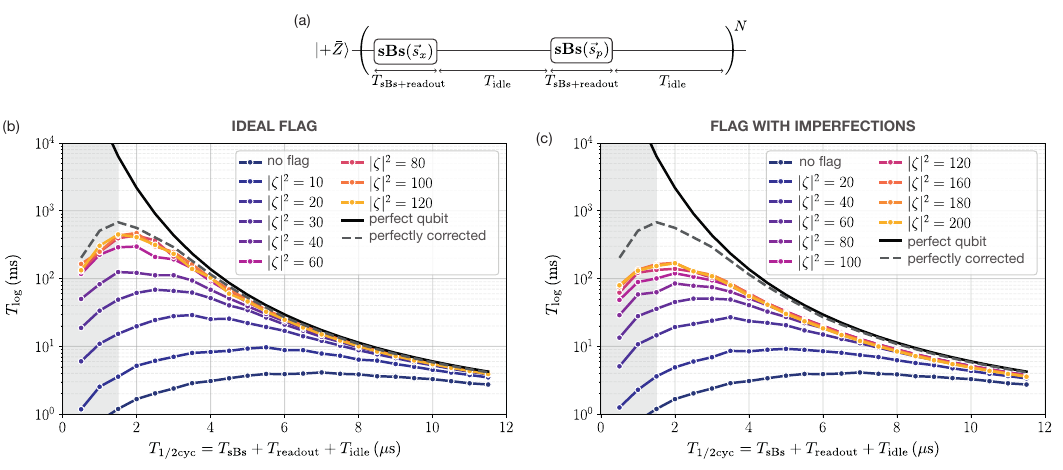}
\caption{
    \textbf{Simulation of the flagged sBs-stabilization protocol on a square GKP code.}
    System parameters as in~\cref{fig:gkp_sbs_protocol}, with memory lifetime $1/\kappa_m = 1~\mathrm{ms}$. The ancilla-decay channels are now the exact ones at finite $\lambda$, see~\cref{eq-apdx:CD_cont_err_channel_finite_lambda}, and the feedback of~\cref{eq:gkp_feedback_lambda} also corrects the residual memory rotations.
    (a) Simulated sequence: $N$ QEC cycles of flagged sBs rounds for the $\vec{s}_x$ and $\vec{s}_p$ stabilizers, separated by idle periods. The decay of $\bar{Z}$ of~\cref{eq:gkp_logicalZ} gives the logical lifetime $T_{\mathrm{log}}$.
    (b) $T_{\mathrm{log}}$ versus the half-cycle duration $T_{1/2\mathrm{cyc}}$, for increasing $|\flag|^2$ and an ideal flag. Black solid, perfect qubit. Grey dashed, perfectly corrected decays. The shaded region corresponds to $T_{1/2\text{cyc}} < T_\text{sBs} + T_\text{readout} = 1.5 \, \mu\mathrm{s}$.
    (c) Same as (b) with flag imperfections, $\eta = 0.5$, $T_1^{\text{flag}} = 1~\mathrm{ms}$ and $K/2\pi = 2~\mathrm{kHz}$.
}
\label{fig:gkp_sbs_simu}
\end{figure*}
 
\subsubsection{Numerical simulation}
 
\emph{Numerical method.} In contrast with the parity measurement of~\cref{subsec:CR_cat_parity}, the logical error per round cannot be read off from a stationary invariant here: the code is stabilized over many rounds, and this stabilization combines with the other noise channels acting in between. We therefore resort to Monte-Carlo simulations, averaging over quantum trajectories, each of which represents a single experimental shot.
 
The logical lifetime is extracted as sketched in~\cref{fig:gkp_sbs_simu}(a). Every trajectory starts from the memory prepared in $\ket{+\bar{Z}}$ and undergoes $N=100$ QEC cycles, each comprising a flagged sBs round for the $\vec{s}_x$ and one for the $\vec{s}_p$ stabilizer, where $\vec{s}_x$ and $\vec{s}_p$ denote the generators of the lattice along the position and momentum quadratures, respectively, separated by idle windows during which the memory relaxes. We record the logical-$Z$ expectation value of~\cref{eq:gkp_logicalZ}, sampled every two cycles only, so as to compensate for the logical Paulis that the sBs sequence applies at each cycle. The ensemble-averaged decay of $\langle\bar{Z}\rangle$ is then fitted by a single exponential, whose time constant defines the logical-$Z$ lifetime $T_{\mathrm{log}}$. By the symmetry of the square lattice under exchange of the two quadratures, the logical-$X$ lifetime is expected to behave analogously.
 
Within a cycle, each sBs round is decomposed into three conditional displacements separated by two qubit rotations, and followed by the qubit reset, the heterodyne measurement of the flag, and the corresponding feedback on the memory. For every conditional displacement, the qubit history is drawn among the four contributions of the channel of~\cref{eq:CD_cont_err_channel}, namely no jump, a single jump before or after the echo, and a double jump, with the jump times sampled from the associated qubit decay law. The memory density matrix is then propagated with the closed-form displacement-and-rotation operator corresponding to the drawn trajectory, evaluated at the finite ECD intermediate amplitude used in the simulation (finite $\lambda$, see~\cref{eq-apdx:CD_cont_err_channel_finite_lambda}). The simulation therefore retains the residual memory rotations that the large-$\lambda$ limit of~\cref{fig:gkp_sbs_protocol}(b)-(d) discards.
 
The flag itself is not propagated as a dynamical mode. Since it remains disentangled from the memory and the qubit after each gate of a given trajectory, its state is entirely determined, classically, by the drawn jump times, and needs only be reconstructed at readout. States and unitaries therefore remain matrices of size $2 N_F \times 2 N_F$, with a Fock truncation $N_F = 150$ of the memory sufficient for the GKP amplitudes considered here. At the end of each round, the flag state is built from the trajectory data, including the deterministic Kerr shearing of~\cref{subsec:kerr}, and the heterodyne outcome $\gamma$ is sampled from the corresponding Husimi distribution.

At finite $\lambda$, the feedback is extended to the two-parameter family $\C_\gamma = \Rm(g(\gamma))\,\D(-f(\gamma)\beta)$. Since rotation errors are now involved, rather than optimizing $\bar{Z}$ as in~\cref{eq:gkp_optimal_feedback}, we directly maximize the fidelity to the finite-energy codeword $\ketbra{-\bar{Z}}=\bar{Z}^-$, with
\begin{equation} \label{eq:gkp_feedback_lambda}
\begin{aligned}
&(f(\gamma), g(\gamma)) \\
&= \underset{(\mu, \vartheta)}{\operatorname{argmax}}
\Tr\!\left[ \bar{Z}^- \, \R(\vartheta)\D(-\mu\beta)\,\hrho_{|\gamma}\,
\D(-\mu\beta)^\dagger \R(\vartheta)^\dagger \right].
\end{aligned}
\end{equation}
This feedback is written for the $\vec{s}_p$ round; that of the $\vec{s}_x$ round follows by rotation symmetry, upon changing the displacement axis.
 
The remaining error channels enter as follows. Flag photon loss is included through the qubit dephasing factor of~\cref{eq:flag_decay_dephasing} rather than at the level of the flag state, and the intrinsic qubit dephasing as a further reduction of the same coherence. Memory relaxation is applied as the exact amplitude-damping channel once per half-cycle, for a duration $T_{1/2\mathrm{cyc}}$ that also accounts for the flagged sBs round and the readout.
 
\emph{Parameters.} We simulate the sBs protocol on a memory of lifetime $T_1^{\text{memory}} = 1/\kappa_m = 1~\mathrm{ms}$, dispersively coupled at $\chi_m/2\pi = 50~\mathrm{kHz}$ to an ancillary qubit of lifetime $T_1^{\text{qubit}} = 1/\kappa = 100~\mu\mathrm{s}$, and stabilizing a finite-energy GKP code of size $\Delta = 0.25$. The Big and small conditional displacements last $500~\mathrm{ns}$ and $250~\mathrm{ns}$ respectively, the Big one being realized with $\lambda = 1/\sin(\chi_m T_{\mathrm{Big}}/2)\simeq 13$, so that one sBs round takes $T_{\mathrm{sBs}} = 1~\mu\mathrm{s}$ and the qubit decays with probability $\simeq 0.5\%$ per round. The flag is coupled at $\chi_f/2\pi = 1~\mathrm{MHz}$, which sets its rotation angle during each of the two conditional rotations of the Big displacement to $\ftheta = \pi/4$, so that the arc of~\cref{eq:CD_flag_position} spans half a turn, as seen in~\cref{fig:gkp_sbs_protocol}(c)-(d).
 
The memory relaxes over the whole half-cycle $T_{1/2\mathrm{cyc}} = T_{\mathrm{sBs}} + T_{\mathrm{readout}} + T_{\mathrm{idle}}$, which we scan from $0.5$ to $11.5~\mu\mathrm{s}$ in steps of $0.5~\mu\mathrm{s}$. At $T_{\mathrm{readout}} = 500~\mathrm{ns}$, the sBs round and the readout already take $1.5~\mu\mathrm{s}$, and the shortest half-cycles of this scan are not experimentally accessible, which we mark by the shaded region of~\cref{fig:gkp_sbs_simu}(b)-(c). The two panels differ only by the flag imperfections included in the simulation, which are analyzed in~\cref{sec:limitations}. \Cref{fig:gkp_sbs_simu}(b) assumes an ideal flag, with unit heterodyne efficiency $\eta = 1$, no flag photon loss ($T_1^{\text{flag}} = \infty$) and no flag self-Kerr ($K = 0$), whereas~\cref{fig:gkp_sbs_simu}(c) uses the more realistic values $\eta = 0.5$, $T_1^{\text{flag}} = 1~\mathrm{ms}$ and $K/2\pi = 2~\mathrm{kHz}$.
 
\emph{Reference curves.} Two flag-free references bracket the flagged results in both panels. The solid black line is the perfect-qubit bound, obtained by running the same simulation with an ancilla free of relaxation and dephasing. The only remaining decoherence is memory photon loss during the idle windows, together with the intrinsic imperfection of the finite-energy sBs pumping, and the logical lifetime is maximized by running the cycles as fast as possible. The grey dashed line, labelled \textit{perfectly corrected}, is instead a trajectory simulation in which the flag is present but tracked classically: rather than sampling a heterodyne outcome from the Husimi distribution, the simulation uses the exactly known flag position, so that every single decay occurring during the Big conditional displacement is identified with certainty and the displacement and rotation it induces are undone exactly. This curve is therefore the numerical realization of the semiclassical limit of the flagging protocol. That it still falls short of the perfect-qubit bound is precisely a measure of what the single flag cannot correct, namely the double-decay trajectories of the Big displacement, whose unresolved weight is the one displayed by $\tilde F_{ee}$ in~\cref{fig:gkp_sbs_protocol}(b), and the decays occurring during the two small conditional displacements, which are not flagged at all.
 
\emph{Competing error mechanisms.} Two mechanisms shape the curves of~\cref{fig:gkp_sbs_simu}(b)-(c). Memory decay acts throughout each half-cycle, so its impact grows with $T_{1/2\text{cyc}}$, and in the limit of a perfect qubit this is the dominant limitation, the lifetime then decreasing monotonically with the cycle duration. Decay of the ancilla qubit, on the other hand, occurs once per sBs cycle with a fixed probability of causing a logical error; its contribution therefore increases as the cycles repeat more frequently, \ie as $T_{1/2\text{cyc}}$ decreases. Their interplay produces an optimal idle time, hence an optimal cycle duration, visible as the peak of each curve.
 
\emph{Effect of the flag size.} Increasing the flag mean photon number widens, in units of the vacuum fluctuations, the arc of~\cref{fig:CD_oneflag}(c) along which the jump time is encoded, so that the heterodyne measurement resolves it more finely and the first-order errors of the Big displacement are corrected more accurately. This is the same sharpening that narrows the peaks of~\cref{fig:gkp_sbs_protocol}(b), and it has two visible consequences in~\cref{fig:gkp_sbs_simu}(b). Because each round is now less likely to produce a logical fault, the protocol can afford to correct more frequently and the optimum shifts toward shorter cycles. The peak logical lifetime correspondingly rises with $|\flag|^2$, from $\simeq 4~\mathrm{ms}$ for the unflagged protocol to $\simeq 450~\mathrm{ms}$ at $|\flag|^2 = 120$, within a factor of $1.5$ of the perfectly corrected curve, and the flagged curves merge with the perfect-qubit bound at long cycles, where memory decay dominates and the qubit contribution is negligible.

The improvement nevertheless saturates before the perfectly corrected limit is reached. The mechanism responsible is again a second-order process, but one specific to the flagged sequence rather than to the bare gate. Although we stressed that the two small conditional displacements ideally act as the identity on the flag, they are not immune to ancilla decay. As long as the flag is in vacuum, which is invariant under conditional rotations, such a decay leaves it unaffected. This is always the case for the leading small displacement, and for the trailing one whenever the ancilla did not decay during the Big conditional displacement. After a decay during the Big conditional displacement, however, the flag is left at the position $z(\tau)$ of~\cref{eq:CD_flag_position} rather than in vacuum, and becomes sensitive to a further decay during the trailing small displacement. Such a second decay leaves the flag with a residual rotation and moves it away from the locus of~\cref{eq:CD_flag_position} on which the inference relies, thereby degrading the accuracy of the feedback correction. This effect is also of second order in $\kappa T$, and once first-order errors are corrected accurately enough, these processes dominate the residual logical error and enlarging the flag brings no further gain. One way to lift them is to replace the last small conditional displacement and reset of the qubit by a qubit measurement and flag readout followed by a feedback on the memory combining both the small displacement and the correction.
 
\emph{Effect of the flag imperfections.} \Cref{fig:gkp_sbs_simu}(c) repeats the analysis with a realistic flag, and shows that the qualitative picture is unchanged while the photon budget is not. A finite heterodyne efficiency blurs the outcome, which by~\cref{eq:efficiency_scaling} is equivalent to an ideal measurement on a flag of amplitude $\sqrt{\eta}\flag$, so that reaching a given encoding quality requires $1/\eta$ times more flag photons, which is why comparable performance is obtained in~\cref{fig:gkp_sbs_simu}(c) only at photon numbers about twice as large as in~\cref{fig:gkp_sbs_simu}(b). Flag photon loss and intrinsic qubit dephasing act, in this context, in the same way. Both reduce the qubit coherence,~\cref{eq:flag_decay_dephasing}, without displacing the memory, and, as discussed in~\cref{subsec:flag_photon_loss}, effectively slow down error correction, which is far less harmful than a first-order qubit decay left uncorrected. The flag self-Kerr, finally, is the imperfection that scales worst with the resource the protocol relies on, as it adds an uncertainty to the heterodyne detection of the flag, see~\cref{subsec:kerr}. Flag photon loss and self-Kerr, which both grow with $|\flag|^2$, together set the level at which the curves of~\cref{fig:gkp_sbs_simu}(c) saturate, below the optimum reached with an ideal flag. Even with this realistic set of parameters, however, the gain remains substantial: the optimal logical lifetime rises from $\simeq 4~\mathrm{ms}$ without flag to $\simeq 170~\mathrm{ms}$ at $|\flag|^2 = 200$, a more than forty-fold improvement, obtained by correcting at first order the very decay events that otherwise dominate the logical error budget of the sBs protocol.

\subsection{All-order protection with two flags}\label{subsec:CD-two-flags}

\begin{figure}[t]
\centering
\includegraphics[width=\columnwidth]{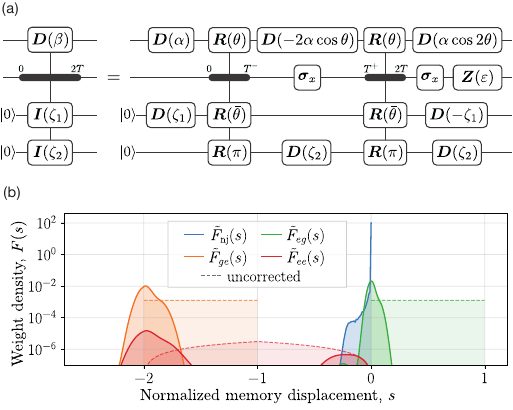}
\caption{
\textbf{Flagging sequence for the $\CD(\beta)$ gate with two flags.}
(a) Flagging sequence of~\cref{eq:protocol_ECD_two_flags}. The first flag follows the one-flag protocol of~\cref{fig:CD_oneflag}(a). The second, $\chi$-matched, stays in vacuum during the first conditional rotation, which leaves it unaffected, and follows the sequence of~\cref{eq:protocol_CR_matching} during the second, thereby recording post-echo decays only.
(b) Weight functions of the memory displacement $s\beta$, before (dashed,~\cref{eq:CD_cont_err_channel_Kraus_Ops}) and after (solid) correction, in the pure-displacement limit $\lambda\to\infty$, for the parameters of~\cref{fig:gkp_sbs_protocol}, $|\flag_1|^2 = 100$, and $|\flag_2|^2 = 20$. Unlike in~\cref{fig:gkp_sbs_protocol}(b), the double-decay weight $\tilde F_{ee}$ is now concentrated around $s=-2$ and $s=0$.
}
\label{fig:CD_twoflags}
\end{figure}

As discussed in~\cref{subsec:CD-one-flag}, the one-flag protocol fails at second order: for double decays, the flag records $s_1+s_2$ while the memory is displaced by $(s_1-s_2)\beta$. The missing information can be retrieved with an additional flag mode monitoring one or both conditional rotations, as in~\cref{sec:CV_flag_for_CR}. Unlike the independent flagging of the two rotations discussed at the beginning of this section, this construction is not defeated by decays adjacent to the echo, since these are already recorded by the first flag. We focus here on the case where only the second rotation is monitored. We assume for simplicity the $\chi$-matching condition $\chi_{f_2}=\pi\chi_m/\theta_\lambda$. As far as ancilla decay alone is concerned, this assumption can be relaxed as in~\cref{subsubsec:CR-no-chi-matching}, see~\cref{subsubsec-apdx:CD_flagging}. The flagging sequence, illustrated in~\cref{fig:CD_twoflags}(a), reads
\begin{equation} \label{eq:protocol_ECD_two_flags}
\begin{split}
    \FCD(\beta,\flag_1,\flag_2) &= \Z_q(\varepsilon_\lambda)\hsigma_x\,\D_{f_1}(-\flag_1) \\
    &\quad \times \ECD_{m f_1 f_2}(\alpha_\lambda,\theta_\lambda)\,\D_{f_1}(\flag_1),
\end{split}
\end{equation}
where $\D_{f_i}$ displaces flag $i$, and $\ECD_{m f_1 f_2}$ denotes the echoed sequence of~\cref{eq:ECD_def} in which each conditional rotation is promoted to the three-mode rotation $\bm{C\!R}_{m f_1 f_2}(\theta_\lambda,\ftheta_\lambda,\pi)$, which extends the joint rotation of~\cref{eq:CRR_def} by a $\pm\pi$ rotation of the second flag. The second flag is only activated during the second conditional rotation, which is accordingly framed, within $\ECD_{m f_1 f_2}$, by the two displacements $\D_{f_2}(\flag_2)$, as in~\cref{eq:protocol_CR_matching}; during the first one, it stays in vacuum and is left unaffected. Without decay, both flags return to vacuum, as in~\cref{fig:CD_oneflag}(b) and~\cref{fig:protocol_CR_matching}(b), so that~\cref{eq:flag_disentangling_condition} holds. As shown in~\cref{subsec-apdx:flagging_channels}, in the limit $\lambda\gg1$ the sequence yields
\begin{widetext}
\begin{equation} \label{eq:2flag_flagging_channel}
    \begin{aligned}
        \mathcal{F}_{D}(\hrho) = {}& F_{\mathrm{nj}}\,\mathcal{D}^{\mathrm{nj}}_{0}(\hrho)\otimes\ketbra{0,0}
        + \int_{-2}^{-1}\! ds\, F_{ge}(s)\,\mathcal{D}^{ge}_{s\beta}(\hrho)\otimes\ketbra{\tilde z_{ge}(s),0} \\
        &+ \int_{0}^{1}\! ds\, F_{eg}(s)\,\mathcal{D}^{eg}_{s\beta}(\hrho)\otimes\ketbra{\tilde z_{eg}(s),\tilde w_{eg}(s)} \\
        &+ \int_{0}^{1}\!\! ds_1\!\int_{1}^{2}\!\! ds_2\, F_{ee}(s_1+s_2)\,\mathcal{D}^{ee}_{(s_1-s_2)\beta}(\hrho)\otimes\ketbra{\tilde z_{ee}(s_1+s_2),\tilde w_{ee}(s_2)},
    \end{aligned}
    \end{equation}
\end{widetext}

with the first-flag amplitudes of~\cref{eq:CD_oneflag_values} ($\flag\to\flag_1$) and the second-flag amplitudes, obtained from~\cref{eq:CR_flag_position},
\begin{equation} \label{eq:2flag_gamma_traj}
\begin{aligned}
\tilde w_{eg}(s)&=\flag_2\left(1-e^{-2i\pi s}\right),\\
\tilde w_{ee}(s_2)&=\flag_2\left(1-e^{2i\pi s_2}\right).
\end{aligned}
\end{equation}
The pair of flag states now determines $s_1+s_2$ and $s_2$, hence the memory displacement, up to $\D(\pm2\beta)$ ambiguities arising from the endpoint degeneracies of the second flag. As $\D(\pm2\beta)$ is a stabilizer, the correlation condition of~\cref{eq:correlation_condition} holds for all trajectories, \ie at all orders in ancilla decay.

Both flags are read out by heterodyne detection and the feedback $\D(-f(\gamma_1,\gamma_2)\beta)$ is optimized as in~\cref{eq:gkp_optimal_feedback}, with the joint likelihood $p(\gamma_1|\tilde{z})\,p(\gamma_2|\tilde{w})$. The resulting weight functions are shown in~\cref{fig:CD_twoflags}(b). The single-decay weights are essentially those of the one-flag protocol of~\cref{fig:gkp_sbs_protocol}(b), but thanks to the all-order encoding, the double-decay weight $\tilde F_{ee}$ is now also concentrated around $s=-2$ and $s=0$, and the continuous error channel of~\cref{eq:CD_cont_err_channel} is fully discretized.

Remarkably, provided the second flag is $\chi$-matched, this all-order protection extends to thermal excitations of the ancilla, see~\cref{subsec-apdx:pm_CD}.

\section{Robustness to experimental imperfections}
\label{sec:limitations}

Until now, the analysis has been based on four idealizations: heterodyne detection with unit efficiency, a flag mode free of photon loss, an ideally harmonic flag mode, and control pulses fast enough to be considered instantaneous. We will now relax these assumptions one by one and show that none of them constitutes a fundamental obstacle. The finite detection efficiency leads to a moderate increase in the number of flag photons; the decay of the flag causes dephasing of the ancilla, which remains less harmful than the logical errors caused by the ancilla; and the Kerr nonlinearity of the flag and the finite duration of the control pulses mostly require a deterministic recalibration of the protocol.

\subsection{Measurement efficiency}

As established in~\cref{sec:CV_flag_for_CR} and~\cref{sec:CV_flag_for_CD}, the performance of the protocols is governed by the flag size $|\flag|^2$, which sets the phase-space separation between the coherent states encoding distinct jump times. So far, the heterodyne measurement resolving them was taken to be ideal, with $\eta = 1$. For a finite detection efficiency $\eta < 1$, the probability density of obtaining the outcome $\gamma$ when the flag is in the coherent state $\ket{\flag}$ reads $p_\eta(\gamma|\flag) = \frac{\eta}{\pi}\, e^{-\eta|\gamma-\flag|^2}$. The outcome remains centered on $\flag$, but the Gaussian broadens from unit variance to $1/\eta$. This blurring amounts to a contraction of the coherent-state amplitude, as follows from the scaling identity
\begin{equation} \label{eq:efficiency_scaling}
p_\eta(\gamma|\flag)\, d^2\gamma = p_{\eta=1}\!\left(\sqrt{\eta}\,\gamma \,\middle|\, \sqrt{\eta}\,\flag\right) d^2(\sqrt{\eta}\,\gamma),
\end{equation}
that is, an inefficient measurement on a coherent state of amplitude $\flag$ has the same statistics as an ideal measurement on a coherent state of amplitude $\sqrt{\eta}\,\flag$.

This contraction can be compensated for at the source: by rescaling every displacement of the flagging sequence by $1/\sqrt{\eta}$, the effective amplitude is restored to its target value, at the cost of increasing the flag photon number by a factor $1/\eta$: a detection efficiency $\eta = 0.5$, for example, doubles the photon budget. Heterodyne inefficiency therefore constitutes more of a resource overhead than an intrinsic obstacle, implying a trade-off between detector quality and flag photon number, with the following caveat: operating the flag at larger amplitudes makes the protocol more sensitive to imperfections of the flag mode itself, which we will discuss next.

\subsection{Flag mode's photon loss} \label{subsec:flag_photon_loss}

We now turn to the intrinsic imperfections of the flag mode, starting with its photon loss at a rate $\kappa_f$. The deterministic amplitude decay $\flag \to \flag e^{-\kappa_f t/2}$ merely rescales the encoding map $\tau \mapsto z(\tau)$ and is absorbed in calibration. The only nontrivial effect is the back-action of the photon loss events themselves: these losses are accompanied by random phase-kicks of the qubit to which it is entangled. Therefore, the flag decay acts as an effective ancilla dephasing channel.

The mechanism is most easily understood at the level of a single ancilla-flag pair. Consider the ancilla disentangled from the flag and prepared in a superposition $c_g\ket{g} + c_e\ket{e}$, with the flag in $\ket{\flag}$. Under the dispersive coupling $-\tfrac{\chi_f}{2}\hf^\dagger\hf\,\hsigma_z$ of~\cref{eq:H_disp_with_flag}, the two ancilla components rotate the flag in opposite directions,
\begin{equation}
    c_g\ket{g}\ket{\flag e^{-i\chi_f t/2}} + c_e\ket{e}\ket{\flag e^{i\chi_f t/2}}.
\end{equation}
A photon-loss event applies the annihilation operator $\hf$ to the flag, which multiplies each branch by its coherent amplitude and thereby imprints a relative phase on the ancilla, leaving the unnormalized state
\begin{equation}
    e^{-i\chi_f t/2}\,c_g\ket{g}\ket{\flag e^{-i\chi_f t/2}} + e^{i\chi_f t/2}\,c_e\ket{e}\ket{\flag e^{i\chi_f t/2}}.
\end{equation}
Each loss event thus kicks the ancilla by a phase determined by the moment it occurs, while between these events, the coherent amplitude decreases as $\flag e^{-\kappa_f t/2}$. Since the times of loss are random, the cumulative phases result, on average, in a loss of coherence of the ancilla, such that the flag manifests itself as pure dephasing channel on the ancilla qubit.

To make this quantitative, recall that a pure-dephasing channel on the ancilla
\begin{equation}
    \mathcal{N}_{\text{deph}}(\hrho) = \K_0^{\text{d}} \hrho (\K_0^{\text{d}})^\dagger + \K_1^{\text{d}} \hrho (\K_1^{\text{d}})^\dagger,
\end{equation}
is fully specified by a single complex dephasing factor $D$ with $|D|\le 1$, through the Kraus operators
\begin{equation} \label{eq:dephasing_kraus}
    \K_0^{\text{d}} = \ketbra{g} + D\ketbra{e}, \quad \K_1^{\text{d}} = \sqrt{1-|D|^2}\,\ketbra{e}.
\end{equation}
This channel multiplies the coherence $\bra{e}\!\hrho\!\ket{g}$ by $D$ while preserving the populations: the modulus $|D|$ determines the strength of the dephasing, and the argument $\arg D$ represents a deterministic phase equivalent to a rotation of the ancilla around the $z$-axis.
Decay of the flag occurs precisely in this way, with a coherence factor $D(t)$ which we now compute, starting from $D(0)=1$. Photon losses form a Poisson process, whose instantaneous rate $\kappa_f|\flag|^2 e^{-\kappa_f t}$ is set by the decaying flag population. During an interval $dt$, either a photon is lost, with probability $\kappa_f|\flag|^2 e^{-\kappa_f t}\,dt$ and multiplying the coherence $\bra{e}\!\hrho\!\ket{g}$ by the phase factor $e^{i\chi_f t}$, or no jump occurs and the coherence is left unchanged. Averaging over these two alternatives yields the recurrence
\begin{equation}
\begin{split}
    D(t+dt) = \big[1 &- \kappa_f|\flag|^2 e^{-\kappa_f t}\,dt \\
    &+ \kappa_f|\flag|^2 e^{-\kappa_f t} e^{i\chi_f t}\,dt \big] D(t).
\end{split}
\end{equation}
Integrating this recurrence up to the gate duration $T$, the coherence factor takes the form $D(T) = e^{i\varphi}\, e^{-r}$, where the deterministic phase $\varphi = \arg D(T)$ is a coherent ancilla rotation that can be calibrated out experimentally, and the modulus $e^{-r} = |D(T)|$ encodes the residual pure dephasing, with
\begin{equation} \label{eq:flag_decay_dephasing}
\begin{split}
    r= |\flag|^2 \bigg[ &\frac{\chi_f^2}{\kappa_f^2+\chi_f^2} - e^{-\kappa_f T} \\
    &+ \frac{\kappa_f e^{-\kappa_f T}}{\kappa_f^2+\chi_f^2}\big(\kappa_f\cos\chi_f T - \chi_f\sin\chi_f T\big) \bigg].
\end{split}
\end{equation}
In the relevant regime where the flag losses are rare, $\kappa_f T \ll 1$ and $\kappa_f \ll \chi_f$, this expression reduces to the compact form
\begin{equation} \label{eq:flag_decay_dephasing_approx}
    r \simeq \frac{\kappa_f|\flag|^2}{\chi_f}\big(2\bar\theta - \sin 2\bar\theta\big),
\end{equation}
where $2\bar\theta = \chi_f T$ is the total flag rotation angle. The dephasing is therefore determined by the ratio of the photon loss rate $\kappa_f|\flag|^2$ to the dispersive rate $\chi_f$ at which the flag rotates in phase space, and, as might be expected, increases linearly with the number of photons in the flag $|\flag|^2$.

In the error correction protocols of the previous sections, such a dephasing is not very harmful. In the parity measurement of rotation-symmetric bosonic codes, ancilla dephasing does not propagate to the memory, and only degrades the correlation between the memory parity and the ancilla state, thereby lowering the fidelity of the syndrome readout. It thus appears as a syndrome-measurement error, which is naturally accounted for by the decoder, rather than as an additional logical-error mechanism. For the finite-energy GKP code stabilized with the sBs protocol, the situation is similar since sBs stabilization is known to be intrinsically robust to ancilla dephasing~\cite{Royer_2020, deNeeve_2022, Sivak_2023}. Here a dephasing event essentially spoils the ongoing stabilization round without displacing the memory, so that its dominant effect is an occasional unsuccessful correction cycle rather than a direct logical error. The impact of such flag-induced dephasing is thus effectively slower error correction. Ultimately, the extra photons required due to limited detector efficiency do not create a hard cap on performance; instead, through flag decay, they simply cause a modest and well-characterized increase in syndrome-measurement errors.

For the parity measurement, this error is readily quantified. Just before the  readout, the dephasing channel leaves the qubit in the state $\hrho_\pm = \frac{1}{2}(\Id \pm D \ketbra{e}{g} \pm D^* \ketbra{g}{e})$, where $\pm$ encodes the parity of the storage mode. Reading out the qubit in the $X$ basis, therefore gives a reduced-contrast parity-measurement. More precisely, the measurement assignment error probability is given by
\begin{equation} \label{eq:decay_infidelity}
    \varepsilon = \frac{1 - |D|}{2} = \frac{1-e^{-r}}{2},
\end{equation}
with the parameters of~\cref{subsec:CR_cat_parity}, where $\chi_f/2\pi = 1~\mathrm{MHz}$ and $T = 1~\mu\mathrm{s}$, such that $\bar\theta=\pi$. A flag lifetime of $T_1^{\mathrm{flag}} = 1/\kappa_f = 1~\mathrm{ms}$ and a flag size of $|\flag|^2 = 100$ photons gives $r \simeq 0.1$, hence $\varepsilon \simeq 5\%$. This accounts for most of the contrast degradation reported in~\cref{fig:cat_parity}(e), the remainder coming from the $|\flag|$-independent floor set by the intrinsic qubit dephasing.

\subsection{Kerr nonlinearity of flag mode}
\label{subsec:kerr}

\begin{figure}[t]
\centering
\includegraphics[width=\columnwidth]{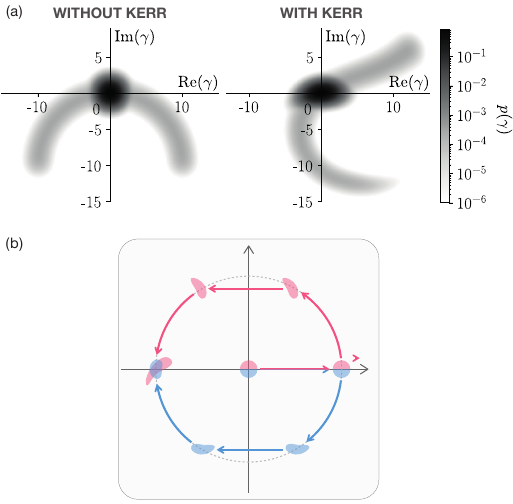}
\caption{
    \textbf{Effect of the Kerr nonlinearity of the flag.}
    (a) Probability density $p(\gamma)$ of the heterodyne outcomes without and with the Kerr term of~\cref{eq:Kerr_Hamiltonian}, for the parameters of~\cref{fig:gkp_sbs_protocol}(c). The distribution is sheared, but the jump times remain resolvable.
    (b) Flag trajectories for the unmatched $\CR(\theta)$ protocol of~\cref{eq:CR_unmatched}. The intermediate displacement does not commute with the Kerr evolution, and the two components no longer recombine exactly.
}
\label{fig:kerr}
\end{figure}

Throughout the previous sections, the qubit-dependent state of the flag mode has been assumed to remain coherent at every stage of the protocol. This holds only as long as the flag cavity behaves as a harmonic oscillator, so that displacements and dispersive rotations map coherent states onto coherent states. Nonetheless, the dispersive qubit-flag interaction usually comes with a residual self-Kerr nonlinearity of the flag mode,
\begin{equation}\label{eq:Kerr_Hamiltonian}
\Ha_K/\hbar = -\frac{K}{2}\,(\hf^\dagger)^2\hf^2.
\end{equation}
The impact of the flag state  distortion due to this Kerr non-linearity depends on the flagging sequence. When the flag undergoes no intermediate displacement, the Kerr evolution commutes throughout the sequence and its impact is merely an additional uncertainty in the flag's heterodyne detection. When intermediate displacements are present, the Kerr evolution additionally leaves a residual entanglement between the flag and the ancilla. We examine the two cases in turn.

The first case covers protocols in which the flag is acted upon only by conditional rotations between its initial and final displacements, such as the matched conditional rotation protocol~\cref{eq:protocol_CR_matching} and the conditional displacement protocol~\cref{eq:protocol_ECD_one_flag}. Because the Kerr Hamiltonian is diagonal in the Fock basis, it commutes with the dispersive interaction: the two ancilla components accumulate the same Kerr distortion over the same total time, recombine into a single common state, and the flag disentangles from the rest of the system as in the ideal case. The main modification is a deterministic recalibration of the protocol: the Kerr term imprints a fixed phase-space rotation whose angle is set by the average photon number in the flag mode. So the displacements and the compensating qubit rotation must be re-adjusted for the two semiclassical trajectories $\langle\hf\rangle_g$ and $\langle\hf\rangle_e$ to meet at the end of the sequence.

The additional uncertainty in the flag's heterodyne detection induced by this Kerr effect can be formulated as a modified probability density of such measurement outcomes. This is illustrated in~\cref{fig:kerr}(a), which compares this probability density $p(\gamma)$ with and without Kerr, for the parameters of the sBs protocol simulations in~\cref{subsec:CD-gkp-sbs} and a flag prepared with $|\flag|^2=100$ photons: the distribution obtained with Kerr is a sheared version of the ideal one, and the various jump-time contributions remain resolvable.

\begin{figure}[t]
\centering
\includegraphics[width=\columnwidth]{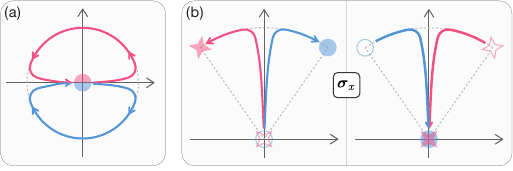}
\caption{
    \textbf{Effect of control pulses of finite duration.}
    Flag trajectories conditioned on the two ancilla states, for displacements and qubit rotations of finite duration. They follow smooth paths set by the drive envelopes, and the protocol remains valid as long as they recombine at the end of the sequence,~\cref{eq:finite_pulse_closure}. (a) $\chi$-matched $\CR(\theta)$ protocol of~\cref{eq:protocol_CR_matching}. (b) $\CD(\beta)$ protocol of~\cref{eq:protocol_ECD_one_flag}.}
\label{fig:finite_disp}
\end{figure}

The second case covers implementations that interleave conditional rotations with intermediate flag displacements, such as the conditional rotation protocol away from the matching condition~\cref{eq:CR_unmatched}, shown in~\cref{fig:kerr}(b). Intermediate displacements no longer commute with the Kerr evolution: the two flag components hence undergo distinct distortions and no longer recombine exactly. This results in a residual entanglement between the flag and the system, which, tracing out the flag, is converted into an effective dephasing of the ancilla.

After the sequence the memory still disentangles from the ancilla and the flag, so that the encoded information is left untouched, and the residual imperfection is confined to the joint qubit-flag state
\begin{equation}\label{eq:qubit_flag_entangled}
    \frac{\ket{g}\ket{\psi_g}\pm\ket{e}\ket{\psi_e}}{\sqrt{2}},
\end{equation}
where $\ket{\psi_g}$ and $\ket{\psi_e}$ are the two Kerr-distorted flag states left by the trajectories conditioned on $\ket{g}$ and $\ket{e}$. The parity is read from the qubit coherence in the $\{\ket{+},\ket{-}\}$ basis, obtained by tracing out the flag,
\begin{equation}\label{eq:reduced_qubit_kerr}
    \hrho_\pm=\frac{1}{2}\left(\Id \pm s\,\ketbra{e}{g} \pm s^*\,\ketbra{g}{e}\right),
    \qquad s=\braket{\psi_g|\psi_e}.
\end{equation}
A perfect recombination, $\ket{\psi_g}=\ket{\psi_e}$, gives $|s|=1$ and leaves the coherence intact, whereas any residual distinguishability of the two flag states makes $|s|<1$ and leads to a parity measurement infidelity
\begin{equation}\label{eq:kerr_infidelity}
    \varepsilon=\frac{1-|\braket{\psi_g|\psi_e}|}{2}.
\end{equation}

This adds to the dephasing produced by flag decay in the previous subsection. The two mechanisms act on independent features of the flag dynamics and, to leading order, reduce the qubit coherence multiplicatively, the first through the factor $e^{-r}$ with $r$ given in~\cref{eq:flag_decay_dephasing}, and the second through the overlap $|\braket{\psi_g|\psi_e}|$. Together they give the parity-measurement infidelity
\begin{equation}\label{eq:combined_infidelity}
    \varepsilon=\frac{1-e^{-r}\,|\braket{\psi_g|\psi_e}|}{2}.
\end{equation}

\subsection{Control pulses of finite duration}
\label{subsec:finite_duration}
Throughout this paper, we treated memory and flag displacements and qubit rotations as instantaneous compared with the slower conditional rotations. This assumption was  made to simplify the presentation. With control pulses of finite duration, the flag simply follows smooth phase-space trajectories set by the drive envelopes instead of the piecewise circular arcs of the instantaneous limit, as illustrated in~\cref{fig:finite_disp}. The protocol remains valid as long as the two conditional trajectories $\langle\hf\rangle_g(t)$ and $\langle\hf\rangle_e(t)$ still recombine at the end of the sequence,
\begin{equation} \label{eq:finite_pulse_closure}
    \langle\hf\rangle_g(T_{\text{tot}}) = \langle\hf\rangle_e(T_{\text{tot}}),
\end{equation}
with $T_{\text{tot}}$ the total sequence duration, so that the no-jump evolution leaves the flag disentangled from the rest of the system.

Control pulses of finite duration also modify the final flag positions $\ket{z(\vec{\tau})}$ associated with the different jump times, but the mapping $\vec{\tau}\mapsto z(\vec{\tau})$ remains deterministic and follows from the known pulse shapes. Therefore, as in the Kerr case, it suffices to evaluate the estimator and the feedback with the updated encoding.

\section{Conclusion} \label{sec:conclusion} 

We have introduced the continuous-variable flag, a protocol that protects ancilla-assisted bosonic operations against spontaneous ancilla decay. Its principle is to record the error induced on the memory by an ancilla decay in the phase-space position of an auxiliary oscillator coupled to the ancilla. A heterodyne measurement of this flag then provides an estimate of the error, and a subsequent feedback undoes it up to a stabilizer of the bosonic code, thereby discretizing the continuous error channel into errors that the code tolerates. We constructed explicit flagging sequences protecting the $\CR(\theta)$ gate to all orders in ancilla decay with a single flag, and the $\CD(\beta)$ gate to first order with one flag and to all orders with two, the $\chi$-matched protocols extending this protection to thermal excitations of the ancilla. Benchmarked with realistic circuit-QED parameters, the flag suppresses the ancilla-induced logical bit-flip probability of the four-component cat-code parity measurement by more than three orders of magnitude, and raises the logical lifetime of a sBs-stabilized finite-energy GKP code towards its memory-limited bound. Finally, we showed that none of the dominant experimental imperfections, namely finite detection efficiency, flag decay, flag self-Kerr and finite control-pulse durations, constitutes a fundamental obstacle: the first amounts to a flag photon-number overhead, and the others to benign ancilla dephasing or deterministic recalibrations of the protocol.

The principle underlying the continuous-variable flag is considerably more general than the specific setting considered here. At its core, it consists in correlating an error occurring on one component of a quantum processor with an auxiliary continuous-variable degree of freedom, in such a way that the computational subsystem is minimally disturbed while a continuous record of the error is written into phase space. Nothing in this construction is tied to spontaneous decay, and the same philosophy could be used to flag other ancilla error mechanisms. Several extensions are worth exploring. First, substituting squeezed or non-Gaussian pointer states for the standard coherent flag states could improve the signal-to-noise ratio of jump-time estimates at fixed photon number. Second, the heterodyne readout could be replaced by measurements tailored to extract the maximal information about the jump time, thereby enhancing the precision of trajectory discrimination. Third, and perhaps most appealing from a hardware standpoint, the measure-and-feedback architecture considered here could be replaced by an \textit{autonomous} feedback scheme, in which the flag oscillator interacts directly with the computational system to correct errors coherently, without explicit measurement or digital processing. Such a scheme would retain the core idea of encoding ancilla error information into an auxiliary continuous-variable mode, while realizing the correction at the hardware level. We leave these directions for future work.

Finally, although we have framed the continuous-variable flag as a tool for bosonic quantum error correction, the underlying idea of continuously recording error information in an auxiliary oscillator may also prove useful in the broader context of hybrid oscillator-qubit quantum processors~\cite{Liu_2024,Crane_2024}. In these architectures, the conditional oscillator-qubit interactions that constitute the native instruction set beyond Gaussian operations are precisely the operations through which ancilla decay propagates, so that supplementing them with a spare bosonic mode acting as a continuous-variable flag could offer a hardware-native means of detecting and correcting such faults where they originate.

\section*{Acknowledgments}

The authors thank Jérémie Guillaud for many enlightening discussions that led to the initial idea of this paper. We also thank Philippe Campagne-Ibarcq and Baptiste Royer for insightful discussions. MM acknowledges support from the Plan France 2030 under project ANR-22-PETQ-0006. MM acknowledges support from 
the Japan Science and Technology Agency (JST) Adopting Sustainable Partnerships for Innovative Research Ecosystem (ASPIRE) under Grant Number JPMJAP2513.
\bibliographystyle{apsrev4-2}
\bibliography{bibliography}

\appendix
\clearpage
\onecolumngrid

\section{Error and flagging channels}
\label{sec-apdx:channels}
 
This appendix derives the quantum channels used throughout the main text: the continuous error channels of the bare conditional gates in~\cref{sec:context}, and the flagging channels of the protocols of~\cref{sec:CV_flag_for_CR,sec:CV_flag_for_CD}. All of them follow from a single construction, the \textit{trajectory} or \textit{Dyson expansion} of the Lindblad dynamics, which we set up first and whose notations are used throughout~\cref{sec-apdx:flag_loss,sec-apdx:decay_excitation}. Along the way we obtain the Kraus operators of the noisy conditional displacement at \emph{finite} intermediate displacement $\lambda$, of which the main text only quotes the limit $\lambda \gg 1$, and which are the ones propagated in the simulations of~\cref{subsec:CD-gkp-sbs}.

\subsection{Notations}
For a Hamiltonian $\Ha(t)$ and a set of jump operators $\{\bL_\sigma\}$ with associated rates $\{\kappa_\sigma\}$, the Lindblad equation reads
\begin{equation} \label{eq-apdx:lindblad}
    \frac{d\hrho}{dt} = \mathcal{L}_t(\hrho) = -\frac{i}{\hbar}\left[\Ha(t),\hrho\right] + \sum_\sigma \kappa_\sigma \left( \bL_\sigma\hrho\bL_\sigma^\dagger - \tfrac{1}{2}\{\bL_\sigma^\dagger\bL_\sigma,\hrho\} \right).
\end{equation}
Its solution $\hrho(t) = \mathcal{T}\exp\left[\int_0^t \mathcal{L}_{t'} \, dt'\right](\hrho_0)$ can be organized as a sum over the number of jump events, through the Dyson expansion~\cite{Holevo_2001}
\begin{equation} \label{eq-apdx:Dyson_exp}
    \hrho(t) = \sum_{p=0}^{+\infty} \mathcal{G}_p(t)(\hrho_0),
    \qquad
    \left\{
    \begin{aligned}
        \mathcal{G}_0(t) &= \mathcal{W}(t,0), \\
        \mathcal{G}_p(t) &= \int\limits_0^{t}\! d\tau_p \int\limits_0^{\tau_p}\! d\tau_{p-1} \dots \int\limits_0^{\tau_2}\! d\tau_1 \;
        \mathcal{W}(t,\tau_p) \, \mathcal{S} \, \mathcal{W}(\tau_p,\tau_{p-1}) \dots \mathcal{S} \, \mathcal{W}(\tau_1,0), \quad p \ge 1 .
    \end{aligned}
    \right.
\end{equation}
Two objects enter this expansion. The first is the superoperator $\mathcal{W}(t_2,t_1)(\hrho) = \bm{W}(t_2,t_1)\,\hrho\,\bm{W}^\dagger(t_2,t_1)$, which describes the deterministic evolution between two consecutive jumps. It is generated by the \textit{no-jump propagator}, the non-unitary operator
\begin{equation} \label{eq-apdx:nojump_propagator}
    \bm{W}(t_2,t_1) = \mathcal{T}\exp\left[-\frac{i}{\hbar}\int_{t_1}^{t_2} \Ha_{\text{eff}}(t) \, dt \right],
    \qquad
    \Ha_{\text{eff}}(t) = \Ha(t) - \frac{i\hbar}{2}\sum_\sigma \kappa_\sigma \bL_\sigma^\dagger \bL_\sigma .
\end{equation}
The second is the \textit{jump superoperator}
\begin{equation} \label{eq-apdx:jump_superoperator}
    \mathcal{S}(\hrho) = \sum_\sigma \kappa_\sigma \, \bL_\sigma \, \hrho \, \bL_\sigma^\dagger ,
\end{equation}
which applies a jump at a given time. Each term of~\cref{eq-apdx:Dyson_exp} therefore has a direct interpretation: $\mathcal{G}_p(t)$ collects the mixture of trajectories undergoing exactly $p$ jumps, at all possible ordered times $0 \le \tau_1 \le \dots \le \tau_p \le t$, with the state evolving under the no-jump propagator in between. In particular, $\mathcal{G}_0(t)$ describes the trajectory without any jump. Comparing with~\cref{sec:cv_flags}, $\mathcal{G}_p(T)$ is nothing but the $p$-jump contribution, of order $(\kappa T)^p$, of the flagged evolution $\mathcal{F} \circ \mathcal{CU}$, or of $\mathcal{N} \circ \mathcal{CU}$ for a bare gate.

When a single jump operator $\bL$, at rate $\kappa$, is retained, as will be the case in this appendix and in~\cref{sec-apdx:flag_loss}, the superoperator form of~\cref{eq-apdx:Dyson_exp} can be unfolded at the level of operators. A trajectory with $p$ jumps at times $\vec{\tau} = (\tau_1,\dots,\tau_p)$ is then described by the trajectory operator $\bm{J}_{\vec{\tau}}$, and the $p$-jump contribution to the dynamics collects all such trajectories:
\begin{equation} \label{eq-apdx:traj_operator}
    \mathcal{G}_p(t)(\hrho) = \!\int d^p\vec{\tau} \; \bm{J}_{\vec{\tau}} \, \hrho \, \bm{J}_{\vec{\tau}}^\dagger,
    \quad \text{where} \quad
    \bm{J}_{\vec{\tau}} = \bm{W}(t,\tau_p) \, \sqrt{\kappa}\bL \, \bm{W}(\tau_p,\tau_{p-1}) \dots \sqrt{\kappa}\bL \, \bm{W}(\tau_1,0).
\end{equation}

\subsection{Continuous error channels induced by ancilla decay}
\label{subsec-apdx:error_channels}
 
We first apply this machinery to the bare gates of~\cref{sec:context}, where the dynamics is governed by the dispersive Hamiltonian $\Ha_{\text{disp}}/\hbar = -\frac{\chi_m}{2} \ha^\dagger \ha \, \hsigma_z$ and the only jump operator retained is the ancilla decay $\bL = \hsigma_- = \ketbra{g}{e}$, at rate $\kappa$. Since $\hsigma_-^\dagger \hsigma_- = \ketbra{e}$ is diagonal in the $\hsigma_z$ basis, the non-unitary part of $\Ha_{\text{eff}}$ commutes with the dispersive interaction and the no-jump propagator factorizes as
\begin{equation} \label{eq-apdx:W_disp}
    \bm{W}(t_2,t_1) = e^{-\frac{\kappa (t_2-t_1)}{2}\ketbra{e}} \; \CR\!\left(\tfrac{\chi_m (t_2-t_1)}{2}\right),
\end{equation}
that is, the ideal conditional rotation of~\cref{eq:conditional_rotation} damped by the survival amplitude of the excited state.
 
Consider now a trajectory in which the ancilla decays at a time $\tau$ within a dispersive interval of duration $T$. Its trajectory operator,~\cref{eq-apdx:traj_operator} with $p=1$, gives,
\begin{equation} \label{eq-apdx:one_jump_traj}
    \bm{J}_\tau \equiv \bm{W}(T,\tau) \, \sqrt{\kappa}\hsigma_- \, \bm{W}(\tau,0)
    = \sqrt{\kappa} e^{-\frac{\kappa \tau}{2}} \; \R\!\left(\tfrac{\chi_m(2\tau - T)}{2}\right) \otimes \ketbra{g}{e} ,
    \qquad \tau \in [0,T].
\end{equation}
The memory is thus left \emph{unconditionally} rotated by $\chi_m(2\tau-T)/2$, which leads to the state of~\cref{eq:traj_decay_during_CR} and the analytical content of~\cref{fig:CR_cont_err_channel}(a)--(b).
 
Trajectories with $p \ge 2$ jumps vanish: $\bm{W}$ is diagonal in $\{\ket{g},\ket{e}\}$, so nothing repopulates the excited state after a decay, and $\hsigma_- \, \bm{W}(t_2,t_1) \, \hsigma_- = 0$. The dispersive evolution under ancilla decay during a time $T$ is therefore described by the superpropagator
\begin{equation} \label{eq-apdx:superpropagator_disp}
    \mathcal{G}(T) = \mathcal{G}_0(T) + \mathcal{G}_1(T),
    \qquad
    \left\{
    \begin{aligned}
        \mathcal{G}_0(T)(\hrho) &= e^{-\frac{\kappa T}{2}\ketbra{e}} \, \CR\!\left(\tfrac{\chi_m T}{2}\right) \hrho \, \CR^\dagger\!\left(\tfrac{\chi_m T}{2}\right) e^{-\frac{\kappa T}{2}\ketbra{e}}, \\
        \mathcal{G}_1(T)(\hrho) &= \int_0^T \, d\tau \; \bm{J}_\tau \, \hrho \, \bm{J}_\tau^\dagger .
    \end{aligned}
    \right.
\end{equation}
Both error channels of~\cref{sec:context} follow from this single result as shown below.
 
\subsubsection{Continuous error channel for the conditional rotation gate}
\label{subsubsec-apdx:CR_error_channel}
 
The $\CR(\theta)$ gate is realized by letting the system evolve under the dispersive interaction for a time $T = 2\theta/\chi_m$, so that its noisy version is exactly $\mathcal{G}(2\theta/\chi_m)$. Factoring out the target unitary, $\mathcal{G}(T) = \mathcal{N}_R \circ \mathcal{CR}_\theta$, the continuous error channel can be written in terms of the jump time $\tau$ as
\begin{equation} \label{eq-apdx:CR_cont_err_channel}
\displaystyle
\mathcal{N}_R(\hrho) = \K_\mathrm{nj} \hrho \K_\mathrm{nj}^\dagger + \int_{0}^{T} d\tau \; \K_\tau \hrho \K_\tau^\dagger,
\end{equation}
where the Kraus operators, obtained by right-multiplying the trajectory operators by $\CR^\dagger(\theta)$, read
\begin{equation} \label{eq-apdx:CR_cont_err_channel_Kraus_Op}
\begin{aligned}
\K_\mathrm{nj} &= \bm{W}(T,0) \CR^\dagger(\theta) =  e^{-\frac{\kappa T}{2} \ketbra{e}} \\
\K_\tau &= \bm{J}_\tau \, \CR^\dagger(\theta) = \sqrt{\kappa} \,  e^{-\frac{\kappa\tau}{2}} \, \R\!\left(2\theta\left(\tfrac{\tau}{T}-1\right)\right) \otimes \ketbra{g}{e}. \\
\end{aligned}
\end{equation}

Since the errors induced on the memory by the ancilla decay reduce here to rotations, it is useful to recast this channel as a continuous rotation-error channel that makes the distribution of these rotation errors explicit. Changing the integration variable to the rotation angle $\phi = 2\theta(\tau/T - 1)$, we obtain
\begin{equation} \label{eq-apdx:CR_cont_err_channel_rot}
\begin{aligned}
\mathcal{N}_{R}(\hrho) &= \R^\mathrm{nj}_0 \hrho (\R^\mathrm{nj}_0)^\dagger + \int_{-2\theta}^{0}  d\phi \; F_{ge}(\phi) \, \R_\phi^{ge} \hrho (\R_\phi^{ge})^\dagger, \\
\text{with} \quad \R^\mathrm{nj}_0 &= \K_{\mathrm{nj}}, \quad
\R^{ge}_\phi = \R(\phi) \otimes \ketbra{g}{e}, \quad
F_{ge}(\phi) = \frac{\kappa T}{2\theta}e^{-\kappa T\left(1+\frac{\phi}{2\theta}\right)},
\end{aligned}
\end{equation}
which corresponds to~\crefrange{eq:CR_cont_err_channel}{eq:weight_function_uncorr_rot} of the main text.
 
\subsubsection{Continuous error channel for the conditional displacement gate}
\label{subsubsec-apdx:CD_error_channel}
 
The conditional displacement is realized by the ECD sequence of~\cref{eq:ECD_def,eq:link_CD_and_ECD}, in which only the two conditional rotations have a finite duration $T = 2\theta_\lambda/\chi_m$. Its noisy version is therefore obtained by substituting $\mathcal{G}(T)$ of~\cref{eq-apdx:superpropagator_disp} for each conditional rotation of the sequence. Throughout this subsection we abbreviate $\alpha = \alpha_\lambda$ and $\theta = \theta_\lambda$, we use the shorthand $\mathrm{c} = \cos\theta$, $\mathrm{c}_2 = \cos 2\theta$ and $\mathrm{s}_2 = \sin 2\theta$, and we write $\bm{W}_{\mathrm{nj}} = \bm{W}(T,0)$ and $\bm{J}_\tau$ for the no-jump and one-jump operators of a single conditional rotation,~\cref{eq-apdx:W_disp,eq-apdx:one_jump_traj}.

Each of the two noisy conditional rotations contributes a no-jump and a one-jump term, so that expanding the sequence yields four families of trajectories. Unlike for a single conditional rotation, two decays are now possible, because the echo pulse repopulates the excited state between the two conditional rotations. These families are the no-jump trajectory, the trajectories with a single jump before the echo, at $\tau \in (0,T)$, those with a single jump after the echo, at $\tau \in (T,2T)$, and those with one jump on each side of the echo, at $\tau_1 \in (0,T)$ and $\tau_2 \in (T,2T)$. Factoring out the target unitary $\CD(\beta)$, the continuous error channel induced by the ancilla decay thus takes the form
\begin{equation}
\label{eq-apdx:CD_cont_err_channel_finite_lambda}
\mathcal{N}_\lambda(\hrho) = \K_{\mathrm{nj}} \hrho \K_{\mathrm{nj}}^\dagger
+ \int_0^{2T} \, d\tau \; \K_\tau \, \hrho \, \K_\tau^\dagger
+ \int_0^{T} \, d\tau_1 \int_T^{2T} \, d\tau_2 \; \K_{\tau_1,\tau_2} \, \hrho \, \K_{\tau_1,\tau_2}^\dagger,
\end{equation}
where the Kraus operators are obtained by right-multiplying the trajectory operators by $\CD^\dagger(\beta)$. We now derive them family by family.

Since the echo swaps $\ket{g}$ and $\ket{e}$ between the two conditional rotations, each ancilla branch spends exactly one conditional rotation in the excited state. The damping factors of $\bm{W}_{\mathrm{nj}}$ are thus symmetrized into the same global factor $e^{-\kappa T/2}$ on both branches. The remaining unitary part is the ideal ECD sequence, which is precisely engineered to realize $\CD(\beta)$. Hence $\K_{\mathrm{nj}} = e^{-\kappa T/2} \, \Id$, and the no-jump trajectory introduces no error on the memory. 

The trajectories with a jump before the echo at $\tau \in (0,T)$ are obtained by replacing, in the ECD sequence, the first conditional rotation by $\bm{J}_\tau$ and the second one by the no-jump evolution $\bm{W}_{\mathrm{nj}}$. Commuting the displacements through the rotations with $\R(\psi)\,\D(\gamma) = \D(\gamma e^{i\psi})\,\R(\psi)$, we obtain
\begin{flalign}
& \underline{\text{Jump before echo } \tau \in (0,T)} \quad & \K_\tau &= \Z_q(\varepsilon_\lambda) \, \hsigma_x \, \D(\alpha \mathrm{c}_2) \, \bm{W}_{\mathrm{nj}} \, \D(-\alpha\mathrm{c}) \, \hsigma_x \, \D(-\alpha\mathrm{c}) \, \bm{J}_\tau \, \D(\alpha) \CD^\dagger(i\alpha\mathrm{s}_2) && \nonumber \\
& & &= \sqrt{\kappa}\, e^{-\frac{\kappa}{2}(\tau+T)}\, \D\big(-\alpha-i \alpha \mathrm{s}_2+\alpha(1-i\mathrm{s}_2) e^{i\chi_m\tau}\big)  \R(\chi_m \tau) \otimes \ketbra{g}{e}. && \label{eq-apdx:Kraus_op_single_jump_before_CD}
\end{flalign}

Similarly, the trajectories with a jump after the echo at $\tau \in (T,2T)$ are obtained by replacing the first conditional rotation by $\bm{W}_{\mathrm{nj}}$ and the second one by the time-shifted jump operator $\bm{J}_{\tau-T}$, which gives
\begin{flalign}
& \underline{\text{Jump after echo } \tau \in (T,2T)} \quad & \K_\tau &= \sqrt{\kappa} \, e^{-\frac{\kappa}{2} (\tau-T)} \, \D\big(\alpha \mathrm{c}_2- \alpha \mathrm{c}_2 e^{i (\chi_m \tau - 4\theta)}\big)  \R(\chi_m \tau - 4\theta) \otimes \ketbra{e}{g}. && \label{eq-apdx:Kraus_op_single_jump_after_CD}
\end{flalign}

Finally, the trajectories with jumps at $\tau_1 \in (0,T)$ and $\tau_2 \in (T,2T)$ are obtained by replacing the first and second conditional rotations by $\bm{J}_{\tau_1}$ and $\bm{J}_{\tau_2-T}$ respectively, yielding
\begin{flalign}
& \underline{\text{Double jumps}} \quad & \K_{\tau_1,\tau_2} &= \kappa e^{-\frac{\kappa}{2} (\tau_1+\tau_2-T)} \, \D\big(\mathrm{c}_2 \alpha-2\mathrm{c}\,\alpha e^{i (\chi_m \tau_2 - 3\theta )}+\alpha(1-i\mathrm{s}_2)  e^{i (\chi_m (\tau_1+\tau_2) - 4\theta )}\big)  \R(\chi_m (\tau_1+\tau_2) - 4\theta)  \otimes \ketbra{e}. && \label{eq-apdx:Kraus_op_double_jump_CD}
\end{flalign}

In each case, the ancilla operator records which initial ancilla states can undergo the corresponding trajectory: a jump before the echo requires the ancilla to start in $\ket{e}$, a jump after the echo requires it to start in $\ket{g}$, and a double jump requires it to start in $\ket{e}$, the echo repopulating $\ket{e}$ after the first decay. The continuous error channel of~\cref{eq-apdx:CD_cont_err_channel_finite_lambda}, with the Kraus operators of~\cref{eq-apdx:Kraus_op_single_jump_before_CD,eq-apdx:Kraus_op_single_jump_after_CD,eq-apdx:Kraus_op_double_jump_CD}, holds for any $\lambda$, and it is the one we use in the GKP simulations of~\cref{subsec:CD-gkp-sbs}. \\

\paragraph{Large intermediate displacements limit.} In the limit of large intermediate displacements, $\lambda \to \infty$, the conditional rotation angle $\theta = \arcsin(1/\lambda)/2$ vanishes, with $\lambda\theta \to 1/2$, while the intermediate displacements $|\alpha| = \lambda|\beta|$ diverge. The phase accumulated by the memory during the conditional rotations, $\chi_m\tau = 2\theta\,\tau/T$, therefore vanishes. As a result, the residual rotations of the Kraus operators vanish as well, since their angles never exceed $2\theta$ in absolute value. The displacements, on the contrary, remain finite, as they combine diverging displacements of order $\lambda\beta$ with vanishing phases. For a jump before the echo, using $\alpha = -i\lambda\beta$ and $\mathrm{s}_2 = 1/\lambda$, together with $\lambda\left(e^{i\chi_m\tau}-1\right) \to i\tau/T$ and $e^{i\chi_m\tau}\to 1$, we find
\begin{equation}
-\alpha(1+i\mathrm{s}_2)+\alpha(1-i\mathrm{s}_2)e^{i\chi_m\tau} = \alpha\left(e^{i\chi_m\tau}-1\right) - i\alpha \mathrm{s}_2\left(1+e^{i\chi_m\tau}\right) \underset{\lambda\to\infty}{\longrightarrow} \beta\,(\tau/T - 2).
\end{equation}
In this limit, a jump before the echo thus displaces the memory by $s\beta$ along the target displacement axis, with $s = \tau/T - 2 \in [-2,-1]$. Rewriting the integral over the jump time as an integral over $s$, the contribution of these trajectories reads
\begin{equation}
\int_0^{T} \, d\tau \; \K_\tau \, \hrho \, \K_\tau^\dagger = \int_{-2}^{-1} ds \; F_{ge}(s) \, \D^{ge}_{s\beta} \hrho (\D^{ge}_{s\beta})^\dagger,
\end{equation}
with $\D^{ge}_{s\beta}=\D(s\beta)\otimes\ketbra{g}{e}$ and the weight function $F_{ge}(s)=\kappa T e^{-\kappa T(s+3)}$. The same limit applies to the other families. A jump after the echo displaces the memory by $\beta(2-\tau/T)$, which leads to $s = 2 - \tau/T \in [0,1]$. A double jump displaces it by $\beta(\tau_1-\tau_2)/T$, which leads to $(s_1, s_2) = (\tau_1/T, \tau_2/T)$. Altogether, we obtain the continuous displacement-error channel
\begin{equation} \label{eq-apdx:CD_cont_err_channel}
\begin{aligned}
\mathcal{N}_{D}(\hrho) &= F_{\mathrm{nj}}~\D^\mathrm{nj}_0 \hrho (\D^\mathrm{nj}_0)^\dagger + \int_{-2}^{-1} ds \; F_{ge}(s) \, \D^{ge}_{s\beta} \hrho (\D^{ge}_{s\beta})^\dagger
+ \int_{0}^{1} ds \;  F_{eg}(s) \, \D^{eg}_{s\beta} \hrho (\D^{eg}_{s\beta})^\dagger \\
&+\int_{0}^{1} ds_1\int_{1}^2 ds_2 \; F_{ee}(s_1+s_2) \, \D^{ee}_{(s_1-s_2)\beta} \hrho (\D^{ee}_{(s_1-s_2)\beta})^\dagger 
\end{aligned}
\end{equation}
with Kraus operators and associated weight functions
\begin{equation} \label{eq-apdx:CD_cont_err_channel_Kraus_Ops}
\begin{array}{rll}
\D^\mathrm{nj}_{\alpha}&=\D(\alpha), & F_\mathrm{nj} = e^{-\kappa T}\\
\D^{ge}_{\alpha}&=\D(\alpha)\otimes\ketbra{g}{e},& F_{ge}(s)=\kappa T e^{-\kappa T(s+3)}\\
\D^{eg}_{\alpha}&=\D(\alpha)\otimes\ketbra{e}{g},& F_{eg}(s)=\kappa T e^{-\kappa T(1-s)}\\
\D^{ee}_{\alpha}&=\D(\alpha)\otimes\ketbra{e}{e},& F_{ee}(s)=(\kappa T)^2 e^{-\kappa T(s-1)},
\end{array}
\end{equation}
which correspond to~\cref{eq:CD_cont_err_channel,eq:CD_cont_err_channel_Kraus_Ops} of the main text.
 
\subsection{Flagging channels under ancilla decay}
\label{subsec-apdx:flagging_channels}
 
We now add the flag mode, so that the dynamics is governed by the two-mode dispersive Hamiltonian of~\cref{eq:H_disp_with_flag}, while the jump operator remains the ancilla decay $\hsigma_-$. Since the flag part of $\Ha_{\text{disp}}$ is again diagonal in the $\hsigma_z$ basis, everything in~\cref{subsec-apdx:error_channels} carries over with the memory rotation replaced by the joint rotation of~\cref{eq:CRR_def}. The no-jump and one-jump trajectory operators become
\begin{equation} \label{eq-apdx:W_disp_flag}
    \bm{W}(t_2,t_1) = e^{-\frac{\kappa(t_2-t_1)}{2}\ketbra{e}} \; \CRR\!\left(\tfrac{\chi_m (t_2-t_1)}{2}, \tfrac{\chi_f (t_2-t_1)}{2}\right),
    \qquad
    \bm{J}_\tau = \sqrt{\kappa}\, e^{-\frac{\kappa\tau}{2}} \; \Rm\!\left(\tfrac{\chi_m(2\tau - T)}{2}\right) \Rf\!\left(\tfrac{\chi_f(2\tau-T)}{2}\right) \otimes \ketbra{g}{e} ,
\end{equation}
and, as before, $\mathcal{G}_p = 0$ for $p \ge 2$ within a single dispersive interval. The superpropagator of~\cref{eq-apdx:superpropagator_disp} therefore still holds, with its no-jump and one-jump trajectory operators replaced by $\bm{W}(T,0)$ and $\bm{J}_\tau$ of~\cref{eq-apdx:W_disp_flag}.

Two consequences follow. First, the flagging sequences are constructed so that the flag disentangles from the memory--ancilla system at the end of each trajectory, as we show below; the Kraus operators appearing in the flagging channels are therefore those of~\cref{subsec-apdx:error_channels}. Second, since the flag is initialized in a coherent state and is only acted upon by rotations and displacements throughout the sequence, the flag state attached to each ancilla component $\ket{g}$ and $\ket{e}$ remains a coherent state along every trajectory. Computing a flagging channel therefore reduces to propagating complex amplitudes through the sequence, trajectory by trajectory.
 
\subsubsection{Flagging channel of the conditional rotation gate}
\label{subsubsec-apdx:CR_flagging}
 
Consider first the conditional rotation, in the general form of~\cref{eq:CR_unmatched}, of which the $\chi$-matched sequence of~\cref{eq:protocol_CR_matching} is the particular case $\ftheta = \pi$. The flag, initialized in vacuum, is displaced to $\flag$, rotated by $\pm\ftheta/2$ during the first half of the gate, translated by $-2\flag\cos\frac{\ftheta}{2}$, rotated by $\pm\ftheta/2$ again and finally displaced by $\flag$, the signs being those of the ancilla state ($\pm \leftrightarrow\ket{e/g}$). In the absence of a decay, the two components therefore end at
\begin{equation} \label{eq-apdx:CR_nojump_flag}
    \flag \left[ 1 - 2\cos\tfrac{\ftheta}{2} \, e^{\pm i \ftheta/2} + e^{\pm i\ftheta} \right] = 0 ,
\end{equation}
since $2\cos\frac{\ftheta}{2}\, e^{\pm i\ftheta/2} = 1 + e^{\pm i\ftheta}$, so that the flag returns to vacuum and disentangles, as required by~\cref{eq:flag_disentangling_condition}. By the relation $\D(\gamma_1)\D(\gamma_2) = e^{(\gamma_1\gamma_2^* - \gamma_1^*\gamma_2)/2} \D(\gamma_1+\gamma_2)$, the composition of the three displacements does leave the ancilla-dependent phase $e^{\pm i |\flag|^2 \sin\ftheta}$, \ie the ancilla operator $e^{i|\flag|^2 \sin\ftheta \, \hsigma_z}$; with the convention $\Z_q(\varepsilon) = e^{-i\varepsilon \hsigma_z/2}$, this is exactly what the final rotation $\Z_q(2|\flag|^2\sin\ftheta)$ of~\cref{eq:CR_unmatched} compensates, leaving the no-jump Kraus operator unchanged.
 
If instead the ancilla decays at $\tau \in [0,T/2]$, \ie during the first half of the gate, the flag rotates by $\chi_f(2\tau - T/2)/2 = 2\ftheta\tau/T - \ftheta/2$ according to~\cref{eq-apdx:W_disp_flag}, is translated, and then rotates by $-\ftheta/2$ only, the ancilla having relaxed to $\ket{g}$. A decay at $\tau \in [T/2,T]$ is treated the same way, the flag now rotating by $+\ftheta/2$ before the intermediate displacement and by $2\ftheta\tau/T - 3\ftheta/2$ after it. Using $\flag e^{\pm i\ftheta/2} - 2\flag\cos\frac{\ftheta}{2} = -\flag e^{\mp i\ftheta/2}$, the two cases give
\begin{equation} \label{eq-apdx:CR_flag_positions}
\left\{
\begin{aligned}
    z(\tau) &= \left[ \flag e^{i\left(2\ftheta\tau/T - \ftheta/2\right)} - 2\flag\cos\tfrac{\ftheta}{2} \right] e^{-i\ftheta/2} + \flag
    = \flag \, e^{i\ftheta\frac{2\tau - T}{T}} - \flag \, e^{-i\ftheta} , && \tau \in [0,T/2], \\
    z(\tau) &= \left[ \flag e^{i\ftheta/2} - 2\flag\cos\tfrac{\ftheta}{2} \right] e^{i\left(2\ftheta\tau/T - 3\ftheta/2\right)} + \flag
    = \flag - \flag \, e^{2i\ftheta\frac{\tau - T}{T}} , && \tau \in [T/2,T],
\end{aligned}
\right.
\end{equation}
which is the locus of~\cref{eq:CR_unmatched_flags} and~\cref{fig:CR_unmatched}(c), the phase of the displacements being global for these rank-one trajectories. The two branches match at $\tau = T/2$ and both vanish at $\tau = 0$ and $\tau = T$, the endpoint degeneracy discussed in~\cref{sec:CV_flag_for_CR}.

The locus of~\cref{eq-apdx:CR_flag_positions}, parametrized by the jump time $\tau$, can equivalently be expressed in terms of the rotation error induced on the memory. Using the memory rotation angle $\phi = 2\theta(\tau/T - 1)$ of~\cref{subsubsec-apdx:CR_error_channel}, the flagging channel of the conditional rotation gate reads
\begin{equation}\label{eq-apdx: Kraus map protocol CR}
    \mathcal{F}(\hrho) = \R^{\mathrm{nj}}_0 \hrho (\R^{\mathrm{nj}}_0)^\dagger \otimes \ketbra{0}
    + \int_{-2\theta}^{0} d\phi \; F_{ge}(\phi)~\R_\phi^{ge} \hrho (\R_\phi^{ge})^\dagger \otimes \ketbra{\tilde z(\phi)},
\end{equation}
where the flag locus is parametrized by
\begin{equation} \label{eq-apdx:map_unmatched_flags}
\tilde z(\phi)=
\begin{cases}
\flag\left(e^{i(\ftheta+\bar\phi)}-e^{-i\ftheta}\right), & \phi\in[-2\theta,-\theta],\\
\flag\left(1-e^{i\bar\phi}\right), & \phi\in[-\theta,0],
\end{cases}
\end{equation}
which is the map of~\cref{eq:map_unmatched_flags}. Here $\bar\phi = \frac{\chi_f}{\chi_m}\,\phi$ is the flag rotation angle that accompanies a memory rotation by $\phi$. The two branches coincide at $\phi = -\theta$, and both vanish at the endpoints $\phi = -2\theta$ and $\phi = 0$. This is the endpoint degeneracy discussed in~\cref{sec:CV_flag_for_CR}.

The $\chi$-matched case of~\cref{eq:protocol_CR_matching} is obtained by setting $\ftheta = \pi$ in all the above equations: the intermediate displacement $\Df(-2\flag\cos\frac{\ftheta}{2})$ and the final ancilla rotation $\Z_q(2|\flag|^2\sin\ftheta)$ reduce to the identity, and both branches of~\cref{eq-apdx:map_unmatched_flags} merge into $\tilde z(\phi) = \flag\left(1-e^{i\pi\phi/\theta}\right)$, the circle of radius $|\flag|$ centered at $\flag$ of~\cref{eq:CR_flag_position,eq:map_chi_match}.
 
\subsubsection{Flagging channel of the conditional displacement gate}
\label{subsubsec-apdx:CD_flagging}

\paragraph{One-flag protocol of~\cref{subsec:CD-one-flag}}
Consider now the flagged conditional displacement of~\cref{eq:protocol_ECD_one_flag}. The memory displacements, the echo $\hsigma_x$ and the final ancilla operations $\Z_q(\varepsilon_\lambda)\hsigma_x$ act trivially on the flag, and $\CRR = \CR_m\CR_f$ factorizes, so that the flag part of the sequence reduces to
\begin{equation} \label{eq-apdx:CD_flag_sequence}
    \Df(-\flag)\,\CR_f(\ftheta)\,\hsigma_x\,\CR_f(\ftheta)\,\Df(\flag) .
\end{equation}
In the absence of a decay, the echo swaps the two ancilla components between the two conditional rotations, so that each of them is rotated by $\pm\ftheta$ and then by $\mp\ftheta$, \ie $\CR_f(\ftheta)\,\hsigma_x\,\CR_f(\ftheta) = \hsigma_x$ for any $\ftheta$. The two framing displacements then compose into the identity, $\Df(-\flag)\Df(\flag) = \Id$, with no residual phase. The flag therefore returns to vacuum and disentangles, as required by~\cref{eq:flag_disentangling_condition}, without any $\chi$-matching condition or additional ancilla rotation: the echo itself supplies the recombination mechanism.

If instead the ancilla decays, each trajectory family of~\cref{subsubsec-apdx:CD_error_channel} leaves the flag rotated by a single net angle $\psi$ between the two framing displacements, and the composition rule of the displacements gives
\begin{equation} \label{eq-apdx:CD_flag_map}
    \Df(-\flag)\,\Rf(\psi)\,\Df(\flag)\ket{0} = e^{i|\flag|^2\sin\psi}\,\ket{\flag\left(e^{i\psi}-1\right)} ,
\end{equation}
where the phase is global for these rank-one trajectories. The net angle follows from~\cref{eq-apdx:W_disp_flag}, the one-jump operator rotating the flag by $\ftheta(2u/T-1)$ for a decay at time $u$ within a conditional rotation of duration $T$, with $\ftheta = \chi_f T/2$. A decay before the echo, at $\tau \in (0,T)$, requires the ancilla to start in $\ket{e}$; the flag rotates by $\ftheta(2\tau/T-1)$ during the first conditional rotation and, the echo having restored $\ket{e}$, by a further $+\ftheta$ during the second. A decay after the echo, at $\tau \in (T,2T)$, requires the ancilla to start in $\ket{g}$; the flag rotates by $-\ftheta$ and then by $\ftheta\left(2(\tau-T)/T-1\right)$. A double decay at $\vec\tau = (\tau_1,\tau_2) \in (0,T)\times(T,2T)$ simply adds the two one-jump contributions, $\ftheta(2\tau_1/T-1)$ and $\ftheta\left(2(\tau_2-T)/T-1\right)$. Hence
\begin{equation} \label{eq-apdx:CD_flag_angles}
    \psi(\tau) =
    \begin{cases}
        2\ftheta\,\tau/T, & \tau \in (0,T), \\[2pt]
        2\ftheta\,(\tau-2T)/T, & \tau \in (T,2T),
    \end{cases}
    \qquad
    \psi(\vec\tau) = 2\ftheta\,\frac{\tau_1+\tau_2-2T}{T} ,
\end{equation}
and~\cref{eq-apdx:CD_flag_map} yields the flag positions $z = \flag\left(e^{i\psi}-1\right)$ of~\cref{eq:CD_flag_position,eq:CD_twojump_flag_position}. All three families lie on the arc of radius $|\flag|$ centered at $-\flag$, of polar angles $\psi \in [-2\ftheta, 2\ftheta]$. Note that these flag positions are exact for any $\lambda$, since the flag is insensitive to the intermediate memory displacements $\alpha_\lambda$.

These flag positions can equivalently be expressed in terms of the displacement error $s\beta$ induced on the memory. In the limit of large intermediary displacement $\lambda \gg 1$, the jump times are related to the parameters of the continuous error channel of~\cref{eq:CD_cont_err_channel} by $s = \tau/T - 2 \in [-2,-1]$ for a decay before the echo, $s = 2 - \tau/T \in [0,1]$ for a decay after it, and $(s_1,s_2) = (\tau_1/T, \tau_2/T) \in [0,1]\times[1,2]$ for a double decay, the memory being then displaced by $(s_1-s_2)\beta$. Combining the Kraus operators and weights of~\cref{eq:CD_cont_err_channel_Kraus_Ops} with the flag states of~\cref{eq-apdx:CD_flag_map}, the flagging channel of the conditional displacement gate reads
\begin{equation} \label{eq-apdx:CD_oneflag_flagging_channel}
\begin{aligned}
\mathcal{F}_{D}(\hrho) &= F_{\mathrm{nj}}\,\mathcal{D}^{\mathrm{nj}}_{0}(\hrho)\otimes\ketbra{0} + \int_{-2}^{-1}\! ds\, F_{ge}(s)\,\mathcal{D}^{ge}_{s\beta}(\hrho)\otimes\ketbra{\tilde z_{ge}(s)} + \int_{0}^{1}\! ds\, F_{eg}(s)\,\mathcal{D}^{eg}_{s\beta}(\hrho)\otimes\ketbra{\tilde z_{eg}(s)} \\
&+ \int_{0}^{1}\!\! ds_1\!\int_{1}^{2}\!\! ds_2\, F_{ee}(s_1+s_2)\,\mathcal{D}^{ee}_{(s_1-s_2)\beta}(\hrho) \otimes\ketbra{\tilde z_{ee}(s_1+s_2)},
\end{aligned}
\end{equation}
where the flag amplitudes are obtained by substituting $s$ into~\cref{eq-apdx:CD_flag_angles}, \ie $\psi = 2\ftheta(s+2)$, $-2\ftheta s$ and $2\ftheta(s_1+s_2-2)$, respectively:
\begin{equation} \label{eq-apdx:CD_oneflag_values}
\tilde z_{ge}(s)=\flag\left(e^{2i\ftheta(s+2)}-1\right), \qquad
\tilde z_{eg}(s)=\flag\left(e^{-2i\ftheta s}-1\right), \qquad 
\tilde z_{ee}(s)=\flag\left(e^{2i\ftheta(s-2)}-1\right).
\end{equation}
This is the channel of~\cref{eq:CD_oneflag_flagging_channel,eq:CD_oneflag_values}. At finite $\lambda$, the flag amplitudes are unchanged and only the memory Kraus operators are replaced by those of~\cref{subsubsec-apdx:CD_error_channel}. \\

\paragraph{Two-flag protocol of~\cref{subsec:CD-two-flags}}
The second flag need not be $\chi$-matched. For an arbitrary coupling $\chi_{f_2}$, it rotates by $\pm\ftheta_2$ during each conditional rotation, with $\ftheta_2 = \theta\chi_{f_2}/\chi_m$, and is activated during the second conditional rotation only, through the unmatched sequence of~\cref{eq:CR_unmatched}. The second conditional rotation of $\ECD_{m f_1 f_2}$ thus becomes
\begin{equation} \label{eq-apdx:2flag_unmatched_block}
\Z_q\!\left(2|\flag_2|^2\sin\ftheta_2\right)\D_{f_2}(\flag_2)\,\bm{C\!R}_{m f_1 f_2}\!\left(\tfrac{\theta}{2},\tfrac{\ftheta}{2},\tfrac{\ftheta_2}{2}\right)\D_{f_2}\!\left(-2\flag_2\cos\tfrac{\ftheta_2}{2}\right)\bm{C\!R}_{m f_1 f_2}\!\left(\tfrac{\theta}{2},\tfrac{\ftheta}{2},\tfrac{\ftheta_2}{2}\right)\D_{f_2}(\flag_2) .
\end{equation}
Splitting the rotation in two halves leaves the memory and the first flag unaffected, their two half-rotations composing into $\CR_m(\theta)$ and $\CR_{f_1}(\ftheta)$. The ancilla rotation compensates the ancilla-dependent phase left by the displacements of the second flag, as in~\cref{subsubsec-apdx:CR_flagging}. Since it only acts on the ancilla, it can be moved through the final $\hsigma_x$ of~\cref{eq:protocol_ECD_two_flags}, using $\hsigma_x\Z_q(\varepsilon) = \Z_q(-\varepsilon)\hsigma_x$. It then merges with the residual rotation into $\Z_q\!\left(\varepsilon_\lambda - 2|\flag_2|^2\sin\ftheta_2\right)$.

During the first conditional rotation, the second flag is still in vacuum, which is invariant under rotations, so that it is left unaffected whatever happens to the ancilla. During the second, it undergoes exactly the conditional-rotation flagging sequence of~\cref{subsubsec-apdx:CR_flagging}, driven by the ancilla state during that rotation only. If no decay occurs there, as for the no-jump and pre-echo trajectories, it returns to vacuum by~\cref{eq-apdx:CR_nojump_flag}. If the ancilla decays at $\tau_2 \in (T,2T)$, as for the post-echo and double-decay trajectories, it ends at the position of~\cref{eq-apdx:CR_flag_positions} with $\tau \to \tau_2 - T$. In terms of the memory rotation angle of~\cref{subsubsec-apdx:CR_flagging}, this decay corresponds to $\phi = 2\theta(\tau_2/T - 2)$, \ie $\phi = -2\theta s$ for a single post-echo decay and $\phi = 2\theta(s_2-2)$ for a double decay. The flagging channel therefore takes the form
\begin{equation} \label{eq-apdx:2flag_flagging_channel}
\begin{aligned}
\mathcal{F}_{D}(\hrho) &= F_{\mathrm{nj}}\,\mathcal{D}^{\mathrm{nj}}_{0}(\hrho)\otimes\ketbra{0,0} + \int_{-2}^{-1}\! ds\, F_{ge}(s)\,\mathcal{D}^{ge}_{s\beta}(\hrho)\otimes\ketbra{\tilde z_{ge}(s),0} \\
&+ \int_{0}^{1}\! ds\, F_{eg}(s)\,\mathcal{D}^{eg}_{s\beta}(\hrho)\otimes\ketbra{\tilde z_{eg}(s),\tilde w_{eg}(s)} \\
&+ \int_{0}^{1}\!\! ds_1\!\int_{1}^{2}\!\! ds_2\, F_{ee}(s_1+s_2)\,\mathcal{D}^{ee}_{(s_1-s_2)\beta}(\hrho) \otimes\ketbra{\tilde z_{ee}(s_1+s_2),\tilde w_{ee}(s_2)},
\end{aligned}
\end{equation}
with the first-flag amplitudes of~\cref{eq-apdx:CD_oneflag_values} ($\flag\to\flag_1$). The second-flag amplitudes, $\tilde w_{eg}(s) = \tilde z(-2\theta s)$ and $\tilde w_{ee}(s_2) = \tilde z\left(2\theta(s_2-2)\right)$, follow from the map of~\cref{eq-apdx:map_unmatched_flags} with $\flag\to\flag_2$ and $\ftheta\to\ftheta_2$:
\begin{equation} \label{eq-apdx:2flag_gamma_traj_unmatched}
\tilde w_{eg}(s)=
\begin{cases}
\flag_2\left(e^{i\ftheta_2(1-2s)}-e^{-i\ftheta_2}\right), & s\in[\tfrac12,1],\\[2pt]
\flag_2\left(1-e^{-2i\ftheta_2 s}\right), & s\in[0,\tfrac12],
\end{cases}, \qquad 
\tilde w_{ee}(s_2)=
\begin{cases}
\flag_2\left(e^{i\ftheta_2(2s_2-3)}-e^{-i\ftheta_2}\right), & s_2\in[1,\tfrac32],\\[2pt]
\flag_2\left(1-e^{2i\ftheta_2(s_2-2)}\right), & s_2\in[\tfrac32,2].
\end{cases}
\end{equation}
As long as $\ftheta_2 < 2\pi$, these maps are injective, so that the second flag records post-echo decays unambiguously. Together with the sum $s_1+s_2$ recorded by the first flag, the pair of flag states then determines both $s_1$ and $s_2$.

The $\chi$-matched protocol of~\cref{eq:protocol_ECD_two_flags} presented in the main text is obtained by setting $\ftheta_2 = \pi$. The intermediate displacement $\D_{f_2}(-2\flag_2\cos\frac{\ftheta_2}{2})$ and the compensating rotation $\Z_q(2|\flag_2|^2\sin\ftheta_2)$ then reduce to the identity. The two branches of each amplitude in~\cref{eq-apdx:2flag_gamma_traj_unmatched} merge into
\begin{equation} \label{eq-apdx:2flag_gamma_traj}
\tilde w_{eg}(s)=\flag_2\left(1-e^{-2i\pi s}\right),
\qquad
\tilde w_{ee}(s_2)=\flag_2\left(1-e^{2i\pi s_2}\right),
\end{equation}
which, together with~\cref{eq-apdx:2flag_flagging_channel}, are~\cref{eq:2flag_flagging_channel,eq:2flag_gamma_traj}.

\section{Flag photon loss and ancilla dephasing}
\label{sec-apdx:flag_loss}

We derive the ancilla dephasing induced by flag photon loss, quoted in~\cref{eq:flag_decay_dephasing}, using the notations of~\cref{sec-apdx:channels}. The relevant dynamics is now the dispersive coupling of the flag to the ancilla, $\Ha/\hbar = -\frac{\chi_f}{2}\hf^\dagger\hf \, \hsigma_z$, and the jump operator is now the flag photon loss $\bL = \hf$, at rate $\kappa_f$. The memory part of the dispersive Hamiltonian commutes with everything below and is omitted. The no-jump propagator factorizes into a unitary and a damping part
\begin{equation} \label{eq-apdx:flag_loss_W}
    \bm{W}(t_2,t_1) = \bm{U}(t_2 - t_1) \, \bm{V}(t_2-t_1),
    \qquad
    \bm{U}(t) = e^{\frac{i}{2}\chi_f t \, \hf^\dagger\hf \, \hsigma_z},
    \qquad
    \bm{V}(t) = e^{-\frac{\kappa_f t}{2}\hf^\dagger \hf} .
\end{equation}
The key step is the commutation of a flag decay through the no-jump evolution. Since $\hf \, g(\hf^\dagger\hf) = g(\hf^\dagger\hf + 1) \, \hf$ for any function $g$, one gets
\begin{equation} \label{eq-apdx:flag_loss_commutation}
    \hf \, \bm{W}(t_2,t_1) = \bm{B}(t_2 - t_1) \, \bm{W}(t_2,t_1) \, \hf ,
    \qquad
    \bm{B}(t) = e^{-\frac{\kappa_f t}{2}} \, e^{\frac{i}{2}\chi_f t \, \hsigma_z} ,
\end{equation}
\ie each loss event, when moved past the deterministic evolution, leaves behind an operator acting on the ancilla alone. This is the back-action anticipated in~\cref{subsec:flag_photon_loss}: a lost photon carries which-path information about the ancilla state, and imprints on it a phase set by the time at which it was emitted.
 
Consider then a trajectory with $p$ losses at times $0 \le \tau_1 \le \dots \le \tau_p \le T$. Moving the loss operator emitted at $\tau_j$ all the way to the right requires commuting it through a total elapsed time $\tau_j$, and $\bm{B}$, being a function of $\hsigma_z$ only, commutes with all the operators it meets on the way. Using $\bm{B}(t_1)\bm{B}(t_2) = \bm{B}(t_1+t_2)$, the trajectory operator of~\cref{eq-apdx:traj_operator} therefore collapses to
\begin{equation} \label{eq-apdx:flag_loss_traj}
    \bm{J}_{\vec{\tau}} = \bm{B}\!\left(\tau_1 + \dots + \tau_p\right) \U(T) \, \bm{V}(T) \, \hf^p ,
\end{equation}
in which the jump times enter only through their sum. Denoting by $\mathcal{B}(t)$, $\mathcal{U}(t)$, $\mathcal{V}(t)$ and $\mathcal{S}_{\hf}$ the superoperators associated with $\bm{B}(t)$, $\bm{U}(t)$, $\bm{V}(t)$ and with the flag decay, the $p$-jump contribution of~\cref{eq-apdx:Dyson_exp} becomes
\begin{equation} \label{eq-apdx:flag_loss_Gp}
    \mathcal{G}_p(T) = \mathcal{U}(T) \left[ \, \int\limits_0^{T}\! d\tau_p \dots \int\limits_0^{\tau_2}\! d\tau_1 \; \mathcal{B}(\tau_1 + \dots + \tau_p) \right] \mathcal{V}(T) \, \mathcal{S}_{\hf}^p ,
\end{equation}
where the unitary part has been pulled out on the left, as it is the only piece that entangles the flag with the ancilla. The nested integrals are evaluated by decomposing $\mathcal{B}$ over the ancilla blocks. Introducing $\mathcal{P}_{\mu\nu}(\bullet) = \ketbra{\mu} \bullet \ketbra{\nu}$ for $\mu,\nu \in \{g,e\}$, which selects the coefficient of $\ketbra{\mu}{\nu}$, one has
\begin{equation} \label{eq-apdx:B_decomposition}
    \mathcal{B}(t) = \sum_{\mu,\nu} e^{c_{\mu\nu} t} \, \mathcal{P}_{\mu\nu},
    \qquad
    c_{gg} = c_{ee} = -\kappa_f, \quad c_{eg} = i\chi_f - \kappa_f, \quad c_{ge} = -i\chi_f - \kappa_f ,
\end{equation}
so that only the scalar integrals
\begin{equation} \label{eq-apdx:nested_integrals}
    f^{(p)}_c(T) = \int\limits_0^{T}\! d\tau_p \dots \int\limits_0^{\tau_2}\! d\tau_1 \; e^{c(\tau_1 + \dots + \tau_p)}
    = \frac{1}{p!} \left( \frac{e^{cT}-1}{c} \right)^{p}
\end{equation}
are needed; the closed form can be obtained for instance inductively on $p$.
 
It remains to apply the result to a flag prepared in a coherent state. Using $\mathcal{S}_{\hf}(\ketbra{\flag}) = \kappa_f |\flag|^2 \ketbra{\flag}$ and $\mathcal{V}(T)(\ketbra{\flag}) = e^{\left(e^{-\kappa_f T}-1\right)|\flag|^2} \ketbra{\flag e^{-\kappa_f T/2}}$, and summing~\cref{eq-apdx:flag_loss_Gp} over $p$ with~\cref{eq-apdx:nested_integrals}, each ancilla block picks up the factor $\exp\left[ \kappa_f|\flag|^2 (e^{c_{\mu\nu}T}-1)/c_{\mu\nu} \right]$. For the populations, $c_{gg} = c_{ee} = -\kappa_f$ and this factor exactly cancels the prefactor brought by $\mathcal{V}(T)$, as it must. For the coherences it does not, and one is left with
\begin{equation} \label{eq-apdx:flag_loss_channel}
    \mathcal{G}(T)\left(\bullet \otimes \ketbra{\flag}\right) = \mathcal{U}(T) \left[ \mathcal{N}_{\text{deph}}(\bullet) \otimes \ketbra{\flag e^{-\kappa_f T/2}} \right],
\end{equation}
where $\mathcal{N}_{\text{deph}}$ is precisely the pure-dephasing channel of~\cref{eq:dephasing_kraus}, with the dephasing factor
\begin{equation} \label{eq-apdx:D_closed_form}
    D(T) = \exp\left[ \left(e^{-\kappa_f T} - 1\right)|\flag|^2 + \frac{\kappa_f |\flag|^2}{i\chi_f - \kappa_f}\left( e^{(i\chi_f - \kappa_f)T} - 1 \right) \right]
\end{equation}
multiplying the coherence $\bra{e}\!\hrho\!\ket{g}$. Flag photon loss thus has exactly the two effects described in~\cref{subsec:flag_photon_loss}: a deterministic contraction $\flag \to \flag e^{-\kappa_f T/2}$ of the flag amplitude, which merely rescales the encoding map $\tau \mapsto z(\tau)$ and is absorbed in the calibration, and a pure dephasing of the ancilla. Writing $D(T) = e^{i\varphi} e^{-r}$ and separating the modulus from the argument of~\cref{eq-apdx:D_closed_form} gives
\begin{equation} \label{eq-apdx:r_phi}
\left\{
\begin{aligned}
    r &= |\flag|^2 \left[ \frac{\chi_f^2}{\kappa_f^2 + \chi_f^2} - e^{-\kappa_f T} + \frac{\kappa_f e^{-\kappa_f T}}{\kappa_f^2 + \chi_f^2}\left( \kappa_f \cos\chi_f T - \chi_f \sin\chi_f T \right) \right], \\
    \varphi &= \frac{\kappa_f |\flag|^2}{\kappa_f^2 + \chi_f^2} \left[ \chi_f - \chi_f \cos(\chi_f T) \, e^{-\kappa_f T} - \kappa_f \sin(\chi_f T) \, e^{-\kappa_f T} \right],
\end{aligned}
\right.
\end{equation}
which is~\cref{eq:flag_decay_dephasing} together with the closed form of the deterministic phase that can be calibrated out.
 
Hence,~\cref{eq-apdx:D_closed_form} applies to a single dispersive interval, during which the flag amplitude keeps a constant modulus $|\flag|$: this is the case of the matched sequence of~\cref{eq:protocol_CR_matching}, relevant for the parity measurement of~\cref{subsec:CR_cat_parity}. Sequences containing an ancilla $\pi$-pulse or intermediate flag displacements, such as~\cref{eq:protocol_ECD_one_flag,eq:CR_unmatched}, are treated by applying the same result interval by interval: the moduli of the dephasing factors multiply, so that the exponents $r$ add, and the resulting phase remains deterministic.

\section{Flagging channels under qubit decay and excitation}
\label{sec-apdx:decay_excitation}

So far, both in the main text and in~\cref{sec-apdx:channels}, continuous-variable flags have been used to protect the memory against the errors induced by ancilla decay. For this jump alone, the trajectory expansion of~\cref{eq-apdx:Dyson_exp} terminates after a single jump per dispersive interval, so that a gate can only suffer a finite number of jumps, at most two for the conditional displacement. Flagging then amounts to encoding the jump time(s) injectively in the phase-space position of the flag. In this appendix, we show that the continuous-variable flag reaches beyond this picture: what it requires is not to resolve the jump times, but to correlate the error induced on the memory with the flag position, as expressed by the correlation condition of~\cref{eq:correlation_condition}.

To this end, we model the ancilla as a two-level system $\{\ket{g},\ket{e}\}$, thereby neglecting leakage, and consider both its decay $\hsigma_-$ and its thermal excitation $\hsigma_+ = \ketbra{e}{g}$. We show that two of the protocols introduced in this paper then flag the induced memory errors to all orders, up to a stabilizer of the code. First, the $\chi$-matched flagging sequence of the $\CR(\theta)$ gate,~\cref{eq:protocol_CR_matching}, encodes the rotation error of the memory in the flag position for any sequence of decay and excitation events, which establishes the claim made at the end of~\cref{subsubsec:chi-match}. Second, the two-flag sequence of the $\CD(\beta)$ gate,~\cref{eq:protocol_ECD_two_flags}, whose second flag is $\chi$-matched, likewise flags decay and excitation errors to all orders, as claimed at the end of~\cref{subsec:CD-two-flags}. By contrast, the one-flag sequence of~\cref{eq:protocol_ECD_one_flag} loses its first-order protection as soon as excitations are allowed, which makes the second flag necessary.

These properties follow from the formalism of~\cref{sec-apdx:channels}, the Lindblad equation~\cref{eq-apdx:lindblad} now containing the two jump operators $\bL_- = \hsigma_-$ and $\bL_+ = \hsigma_+$, at rates $\kappa_-$ and $\kappa_+$. The decay rate $\kappa_-$ is the rate $\kappa$ of the main text and of the previous appendices, and all the results derived so far are recovered by setting $\kappa_+ = 0$. Unlike in the decay-only case, however, the trajectory expansion no longer terminates after a single jump: at any time, an ancilla in $\ket{e}$ can decay to $\ket{g}$, and an ancilla in $\ket{g}$ can be excited back to $\ket{e}$, so that trajectories with arbitrarily many jumps contribute.

The appendix is organized as follows. We first derive the dispersive dynamics in the presence of both jumps, and resum the trajectory expansion to all orders (\cref{subsec-apdx:pm_expansion}). We then treat the $\CR(\theta)$ gate (\cref{subsec-apdx:pm_CR}) and the $\CD(\beta)$ gate (\cref{subsec-apdx:pm_CD}) in turn: for each, we generalize its continuous error channel,~\cref{eq-apdx:CR_cont_err_channel_rot,eq-apdx:CD_cont_err_channel}, and follow the flag along the trajectories of its flagging protocols to show that its final position encodes the error induced on the memory.

\subsection{Dispersive dynamics under qubit decay and excitation}
\label{subsec-apdx:pm_expansion}
 
Consider a single dispersive interval of duration $T$, governed by the two-mode dispersive Hamiltonian of~\cref{eq:H_disp_with_flag}. Since $\hsigma_+\hsigma_- = \ketbra{e}$ and $\hsigma_-\hsigma_+ = \ketbra{g}$ are both diagonal in the $\hsigma_z$ basis, the non-unitary part of $\Ha_{\text{eff}}$ in~\cref{eq-apdx:nojump_propagator} still commutes with the dispersive interaction, and the no-jump propagator of~\cref{eq-apdx:W_disp_flag} merely acquires a supplementary damping of the ground state,
\begin{equation} \label{eq-apdx:W_disp_flag_pm}
    \bm{W}(t_2,t_1) = e^{-\frac{t_2-t_1}{2}\left(\kappa_-\ketbra{e} + \kappa_+\ketbra{g}\right)} \; \CRR\!\left(\tfrac{\chi_m (t_2-t_1)}{2}, \tfrac{\chi_f (t_2-t_1)}{2}\right).
\end{equation}

Unlike in the decay-only case, however, the excited state is repopulated by every excitation $\hsigma_+$, so that trajectories with arbitrarily many jumps contribute, and the truncation $\mathcal{G}_p = 0$ for $p \ge 2$, used throughout~\cref{subsec-apdx:error_channels}, no longer holds. What survives instead is a rigid \emph{alternation}: since $\hsigma_\pm \, \bm{W}(t_2,t_1) \, \hsigma_\pm = 0$, a decay can only be followed by an excitation, and conversely. A trajectory with $p$ jumps at ordered times $\vec{\tau} = (\tau_1,\dots,\tau_p)$ is therefore entirely specified by $\vec{\tau}$ and by the ancilla state $\ket{\nu}$ it starts from. Along it, the ancilla follows a telegraph signal that starts in $\ket{\nu}$, flips at each $\tau_j$, and ends in a state $\ket{\mu}$ fixed by the parity of $p$, with $\mu = \nu$ for even $p$ and $\mu \neq \nu$ for odd $p$. A jump trajectory can thus only leave behind one of the four ancilla operators $\ketbra{g}{e}$, $\ketbra{e}$, $\ketbra{e}{g}$ and $\ketbra{g}$, and the $p$-jump contribution to the superpropagator decomposes as
\begin{equation}
\label{eq-apdx:superpropagator_p_jump_decay_exc}
\mathcal{G}_p(T)(\hrho) = \sum_{\mu,\nu \in\{g,e\}}\;\int\limits_0^{T}\! d\tau_p \dots \int\limits_0^{\tau_2}\! d\tau_1 \; \bm{J}^{\mu\nu}_{\vec{\tau}} \, \hrho \, (\bm{J}^{\mu\nu}_{\vec{\tau}})^\dagger .
\end{equation}
Here $\bm{J}^{\mu\nu}_{\vec{\tau}}$ is the trajectory operator of~\cref{eq-apdx:traj_operator} for a trajectory starting in $\ket{\nu}$, with $\sqrt{\kappa}\bL$ replaced at each jump time by $\sqrt{\kappa_-}\hsigma_-$ or $\sqrt{\kappa_+}\hsigma_+$ according to the ancilla state at that time. By convention, $\bm{J}^{\mu\nu}_{\vec{\tau}} = 0$ whenever $\mu$ is incompatible with the parity of $p$.
 
Importantly, the memory and the flag see this telegraph history only through the time $\tau_e(\vec{\tau}) \in [0,T]$ that the ancilla spends in $\ket{e}$, the time spent in $\ket{g}$ being $T - \tau_e$. Indeed, all the rotations generated by~\cref{eq-apdx:W_disp_flag_pm} share the same axis and simply add up, so that the memory and the flag are rotated by $\chi_m(2\tau_e - T)/2$ and $\chi_f(2\tau_e - T)/2$ respectively, while the damping factors accumulate into $e^{-\frac{1}{2}\left[\kappa_- \tau_e + \kappa_+ (T-\tau_e)\right]}$. Each trajectory operator therefore collapses to an expression in which the jump times enter through the single quantity $\tau_e$,
\begin{equation} \label{eq-apdx:pm_traj_operator}
\bm{J}^{\mu\nu}_{\vec{\tau}} = \sqrt{\kappa_-^{n_-}\kappa_+^{n_+}} \; e^{-\frac{1}{2}\left[\kappa_- \tau_e + \kappa_+ (T-\tau_e)\right]} \; \bm{J}^{\mu\nu}(\tau_e),
\qquad \text{where} \qquad
\bm{J}^{\mu\nu}(t) = \Rm\!\left(\tfrac{\chi_m(2t - T)}{2}\right) \Rf\!\left(\tfrac{\chi_f(2t-T)}{2}\right) \otimes \ketbra{\mu}{\nu}.
\end{equation}
Here, $n_-$ and $n_+$ denote the numbers of decays and excitations along the trajectory. Because the jumps alternate, they are fixed by $p$ and by the initial state: $n_- = n_+$ for the diagonal families $\ketbra{e}$ and $\ketbra{g}$, while $n_- = n_+ + 1$ for $\ketbra{g}{e}$ and $n_+ = n_- + 1$ for $\ketbra{e}{g}$. For a single decay and $\kappa_+ = 0$, one has $\tau_e = \tau$, and $\bm{J}^{ge}_{\vec{\tau}}$ reduces to the one-jump operator $\bm{J}_\tau$ of~\cref{eq-apdx:W_disp_flag}.

Since all the trajectories of a given family $\ketbra{\mu}{\nu}$ depend on the jump times only through $\tau_e$, the $p$-fold nested integrals of~\cref{eq-apdx:superpropagator_p_jump_decay_exc} reduce to a single integral over $\tau_e$, weighted by the density of $\tau_e$ at fixed $p$. For trajectories starting in $\ket{e}$, the ancilla is excited during the intervals $[0,\tau_1], [\tau_2,\tau_3], \dots$, the last of which ends at $\tau_p$ for odd $p$ and at $T$ for even $p$. Summing their lengths gives $\tau_e(\vec{\tau}) = -\sum_{j=1}^p (-1)^j \tau_j + \tfrac{1+(-1)^p}{2}\,T$, and one shows by induction on $p$, using $\int_a^b du \int_a^u dv = \int_a^b dv \int_v^b du$, that for any function $h$,
\begin{equation} \label{eq-apdx:nested_integrals_pm}
\int\limits_0^{T}\! d\tau_p \dots \int\limits_0^{\tau_2}\! d\tau_1 \; h\big(\tau_e(\vec{\tau})\big) = \int_0^T\! dt \; h(t) \, \varrho_p(t),
    \quad \text{where} \quad
    \left\{
    \begin{aligned}
        \varrho_{2n}(t) &= \frac{t^{n} \, (T-t)^{n-1}}{n! \, (n-1)!}, & n \ge 1, \\
        \varrho_{2n+1}(t) &= \frac{\left[t \, (T-t)\right]^{n}}{(n!)^2}, & n \ge 0 .
    \end{aligned}
    \right.
\end{equation}
For trajectories starting in $\ket{g}$, the same expression gives the time $T - \tau_e$ spent in $\ket{g}$, so that their density is obtained by exchanging $t \leftrightarrow T-t$.

Every $p$-jump contribution thus reduces to an integral over the same variable $\tau_e \in [0,T]$, so that the sum over $p$ can be carried out inside this integral, family by family. The full superpropagator $\mathcal{G}(T) = \mathcal{G}_0(T) + \sum_{p=1}^{+\infty} \mathcal{G}_p(T)$, whose no-jump contribution is $\mathcal{G}_0(T) = \mathcal{W}(T,0)$, therefore decomposes as
\begin{equation} \label{eq-apdx:superpropagator_pm}
\begin{aligned}
    \mathcal{G}(T)(\hrho) = \bm{W}(T,0)\,\hrho\,\bm{W}^\dagger(T,0)
    &+ \int_0^T \! dt \; G_{ge}(t) \, \bm{J}^{ge}(t) \, \hrho \, (\bm{J}^{ge}(t))^\dagger
    + \int_0^T \! dt \; G_{ee}(t) \, \bm{J}^{ee}(t) \, \hrho \, (\bm{J}^{ee}(t))^\dagger \\
    &+ \int_0^T \! dt \; G_{eg}(t) \, \bm{J}^{eg}(t) \, \hrho \, (\bm{J}^{eg}(t))^\dagger
    + \int_0^T \! dt \; G_{gg}(t) \, \bm{J}^{gg}(t) \, \hrho \, (\bm{J}^{gg}(t))^\dagger ,
\end{aligned}
\end{equation}
with the family operators $\bm{J}^{\mu\nu}(t)$ of~\cref{eq-apdx:pm_traj_operator}, and where the weight densities $G_{\mu\nu}(t)$ are obtained from summing on $p$ the prefactor terms, with the prescribed pairs $(n_-,n_+)$ described above. Combining~\cref{eq-apdx:pm_traj_operator,eq-apdx:nested_integrals_pm}, one finds
\begin{equation} \label{eq-apdx:transfer_weights}
\begin{aligned}
G_{ge}(t) &= \sum_{n=0}^{\infty} \kappa_-^{n+1}\kappa_+^{n} \, e^{-\left[\kappa_- t + \kappa_+ (T-t)\right]} \, \varrho_{2n+1}(t) = \kappa_- \, e^{-\left[\kappa_- t + \kappa_+ (T-t)\right]} \, I_0(\xi_t), \\ 
G_{ee}(t) &= \sum_{n=1}^{\infty} \kappa_-^{n}\kappa_+^{n} \, e^{-\left[\kappa_- t + \kappa_+ (T-t)\right]} \, \varrho_{2n}(t) = \sqrt{\kappa_-\kappa_+}\,\sqrt{\tfrac{t}{T-t}} \; e^{-\left[\kappa_- t + \kappa_+ (T-t)\right]} \, I_1(\xi_t), \\
G_{eg}(t) &= \sum_{n=0}^{\infty} \kappa_-^{n}\kappa_+^{n+1} \, e^{-\left[\kappa_- t + \kappa_+ (T-t)\right]} \, \varrho_{2n+1}(T-t) = \kappa_+ \, e^{-\left[\kappa_- t + \kappa_+ (T-t)\right]} \, I_0(\xi_t), \\ 
G_{gg}(t) &= \sum_{n=1}^{\infty} \kappa_-^{n}\kappa_+^{n} \, e^{-\left[\kappa_- t + \kappa_+ (T-t)\right]} \, \varrho_{2n}(T-t) = \sqrt{\kappa_-\kappa_+}\,\sqrt{\tfrac{T-t}{t}} \; e^{-\left[\kappa_- t + \kappa_+ (T-t)\right]} \, I_1(\xi_t),
\end{aligned}
\end{equation}
where $I_\nu(x) = \sum_{m \ge 0} \frac{1}{m! \, \Gamma(m+\nu+1)}\left(\frac{x}{2}\right)^{2m+\nu}$ denote the modified Bessel functions of the first kind, and where $\xi_t = 2\sqrt{\kappa_-\kappa_+t(T-t)}$. The second equalities follow from this series, since $(\xi_t/2)^2 = \kappa_-\kappa_+ t(T-t)$.

\Cref{eq-apdx:superpropagator_pm,eq-apdx:transfer_weights} generalize the decay-only superpropagator of~\cref{eq-apdx:superpropagator_disp} to all orders in both jump processes. Indeed, for $\kappa_+ \to 0$, the no-jump propagator of~\cref{eq-apdx:W_disp_flag_pm} reduces to that of~\cref{eq-apdx:W_disp_flag}, and since $\xi_t \to 0$, one has $I_0(\xi_t) \to 1$ and $\sqrt{\kappa_-\kappa_+}\,I_1(\xi_t) \to 0$, so that $G_{ee}$, $G_{eg}$ and $G_{gg}$ vanish while $G_{ge}(t) \to \kappa\, e^{-\kappa t}$. The only surviving family then gives $G_{ge}(t)\,\bm{J}^{ge}(t)\,\hrho\,\bm{J}^{ge}(t)^\dagger = \bm{J}_t\,\hrho\,\bm{J}_t^\dagger$, with $\bm{J}_t$ the one-jump operator of~\cref{eq-apdx:W_disp_flag}, and~\cref{eq-apdx:superpropagator_pm} reduces to $\mathcal{G}_0(T) + \mathcal{G}_1(T)$ of~\cref{eq-apdx:superpropagator_disp}.

\subsection{Conditional rotation gate under qubit decay and excitation}
\label{subsec-apdx:pm_CR}
 
\subsubsection{Continuous error channel}
\label{subsubsec-apdx:pm_CR_error_channel}
 
The $\CR(\theta)$ gate is realized by a single dispersive interval of duration $T = 2\theta/\chi_m$, so that its noisy version is directly given by the superpropagator $\mathcal{G}(T)$ of~\cref{eq-apdx:superpropagator_pm}, the flag being ignored. As in~\cref{subsubsec-apdx:CR_error_channel}, we factor out the target unitary, $\mathcal{G}(T) = \mathcal{N}_R \circ \mathcal{CR}_\theta$, and obtain the Kraus operators by right-multiplying the no-jump propagator and the family operators by $\CR^\dagger(\theta) = \R(-\theta)\otimes\ketbra{e} + \R(\theta)\otimes\ketbra{g}$. The no-jump propagator $\bm{W}(T,0)$ of~\cref{eq-apdx:W_disp_flag_pm} then reduces to its damping part, and introduces no error on the memory. The family operator $\bm{J}^{\mu\nu}(\tau_e)$ of~\cref{eq-apdx:pm_traj_operator}, associated with a trajectory spending a time $\tau_e$ in $\ket{e}$, rotates the memory by $\theta(2\tau_e/T - 1)$. This angle is shifted by $-\theta$ for the two families $\ketbra{\mu}{e}$ starting in $\ket{e}$, and by $+\theta$ for the two families $\ketbra{\mu}{g}$ starting in $\ket{g}$, so that the memory is left rotated by
\begin{equation} \label{eq-apdx:pm_residual_angle}
    \phi = 2\theta\left(\tfrac{\tau_e}{T} - 1\right) \in [-2\theta, 0] \quad \text{for } \nu = e,
    \qquad
    \phi = 2\theta\,\tfrac{\tau_e}{T} \in [0, 2\theta] \quad \text{for } \nu = g.
\end{equation}
For $\nu = e$, this is the rotation angle of~\cref{subsubsec-apdx:CR_error_channel}, with the jump time $\tau$ replaced by the time $\tau_e$ spent in $\ket{e}$. Changing the integration variable from $\tau_e$ to $\phi$, as in~\cref{eq-apdx:CR_cont_err_channel_rot}, the continuous error channel under ancilla decay and excitation reads
\begin{equation} \label{eq-apdx:CR_cont_err_channel_pm}
    \mathcal{N}_R(\hrho) = \R^{\mathrm{nj}}_0 \hrho (\R^{\mathrm{nj}}_0)^\dagger
    + \sum_{\mu \in \{g,e\}} \left[ \int_{-2\theta}^{0} d\phi \; F_{\mu e}(\phi) \, \R^{\mu e}_\phi \hrho (\R^{\mu e}_\phi)^\dagger
    + \int_{0}^{2\theta} d\phi \; F_{\mu g}(\phi) \, \R^{\mu g}_\phi \hrho (\R^{\mu g}_\phi)^\dagger \right],
\end{equation}
with Kraus operators and associated weight functions
\begin{equation} \label{eq-apdx:CR_Kraus_Op_pm}
\begin{aligned}
    \R^{\mathrm{nj}}_0 &= e^{-\frac{T}{2}\left(\kappa_-\ketbra{e} + \kappa_+\ketbra{g}\right)},
    & \R^{\mu\nu}_\phi &= \R(\phi) \otimes \ketbra{\mu}{\nu}, \\
    F_{\mu e}(\phi) &= \frac{T}{2\theta}\, G_{\mu e}\!\left(T\left(1+\tfrac{\phi}{2\theta}\right)\right),
    & F_{\mu g}(\phi) &= \frac{T}{2\theta}\, G_{\mu g}\!\left(\tfrac{T\phi}{2\theta}\right),
\end{aligned}
\end{equation}
where $G_{\mu\nu}$ are the weight densities of~\cref{eq-apdx:transfer_weights}. Setting $\kappa_+ = 0$ recovers~\cref{eq-apdx:CR_cont_err_channel_rot}: $G_{ee}$, $G_{eg}$ and $G_{gg}$ vanish, and $G_{ge}(t) \to \kappa\,e^{-\kappa t}$ gives back the weight function $F_{ge}$. With excitations, the continuous error channel is significantly worse. The residual rotation now spans $[-2\theta,2\theta]$ rather than $[-2\theta,0]$, and, since the weights collect trajectories with any number of jumps, the channel cannot be truncated at any finite order in $\kappa_\pm T$.
 
\subsubsection{Flagging channel of the $\chi$-matched sequence}
\label{subsubsec-apdx:pm_CR_flagging}
 
The flag, however, sees the ancilla history through the same quantity $\tau_e$ as the memory. Consider the $\chi$-matched sequence of~\cref{eq:protocol_CR_matching}, for which $\ftheta = \chi_f T/2 = \pi$. The flag, initialized in vacuum, is displaced to $\flag$, rotated by $\ftheta(2\tau_e/T - 1)$ according to~\cref{eq-apdx:pm_traj_operator}, and displaced by $\flag$ again. For a trajectory spending a time $\tau_e$ in $\ket{e}$, it therefore ends in the coherent state
\begin{equation} \label{eq-apdx:pm_flag_position}
    z(\tau_e) = \flag\left(1 + e^{i\pi\frac{2\tau_e - T}{T}}\right) = \flag\left(1 - e^{2i\pi \tau_e/T}\right),
\end{equation}
which is~\cref{eq:CR_flag_position} with the jump time $\tau$ replaced by $\tau_e$. By the composition rule of the displacements recalled in~\cref{subsubsec-apdx:CR_flagging}, the sequence also leaves the phase $e^{-i|\flag|^2\sin\left[\pi(2\tau_e-T)/T\right]}$. This phase vanishes on the two branches of the no-jump trajectory, which correspond to $\tau_e = T$ for the ancilla in $\ket{e}$ and $\tau_e = 0$ for the ancilla in $\ket{g}$, so that no compensating ancilla rotation is needed in~\cref{eq:protocol_CR_matching}. For the jump trajectories, it is global and therefore unphysical.
 
Two features of~\cref{eq-apdx:pm_flag_position} are essential. First, the flag returns to vacuum only for $\tau_e \in \{0,T\}$, \ie for the no-jump trajectory and for a measure-zero set of jump trajectories. Second, the flag position and the residual memory rotation of~\cref{eq-apdx:pm_residual_angle} are two functions of the \emph{same} quantity $\tau_e$, so that the former determines the latter. Eliminating $\tau_e$ gives a single encoding map, common to the four families,
\begin{equation} \label{eq-apdx:pm_encoding_map}
    \tilde z(\phi) = \flag \left( 1 - e^{i\pi\phi/\theta} \right),
    \qquad \phi \in [-2\theta, 2\theta],
\end{equation}
which is the map of~\cref{eq:map_chi_match}, now extended to positive residual angles. Attaching these flag states to the trajectories of~\cref{eq-apdx:CR_cont_err_channel_pm}, the flagging channel of the $\chi$-matched sequence under ancilla decay and excitation reads
\begin{equation} \label{eq-apdx:Kraus_map_protocol_CR_pm}
\begin{aligned}
    \mathcal{F}(\hrho) = \R^{\mathrm{nj}}_0 \hrho (\R^{\mathrm{nj}}_0)^\dagger \otimes \ketbra{0}
    + \sum_{\mu \in \{g,e\}} \Bigg[ &\int_{-2\theta}^{0} d\phi \; F_{\mu e}(\phi) \, \R^{\mu e}_\phi \hrho (\R^{\mu e}_\phi)^\dagger \otimes \ketbra{\tilde z(\phi)} \\
    + &\int_{0}^{2\theta} d\phi \; F_{\mu g}(\phi) \, \R^{\mu g}_\phi \hrho (\R^{\mu g}_\phi)^\dagger \otimes \ketbra{\tilde z(\phi)} \Bigg].
\end{aligned}
\end{equation}
This is the channel of~\cref{eq-apdx: Kraus map protocol CR} at $\ftheta = \pi$, with its single jump term replaced by the four families of~\cref{eq-apdx:superpropagator_pm}, and it reduces to it for $\kappa_+ = 0$.
 
The map $\tilde z$ of~\cref{eq-apdx:pm_encoding_map} is $2\theta$-periodic and injective on $(-2\theta,0]$. The flag therefore determines the residual memory rotation \emph{modulo} $2\theta$, and nothing more. A trajectory starting in $\ket{e}$ with residual angle $\phi$ and a trajectory starting in $\ket{g}$ with residual angle $\phi + 2\theta$ share the same $\tau_e$, and leave the flag in the same state. Likewise, the jump trajectories with $\tau_e \in \{0,T\}$, which leave the memory rotated by $0$ or $\pm 2\theta$, cannot be distinguished from the no-jump one. These degeneracies are harmless. As discussed in~\cref{sec:cv_flags}, the correction only has to return the state to the code space up to a stabilizer, and $\R(\pm 2\theta)$ typically corresponds to a stabilizer in the context of rotation-symmetric codes, see~\cref{subsec:intro_CR}. Up to this stabilizer, a heterodyne outcome distributed around $\tilde z(\phi)$ thus carries the same information about the memory rotation as in the decay-only case, and the measure-and-feedback protocol of~\cref{sec:cv_flags} applies unchanged. The flag does not resolve the final ancilla state $\mu$ either, which is irrelevant as well: at fixed $\nu$ and $\phi$, the families $\ketbra{g}{\nu}$ and $\ketbra{e}{\nu}$ act identically on the memory and only differ through their ancilla operator. Once the rotation has been undone up to $\R(2\theta)$, the remaining noise therefore acts on the ancilla alone, which is the discretization announced in~\cref{sec:cv_flags}. The correlation condition of~\cref{eq:correlation_condition} thus holds for any sequence of ancilla decays and excitations, which establishes the statement made at the end of~\cref{subsubsec:chi-match}: the $\chi$-matched flag encodes the memory rotation to all orders in $\kappa_- T$ and $\kappa_+ T$.
 
This all-order property is specific to the $\chi$-matched sequence. Away from $\ftheta = \pi$, the intermediate displacement of~\cref{eq:CR_unmatched} breaks the above argument at second order. The flag position then depends separately on the times spent in $\ket{e}$ during each half of the gate, and not only on their sum $\tau_e$, while the memory rotation still depends on $\tau_e$ alone. The resulting two-parameter family of flag positions folds onto itself, so that multi-jump trajectories with different residual rotations can leave the flag in the same state. The single-decay encoding of~\cref{eq-apdx:CR_flag_positions}, corresponding to the $p=1$ term of $G_{ge}$, is unaffected.
 
\subsection{Conditional displacement gate under qubit decay and excitation}
\label{subsec-apdx:pm_CD}
 
\subsubsection{Continuous error channel}
\label{subsubsec-apdx:pm_CD_error_channel}
 
The ECD sequence of~\cref{eq:ECD_def} contains two dispersive intervals separated by the echo, which changes the picture qualitatively. We keep the notations of~\cref{subsubsec-apdx:CD_error_channel}, $\alpha = \alpha_\lambda$, $\theta = \theta_\lambda$, $\mathrm{c} = \cos\theta$ and $\mathrm{c}_2 = \cos 2\theta$, with $T = 2\theta/\chi_m$, and ignore the flag until~\cref{subsubsec-apdx:pm_CD_flagging}. By~\cref{eq-apdx:superpropagator_pm}, each conditional rotation contributes its no-jump propagator and the four jump families, the latter depending on the jump times only through the time spent in $\ket{e}$ during that rotation. We denote by $t_1$ and $t_2$ these times for the first and second conditional rotations, respectively.
 
The no-jump propagator of a single conditional rotation is not of the form of the family terms of~\cref{eq-apdx:superpropagator_pm}: since it is diagonal in $\{\ket{g},\ket{e}\}$, it reads $\bm{W}(T,0) = e^{-\kappa_- T/2}\,\bm{J}^{ee}(T) + e^{-\kappa_+ T/2}\,\bm{J}^{gg}(0)$, a coherent superposition of the two branches, whose cross terms carry the ancilla coherence preserved by the ideal gate. On an input with a definite ancilla state $\ket{\nu}$, however, only one branch survives, $\bm{W}(T,0)\,(\Id\otimes\ketbra{\nu}) = e^{-\kappa_\nu T/2}\,\bm{J}^{\nu\nu}(t^{\mathrm{nj}}_\nu)$, with $\kappa_e = \kappa_-$, $\kappa_g = \kappa_+$, $t^{\mathrm{nj}}_e = T$ and $t^{\mathrm{nj}}_g = 0$. On such inputs, the no-jump term of~\cref{eq-apdx:superpropagator_pm} thus becomes a fifth contribution to the family $\ketbra{\nu}$, with the singular density
\begin{equation} \label{eq-apdx:nj_densities}
    G^{\mathrm{nj}}_e(t) = e^{-\kappa_- T}\,\delta_{t - T},
    \qquad
    G^{\mathrm{nj}}_g(t) = e^{-\kappa_+ T}\,\delta_t,
\end{equation}
where the delta functions are fully counted at the boundaries of $[0,T]$. In the ECD sequence, this applies to every trajectory with at least one jump: the rank-one ancilla operator of the conditional rotation containing the jump fixes the ancilla state seen by the other one. Only the trajectory without any jump retains the coherence between the two branches, and is treated separately below.
 
Consider first the memory. By~\cref{eq-apdx:pm_traj_operator}, the two conditional rotations rotate the memory by $\theta(2t_1-T)/T$ and $\theta(2t_2-T)/T$. Commuting the displacements through the rotations with $\R(\psi)\D(\gamma) = \D(\gamma e^{i\psi})\R(\psi)$, as in~\cref{subsubsec-apdx:CD_error_channel}, the memory part of the trajectory operator reads, up to a global phase,
\begin{equation} \label{eq-apdx:CD_pm_memory}
    \D(\alpha \mathrm{c}_2)\,\R\!\left(\theta\tfrac{2t_2-T}{T}\right)\D(-2\alpha \mathrm{c})\,\R\!\left(\theta\tfrac{2t_1-T}{T}\right)\D(\alpha)
    \;\propto\;
    \D\!\left(\alpha\left[\mathrm{c}_2 - 2\mathrm{c}\,e^{i\theta\frac{2t_2-T}{T}} + e^{2i\theta\frac{t_1+t_2-T}{T}}\right]\right)\R\!\left(2\theta\,\tfrac{t_1+t_2-T}{T}\right),
\end{equation}
where the two displacements $\D(-\alpha c)$ framing the echo have been merged, since the echo acts trivially on the memory. Unlike for the conditional rotation gate, the trajectory thus enters the memory error through two independent quantities, $t_2$ and $t_1+t_2$. The damping factors, on the other hand, accumulate over the total time $t_1+t_2$ spent in $\ket{e}$ and $2T - t_1 - t_2$ spent in $\ket{g}$.
 
Consider next the ancilla. A trajectory starting in $\ket{\nu}$ is taken to $\ket{\mu}$ by the first conditional rotation, the echo flips it to $\ket{\bar\mu}$, the second conditional rotation takes it to $\ket{\bar\omega}$, and the closing $\hsigma_x$ of~\cref{eq:link_CD_and_ECD} flips it to $\ket{\omega}$, with $\bar g = e$ and $\bar e = g$. The trajectory operator is thus proportional to $\ketbra{\omega}{\nu}$, which defines the four families of the gate. Since the rate prefactors and damping factors of~\cref{eq-apdx:pm_traj_operator} multiply over the two conditional rotations, the weight of a trajectory factorizes. Distinguishing the trajectories with jumps during the first conditional rotation only, during the second only, or during both, the joint density of $(t_1,t_2)$ in the family $\ketbra{\omega}{\nu}$ is obtained by composing the densities of~\cref{eq-apdx:transfer_weights,eq-apdx:nj_densities},
\begin{equation} \label{eq-apdx:CD_pm_joint_weight}
    F_{\omega\nu}(t_1,t_2) = G^{\mathrm{nj}}_{\bar\omega}(t_2) \, G_{\omega\nu}(t_1) + G_{\bar\omega\bar\nu}(t_2) \, G^{\mathrm{nj}}_{\nu}(t_1) + \sum_{\mu \in \{g,e\}} G_{\bar\omega\bar\mu}(t_2) \, G_{\mu\nu}(t_1).
\end{equation}
The trajectory without any jump is the composition of the two no-jump propagators. Since the echo swaps the two branches, each of them spends one conditional rotation in $\ket{e}$ and one in $\ket{g}$, and the damping is symmetrized as in~\cref{subsubsec-apdx:CD_error_channel}. This trajectory thus contributes the Kraus operator $e^{-\frac{(\kappa_-+\kappa_+)T}{2}}\,\Id$, and introduces no error on the memory.
 
\paragraph{Large intermediate displacements limit.} In the limit $\lambda \to \infty$, the bracket of~\cref{eq-apdx:CD_pm_memory} vanishes at zeroth order in $\theta$, and its first-order term is $2i\theta(t_1-t_2)/T$. Using $\alpha = -i\lambda\beta$ and $\lambda\theta \to 1/2$, as in~\cref{subsubsec-apdx:CD_error_channel}, we find
\begin{equation} \label{eq-apdx:CD_pm_large_lambda}
    \alpha\left[\mathrm{c}_2 - 2\mathrm{c}\,e^{i\theta\frac{2t_2-T}{T}} + e^{2i\theta\frac{t_1+t_2-T}{T}}\right]
    \xrightarrow[\lambda\to\infty]{} \beta \, \frac{t_1 - t_2}{T},
    \qquad
    2\theta\,\frac{t_1+t_2-T}{T} \xrightarrow[\lambda\to\infty]{} 0,
\end{equation}
so that the memory displacement depends on the difference $t_1 - t_2$ only, while the residual rotation vanishes. Factoring out the target unitary, $\CD^\dagger(\beta)$ contributes $\D(\mp\beta)$ for $\nu = e/g$, and commutes with the residual rotation since the latter vanishes. A trajectory of the family $\ketbra{\omega}{\nu}$ thus displaces the memory by $s\beta$, with
\begin{equation} \label{eq-apdx:CD_pm_s}
    s = \frac{t_1 - t_2}{T} - 1 \quad \text{for } \nu = e,
    \qquad
    s = \frac{t_1 - t_2}{T} + 1 \quad \text{for } \nu = g.
\end{equation}
Altogether, the continuous error channel of the conditional displacement under ancilla decay and excitation reads
\begin{equation} \label{eq-apdx:CD_cont_err_channel_pm}
    \mathcal{N}_{D}(\hrho) = F_{\mathrm{nj}} \, \D^{\mathrm{nj}}_0 \hrho (\D^{\mathrm{nj}}_0)^\dagger
    + \sum_{\omega,\nu \in \{g,e\}} \int_0^T \! dt_1 \int_0^T \! dt_2 \; F_{\omega\nu}(t_1,t_2) \, \D^{\omega\nu}_{s\beta} \hrho (\D^{\omega\nu}_{s\beta})^\dagger,
    \qquad
    F_{\mathrm{nj}} = e^{-(\kappa_-+\kappa_+)T},
\end{equation}
with $s$ given by~\cref{eq-apdx:CD_pm_s}, and where the Kraus operators $\D^{\omega\nu}_{s\beta} = \D(s\beta)\otimes\ketbra{\omega}{\nu}$ extend those of~\cref{eq-apdx:CD_cont_err_channel_Kraus_Ops} to the family $\ketbra{g}{g}$.
 
Setting $\kappa_+ = 0$ recovers~\cref{eq-apdx:CD_cont_err_channel}. $G_{eg}$, $G_{ee}$ and $G_{gg}$ then vanish, so that $F_{gg} = 0$, and the three surviving families are
\begin{equation} \label{eq-apdx:CD_pm_decay_only}
\begin{array}{lll}
    F_{ge}(t_1,t_2) = G_{ge}(t_1)\,G^{\mathrm{nj}}_e(t_2), & t_2 = T, \quad s = t_1/T - 2 \in [-2,-1], & \text{(jump before echo)} \\[2pt]
    F_{eg}(t_1,t_2) = G^{\mathrm{nj}}_g(t_1)\,G_{ge}(t_2), & t_1 = 0, \quad s = 1 - t_2/T \in [0,1], & \text{(jump after echo)} \\[2pt]
    F_{ee}(t_1,t_2) = G_{ge}(t_1)\,G_{ge}(t_2), & s = (t_1-t_2)/T - 1 \in [-2,0], & \text{(double jumps)}
\end{array}
\end{equation}
These are the three jump terms of~\cref{eq-apdx:CD_cont_err_channel}, with the weight functions of~\cref{eq-apdx:CD_cont_err_channel_Kraus_Ops}, once the jump times are restored through $\tau = t_1$ before the echo, $\tau = T + t_2$ after it, and $(\tau_1,\tau_2) = (t_1, T + t_2)$ for a double jump.
 
\subsubsection{Flagging channels of the conditional displacement gate}
\label{subsubsec-apdx:pm_CD_flagging}
 
\paragraph{One-flag protocol of~\cref{subsec:CD-one-flag}}
The flag of~\cref{eq:protocol_ECD_one_flag} is displaced only before and after the whole sequence, and, as in~\cref{subsubsec-apdx:CD_flagging}, it simply picks up the net rotation accumulated over the two conditional rotations, $\psi = \ftheta(2t_1-T)/T + \ftheta(2t_2-T)/T$. By~\cref{eq-apdx:CD_flag_map}, it ends in the coherent state
\begin{equation} \label{eq-apdx:CD_pm_flag_position}
    z(t_1,t_2) = \flag\left(e^{2i\ftheta\frac{t_1+t_2-T}{T}} - 1\right),
\end{equation}
exactly for any $\lambda$, the flag being insensitive to the intermediate memory displacements. Attaching these flag states to the trajectories of~\cref{eq-apdx:CD_cont_err_channel_pm}, the flagging channel of the one-flag protocol, in the limit $\lambda \gg 1$ and to all orders in $\kappa_- T$ and $\kappa_+ T$, reads
\begin{equation} \label{eq-apdx:CD_oneflag_flagging_channel_pm}
    \mathcal{F}_{D}(\hrho) = F_{\mathrm{nj}} \, \mathcal{D}^{\mathrm{nj}}_0(\hrho) \otimes \ketbra{0}
    + \sum_{\omega,\nu \in \{g,e\}} \int_0^T \! dt_1 \int_0^T \! dt_2 \; F_{\omega\nu}(t_1,t_2) \, \mathcal{D}^{\omega\nu}_{s\beta}(\hrho) \otimes \ketbra{z(t_1,t_2)},
\end{equation}
with the displacement channels $\mathcal{D}^{k}_{s\beta}(\hrho) = \D^{k}_{s\beta}\,\hrho\,(\D^{k}_{s\beta})^\dagger$ of~\cref{eq:CD_oneflag_flagging_channel}. This is~\cref{eq-apdx:CD_oneflag_flagging_channel} with its three jump terms replaced by four two-dimensional families, and it reduces to it for $\kappa_+ = 0$.
 
\Cref{eq-apdx:CD_pm_s,eq-apdx:CD_pm_flag_position} contain the key difference with the conditional rotation gate: \emph{the memory displacement is governed by $t_1 - t_2$, while the flag records $t_1 + t_2$}. These are two independent quantities, so that a single flag cannot determine the memory displacement in general. By contrast, the residual rotation of~\cref{eq-apdx:CD_pm_memory} and the damping depend on $t_1+t_2$ only, and are therefore always flagged.
 
With decay only, the flag nevertheless determines the memory displacement at first order, because a single decay saturates one of the two times, see~\cref{eq-apdx:CD_pm_decay_only}. A decay before the echo imposes $t_2 = T$, hence $s = t_1/T - 2$ and $\psi = 2\ftheta\,t_1/T \ge 0$, whereas a decay after the echo imposes $t_1 = 0$, hence $s = 1 - t_2/T$ and $\psi = 2\ftheta(t_2 - T)/T \le 0$, which are the flag angles of~\cref{eq-apdx:CD_flag_angles}. As long as $\ftheta < \pi/2$, the two branches are distinguished by the sign of $\psi$, and each of them is injective. It is the double-decay family, for which $t_1$ and $t_2$ are both free, that spoils the encoding at second order, as discussed in~\cref{subsec:CD-one-flag}.
 
Excitation events remove the first-order property as well. A single excitation before the echo imposes $t_2 = 0$, and leaves the same ancilla operator $\ketbra{e}{g}$ as a single decay after the echo, which imposes $t_1 = 0$. Both trajectories leave the flag in the same state when their nonzero times are equal, $t_1 = t_2 \equiv t$, while displacing the memory by
\begin{equation} \label{eq-apdx:CD_pm_degeneracy}
    s = 1 + \frac{t}{T} \quad \text{(excitation before the echo)},
    \qquad
    s = 1 - \frac{t}{T} \quad \text{(decay after the echo)}.
\end{equation}
The two trajectories thus share the same arc of~\cref{eq-apdx:CD_pm_flag_position}, but their memory displacements differ by $2t\beta/T$, which vanishes or is the stabilizer $\D(2\beta)$ only at the endpoints $t \in \{0,T\}$. The mirror pair, a single excitation after the echo ($t_1 = T$, $s = -t_2/T$) and a single decay before it ($t_2 = T$, $s = t_1/T - 2$), populates the other half of the arc in the same way. The single-jump encoding of~\cref{eq:CD_flag_position} therefore relies on the jump being a decay, which fixes which of $t_1$ and $t_2$ is saturated. Once excitations are allowed, the arc is doubly covered, and the one-flag protocol no longer protects the gate even at first order. \\
 
\paragraph{Two-flag protocol of~\cref{subsec:CD-two-flags}}
The two-flag protocol of~\cref{eq:protocol_ECD_two_flags} restores the encoding, at all orders. Its second flag stays in vacuum during the first conditional rotation, and undergoes the $\chi$-matched sequence of~\cref{eq:protocol_CR_matching} during the second one, with $\chi_{f_2} = \pi\chi_m/\theta$. By~\cref{eq-apdx:pm_flag_position}, with $t \to t_2$ and $\flag \to \flag_2$, it therefore ends in the coherent state
\begin{equation} \label{eq-apdx:CD_pm_secondflag}
    w(t_2) = \flag_2 \left( 1 - e^{2i\pi t_2/T} \right),
\end{equation}
the phase left by its displacements vanishing on the no-jump trajectory and being global otherwise, as in~\cref{subsubsec-apdx:pm_CR_flagging}. The map $t_2 \mapsto w(t_2)$ is injective on $[0,T)$, and generalizes the second-flag amplitudes of~\cref{eq-apdx:2flag_gamma_traj} to any sequence of jumps. For $\ftheta < \pi/2$, the first flag, of amplitude $\flag_1$, determines $t_1 + t_2$ through~\cref{eq-apdx:CD_pm_flag_position}, so that the pair of flag states supplies both $t_1 + t_2$ and $t_2$, hence
\begin{equation} \label{eq-apdx:CD_pm_twoflag_inversion}
    t_1 - t_2 = (t_1 + t_2) - 2t_2 .
\end{equation}
The memory displacement of~\cref{eq-apdx:CD_pm_s} is thus determined by the joint flag record, whatever the number and type of ancilla jumps. Two residual ambiguities remain, and both are harmless for a translation-symmetric code. First, the initial ancilla state $\nu$ is not recorded, so that $s$ is only known up to the shift by $2$ between the two cases of~\cref{eq-apdx:CD_pm_s}. Second, the endpoint degeneracy of~\cref{eq-apdx:CD_pm_secondflag}, $t_2 \in \{0,T\}$ both returning the second flag to vacuum, shifts $t_1 - t_2$ by $2T$ in~\cref{eq-apdx:CD_pm_twoflag_inversion}, and hence $s$ by $2$ as well. In both cases, the memory displacement is known up to $\D(2\beta)$, which typically corresponds to a stabilizer in the context of translation-symmetric codes, see~\cref{subsec:intro CD}. Since the feedback of~\cref{sec:cv_flags} only needs the displacement modulo this stabilizer, the correlation condition of~\cref{eq:correlation_condition} holds at all orders in $\kappa_- T$ and $\kappa_+ T$, as claimed at the end of~\cref{subsec:CD-two-flags}. The corresponding flagging channel is~\cref{eq-apdx:CD_oneflag_flagging_channel_pm}, with $\ketbra{0}$ replaced by $\ketbra{0,0}$ and $\ketbra{z(t_1,t_2)}$ by $\ketbra{z(t_1,t_2),\,w(t_2)}$, and it generalizes~\cref{eq-apdx:2flag_flagging_channel} beyond second order and beyond ancilla decay.
 
As for the conditional rotation gate, this all-order property relies on the $\chi$-matching of the second flag. With the unmatched block of~\cref{eq-apdx:2flag_unmatched_block}, the second flag depends separately on the times spent in $\ket{e}$ during the two halves of the second conditional rotation, and the argument of~\cref{subsubsec-apdx:pm_CR_flagging} shows that it no longer determines $t_2$ once multiple jumps are allowed.

\end{document}